\documentclass[referee]{raa}            % referee version: for submission
\usepackage{graphicx,times}             %for PS/EPS graphics inclusion, new
\input{epsf.sty}                        %for PS/EPS graphics inclusion, old
\input{psfig.sty}                       %for PS/EPS graphics inclusion, old

\begin{document}
 \title{A search for massive young stellar objects towards 98 CH$_{3}$OH maser sources}
%   \subtitle{I. Place Your Subtitle Here}

   \volnopage{Vol.0 (200x) No.0, 000--000}      %%preserved for Editor. DOn't remove!
   \setcounter{page}{1}          %%starting page, preserved for Editor. DOn't remove!

   \author{Tie Liu
      \and Yue-Fang Wu
      \mailto{}
%% Please move "\mailto{}" to the corresponding author of the paper.
   \and Ke Wang
            }
    \offprints{Tie Liu}                   %% is disabled in fact

   \institute{Department of Astronomy, Peking University,
             Beijing 100871, China\\
             \email{yfwu@vega.bac.pku.edu.cn}
%% offprint requests will be sent. Note to pair \mailto{} with \email{}
          }

   \date{Received~~200x month day; accepted~~200x~~month day}

   \abstract{
Using the 13.7 m telescope of Purple Mountain Observatory (PMO), a
survey of J=1-0 lines of CO and its isotopes was carried out towards
98 methanol maser sources in January 2008. Eighty-five sources have
infrared counterparts within one arcmin. In the survey, except 43
sources showing complex or multiple-peak profiles, almost all the
$^{13}$CO line profiles of the other 55 sources have large line
widths of 4.5 km~s$^{-1}$  on average and are usually asymmetric.
Fifty corresponding Infrared Astronomical Satellite (IRAS) sources
of these 55 sources are with $L_{bol}$ larger than
$10^{3}L_{\odot}$, which can be identified as possible high-mass
young stellar sources. Statistics show that the $^{13}$CO line
widths correlate with the bolometric luminosity of the associated
IRAS sources. We also report the mapping results of two sources:
IRAS 06117+1350 and IRAS 07299-1651 here. Two cores were found in
IRAS 06117+1350 and one core was detected in IRAS 07299-1651. The
northwest core in IRAS 06117+1350 and the core in IRAS 07299-1651
can be identified as precursors of UC~H{\sc ii} regions or high-mass
protostellar objects (HMPOs). The southeast core of IRAS 06117+1350
has no infrared counterpart, seeming to be on younger stages than
pre-UC~H{\sc ii} phase.
   }

   \authorrunning{Tie Liu, Y. F. Wu \& Ke Wang }            %author_head in even pages
   \titlerunning{A search for massive star formation regions towards 98 methanol sources}  % title_head in odd pages

   \maketitle
%% The author head (on even pages) and the title head (on odd pages) will be
%% automatically extracted from \author{} and \title{}. Whenever the title is too long,
%% you will be asked to supply a shorter one by inserting either \authorrunning{} or
%% \titlerunning{} before \maketitle. Anyway, you can specify your own heads in advance.
%%
%%
%% Note: In the following text body of your manuscript, please note several differences from
%%       other major journals:
%% (1) \subsection{Please Capitalize the First Letter of Each Notional Word in Subsection Title}
%% (2) Please Capitalize the First Letter of Each Notional Word in table's caption

%
%________________________________________________ sections below
%

\section{Introduction}           %% first-level sections will be auto-capitalized
\label{sect:intro}
%\hspace{15pt}%                   %% preserved for Editor
The formation and evolution of massive stars is still in mystery and
our understanding of how high-mass stars form and evolve lags behind
that of low-mass stars, which is understood better using the theory
constructed by Shu et al.~\cite{shu87}. High-mass stars can produce
an enormous impact on their local environment and the evolution of
the whole Galaxy. However, due to their far-away location,
generation in clusters and short evolutionary time scales, it is
difficult to find samples of high-mass young sources to investigate
their forming processes.

In the past, interstellar H$_{2}$O maser (e.g. Plume et
al.~\cite{plu92}; Wu et al.~\cite{wu06}) was used as tracers for
high-mass star formation regions in early evolutionary phase.
However H$_{2}$O maser can also be found in low-mass star formation
regions (e.g.,Wu et al.~\cite{wu04}), while methanol (CH$_{3}$OH)
masers seem to exclusively associate with high-mass star formation
regions (e.g. Minier et al.~\cite{min03}). Methanol masers are
traditionally divided into two classes. Class $\rm II$ methanol
masers are found in the vicinity of the high-mass young stellar
objects, while Class I methanol masers are believed to trace distant
parts of the outflows from these high-mass star formation regions (
Sobolev et al.~\cite{sob05}). With the motivation of finding more
pre-UC~H{\sc ii} regions or even more earlier high-mass stellar
objects, we carried out a survey of J=1-0 lines of ${}^{12}$CO and
its isotopes ${}^{13}$CO and C$^{18}$O towards 98 methanol maser
sources. Next section describes the observations. Survey results and
mapping results will be given and discussed in section 3. Section 4
summarizes the paper.

%% ChJAA editors DID NOT use \cite{} for citation, \ref and \label for
%% cross-references of Table/Figure in publication version.
%% ChJAA editors prefered you giving a citation as 'Michel et al. 1992', and
%% writting Table~1 or Fig.~1 and so forth. However, that will make authors
%% inconvenient in adjusting/adding/removing text, tables or figures. Anyway,
%% authors can use \cite, \citep and \citet as widely used in other journals.
%% ChJAA editors are moving to use a more flexible LaTeX source.

\section{Observations}
\label{sect:Obs}
%\hspace{15pt}%                   %% preserved for Editor

The observations were made in January 2008 with the 13.7 m telescope
of PMO at Qinghai Station. The ${}^{12}$CO, ${}^{13}$CO and
C$^{18}$O (J=1-0) lines were observed simultaneously by a
superconductor receiver. The back-end is equipped with three
acousto-optic spectrometers (AOSs). Every spectrometer has 1024
channels. The total bandwidths were 145.330 MHz, 42.762 MHz and
43.097 MHz for ${}^{12}$CO, ${}^{13}$CO and C$^{18}$O lines,
corresponding velocity resolutions of 0.37, 0.11 and 0.12
km~s$^{-1}$, respectively. The system temperatures during the
observation ranged from 200 K to 350 K (SBD), depending on the
weather. The half-power beamwidth (HPBW) at 112 GHz was $\sim$
$59^{\prime\prime}$. The pointing accuracy of the telescope was
better than $8^{\prime\prime}$, and the main beam efficiency at the
zenith was about 0.67.

We adopted the position switch mode for all the spectral
measurement. Our mapping steps toward right ascension and
declination directions were both 1 arcmin. The integrating time was
$\sim$3 minutes per position. The noise level of antenna temperature
$T_{A}^{\phantom{A}*}$ was usually about 0.4 K for the $^{12}$CO,
0.3 K for the $^{13}$CO and 0.2 K for the C$^{18}$O (J=1-0) band,
respectively. For the data analysis, the GILDAS software package
including CLASS and GREG was employed (Guilloteau \&
Lucas~\cite{gui00}).

\section{results and discussion}
\subsection{Sample analysis}

Ninety-eight sources were observed, 29 of which belong to the Class
I methanol masers. Table 1 presents basic parameters of all the
sources surveyed. The associated IRAS sources within a
$3^{\prime}$$\times$$3^{\prime}$ are presented in column 2.
Eighty-five have IR counterpart candidates within one arcmin.
Columns 3-6 give the equatorial coordinates and the Galactic
coordinates, respectively. The color indices log($F_{25}$/$F_{12}$)
and log($F_{60}$/$F_{12}$) are presented in columns 7-8 ($F_{12}$,
$F_{25}$, and $F_{60}$ represent the flux at 12$\mu$m, 25$\mu$m and
60$\mu$m, respectively.). Column 9 lists the fluxes at 100$\mu$m.
The types of these methanol masers and the related references were
placed in columns 10-11.

Figure 1 shows the distribution of the sources surveyed in Galactic
coordinates. The sample sources concentrate on the Galactic plane
($|b|<$$2^{\circ}$) and about 90\% are found in the first quadrant.
Only the Orion's source is with $|b|>$$15^{\circ}$, and 72 sources
are located in $0<l<45^{\circ}$.

The locations of the associated IRAS sources in the color-color
planes are plotted in Figure 2. The distributions show that about
75\% of the Class I and 63\% of the Class $\rm II$ methanol maser
sources observed are with the IRAS color indices satisfying the
criteria of the UC~H{\sc ii} regions established by Wood \&
Churchwell~\cite{woo89}. All of these UC~H{\sc ii} region candidates
except Orion's source are with flux larger than 100 Jy at 100
$\mu$m.

\subsection{Survey results and discussion}
 \label{sect:Results}
%\hspace{15pt}%                   %% preserved for Editor
All the 98 sources are detected with J=1-0 lines of the $^{12}$CO,
$^{13}$CO and C$^{18}$O. Forty-three sources hold too complex line
profiles and will be studied further. The other fifty-five sources
were analyzed, including 17 Class $\rm I$ methanol maser sources.
Figure 3 presents the spectra of all the 55 sources and one source
IRAS 18403-0417 with blended lines. The grey, green and red lines
represent $^{12}$CO, $^{13}$CO and C$^{18}$O respectively. The
$^{12}$CO spectra of IRAS 18414-1723, IRAS 18353-0628, G32.74-0.07
and G39.10+0.48 are blended, but the relevant $^{13}$CO emission
spectra can be well distinguished. So these four sources are also
analyzed. In the 55 sources, we detected significant $^{12}$CO and
$^{13}$CO signals, but only 60\% in them showed obvious C$^{18}$O
signals. No differences are found in the detection rate of J=1-0
lines of CO isotopes between class $\rm I$ and class $\rm II$
methanol maser sources. These 55 sources present profuse $^{13}$CO
line profiles. About 40\% of them hold wings and nearly 20 sources
have more than one components. We fitted these $^{13}$CO lines with
a Gaussian function. We distinguish between several different
non-Gaussian $^{13}$CO line profiles with the following characters:

(a)wings; (b)red wing; (c)blue wing; (d)red shoulder; (e)red
asymmetry; (f)blue asymmetry; (g)flat top; (h)two or three
components.

The identities of these $^{13}$CO spectra characters are presented
in column 12 of Table 2. Possible clues of high velocity gas were
detected in five sources of Class $\rm I$ and nine sources of Class
$\rm II$ methanol maser sources (see the last column of Table 2).
There seem no differences in high velocity gas detection rate
between Class $\rm I$ and Class $\rm II$ methanol maser sources.
Detailed investigations with denser molecular probers such as HCN
and CS are needed.

Observation parameters including the antenna temperature
$T_{A}^{\phantom{A}*}$, $V_{LSR}$ and $^{13}$CO line width (FWHM) of
each components were obtained and shown in columns 2-4 of Table 2.
Columns 5-6 list the distances from the Galactic center (R) and the
Sun (D). Nine components cannot provide available distances from
fitting Galactic rotational curve. Column 7 presents the bolometric
luminosity calculated with the formation in Casoli et
al.~\cite{cas86}:
\begin{equation}
  L_{(5-1000\mu~m)}=4\pi~D^{2}\times1.75\times(F_{12}/0.79+F_{25}/2+F_{60}/3.9+F_{100}/9.9)
\end{equation}
where D is the distance. Fifty of the 55 sources are with
$L_{bol}$$>$$10^{3}L_{\odot}$, supposed to be massive star formation
regions.

Assuming that ${}^{12}$CO is optically thick, we derived the excited
temperatures $T_{ex}$ following Garden et al.~(\cite{gar91}).
Assuming ${}^{13}$CO is optically thin, then the optical depth and
column density of ${}^{13}$CO can be straightforwardly obtained
under local thermal equilibrium assumption (LTE). With the abundance
ratio [H$_{2}$]/[$^{13}$CO]=$8.9\times10^{5}$, the column density of
H$_{2}$ was calculated. The results are listed in columns 8-11.

Figure 4 presents the plot of bolometric luminosity $L_{bol}$ vs.
$^{13}$CO line widths of the 55 sources. The linear fit gives:
$log(L_{bol}/L_{\odot})=2.5log(FWHM/km~s^{-1})+3.2$; the correlation
coefficient r=0.52. From column 4 of Table 2 we can see almost all
the $^{13}$CO lines except one component of the source G32.74-0.07
are with line widths larger than 1.3 km~s$^{-1}$, the typical line
width for low-mass sources according to Myers et al.(\cite{mye83}).
And 57 out of these 79 resolved components are with $^{13}$CO line
widths larger than 3 km~s$^{-1}$, indicating possible high-mass star
formation regions ( Wu et al.~\cite{wu03}). The average line width
of $^{13}$CO is 4.5 km~s$^{-1}$, similar to that detected by Purcell
et al.~\cite{pur09}. These results indicate that sources with large
line widths tend to form high-mass stars. This trend is consistent
with the results of Wang et al.~\cite{Wan09}. We also fitted the
C$^{18}$O lines of 30 sources and found they hold line widths of 4
km~s$^{-1}$ on average, smaller than the $^{13}$CO lines,
respectively. The $^{13}$CO lines antenna temperatures of about one
half of these 30 sources are larger than 5.5 times of C$^{18}$O
lines, and about 75\% of the integrated intensities of $^{13}$CO
lines are larger than 5 times of C$^{18}$O lines, respectively.
However, the center velocities of C$^{18}$O lines match with
$^{13}$CO lines very well, suggesting that $^{13}$CO emissions and
C$^{18}$O emissions may be from the same region in the cloud.

\subsection{Mapping results and discussion}
%\hspace{15pt}%                   %% preserved for Editor
Two sources IRAS 06117+1350 and IRAS 07299-1651 were mapped. Figure
5 presents the contours of the integrated intensities of $^{13}$CO
lines. IRAS 06117+1350 has two cores and we mark them with
06117+1350-SE and 06117+1350-NW. One core was detected in IRAS
07299-1651. The positions of IRAS sources are marked with "+" and
the Midcourse Space Experiment ( MSX ) sources are visualized with
triangles.

We calculated the masses of the cores under the LTE assumption. We
also calculated the virial mass using the equation from Ungerechts
et al.~\cite{ung00}. The physical results of these cores are listed
in Table 3. The two cores in IRAS 06117+1350 have virial mass
smaller than the core mass, indicating they are stable and the gas
motions are gravitationally bound. Specially, all of our cores are
far away from UC~H{\sc ii} regions. No Spitzer / IRAC data are
available for the mapped sources. The core in IRAS 07299-1651 is
with the IRAS source and three MSX sources. The northwest core of
IRAS 06117+1350 is far away from any IRAS source but associated with
one MSX source about $0.5^{\prime}$ from the $^{13}$CO emission
peak. Since no 6 cm emissions associate with these two cores, they
can be identified as HMPOs or UC~H{\sc ii} precursors (e.g. Wu et
al.~\cite{wu06} and the references there in). The southeast core of
IRAS 06117+1350 is separated from any IRAS source and MSX source,
indicating an evolutionary stages earlier than pre-UC~H{\sc ii}.

\section{summary}
\label{sect:Discussion and Conclusion}
%\hspace{15pt}%                   %% preserved for Editor

A survey in CO and its isotopes to the locations traced by methanol
masers was performed. Fifty-five sources are further studied. They
all have J=1-0 emissions of $^{12}$CO and $^{13}$CO, but only 60\%
of them have significant C$^{18}$O emissions. No differences in both
CO detection rate and high velocity gas detection rate were found
between Class $\rm I$ and Class $\rm II$ methanol maser sources.
Fifty of these 55 sources are associated with luminous IRAS sources
($L_{bol}>10^{3}L_{\odot}$). The relationship between the bolometric
luminosity of the associated IRAS sources and $^{13}$CO line widths
can be well described as:
$log(L_{bol}/L_{\odot})=2.5log(FWHM/km~s^{-1})+3.2$. About three
quarters of the resolved components have $^{13}$CO line widths much
larger than 3 km~s$^{-1}$. These results show that molecular cores
with large line widths tend to form high-mass stellar objects. In 30
sources, $^{13}$CO lines have larger line widths and antenna
temperatures than C$^{18}$O lines. But the center velocities of
C$^{18}$O lines match with $^{13}$CO lines very well, suggesting
that $^{13}$CO emissions and C$^{18}$O emissions may be from the
same region in the cloud.

Two cores are detected in IRAS 06117+1350 and one core in IRAS
07299-1651. The two cores in IRAS 06117+1350 have virial mass less
than the core mass, but the core in IRAS 07299-1651 has virial mass
larger than the core mass. The core in IRAS 07299-1651 and the
northwest core of the source IRAS 06117+1350 may be pre-UC~H{\sc ii}
or HMPOs and the southeast core of IRAS 06117+1350 revealed seems to
be on an earlier evolutionary stage than pre-UC~H{\sc ii} phase.

\begin{acknowledgements}
We are grateful to all the staff of Qinghai station of PMO for their
assistances during the observations. N.C. Sun and Y.A. Mao are also
deserved our thanks for their help and discussions. This project is
supported by NSFC 10733030 and 10873019.
\end{acknowledgements}

\clearpage

\clearpage
\begin{figure}
  \centering
  \includegraphics[width=7cm]{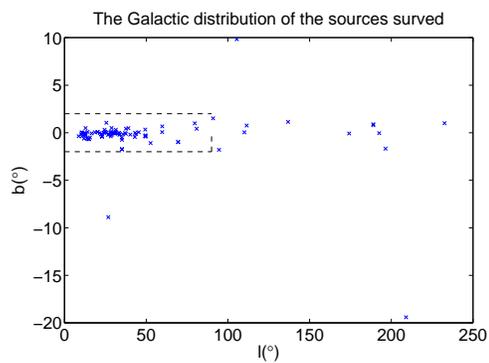}
\caption{The Galactic distribution of the sources surveyed. The
dashed lines indicate the region with $|b|<$$2^{\circ}$ and
$0<l<90^{\circ}$. Nearly all the sources crowd to the Galactic
plane, and in the longitude range $0^{\circ}$ to $60^{\circ}$ the
plane is so densely populated.}
\end{figure}

\begin{figure}
\begin{minipage}[c]{0.45\textwidth}
  \centering
  \includegraphics[width=6.5cm]{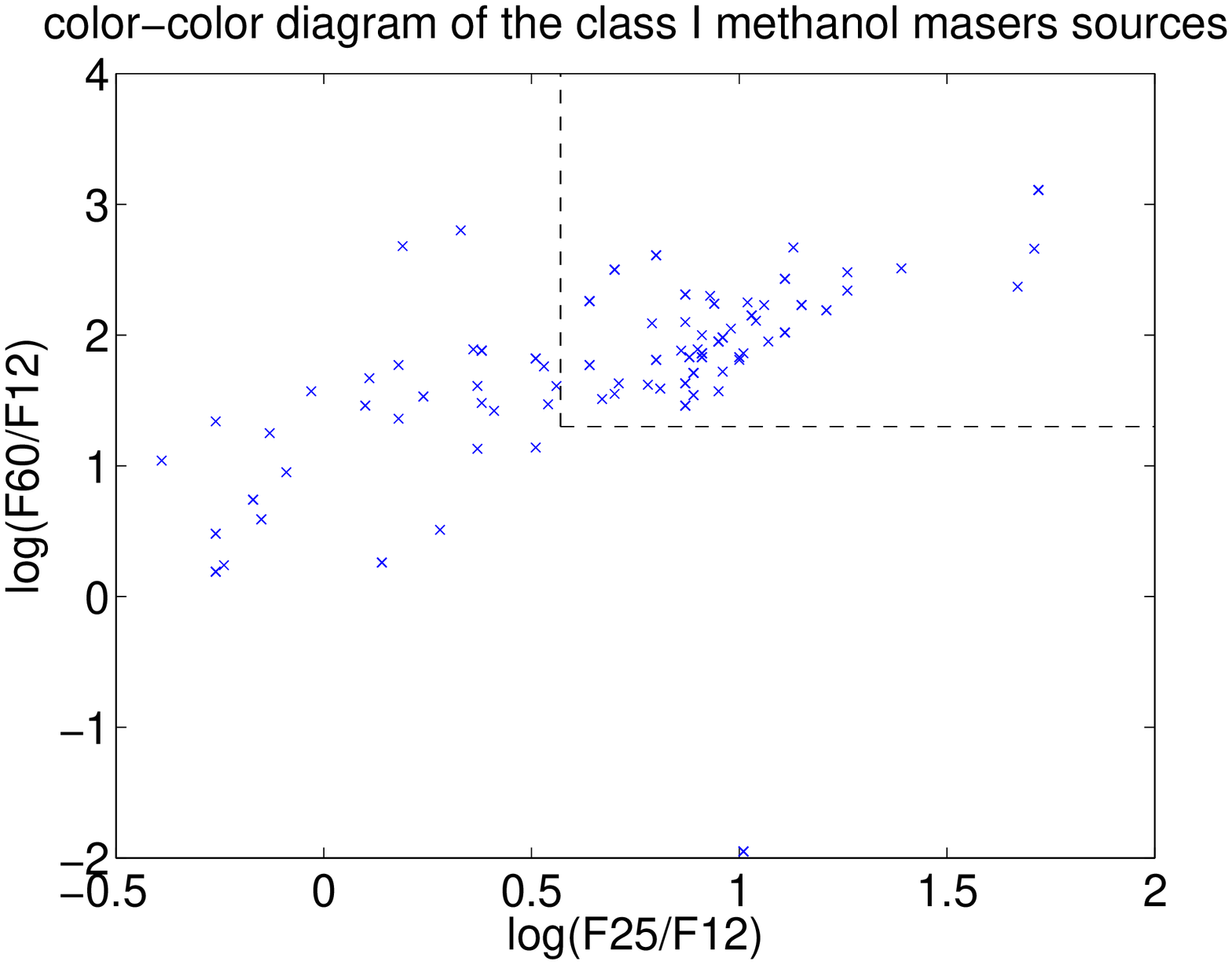}
\end{minipage}
\begin{minipage}[c]{0.45\textwidth}
  \centering
  \includegraphics[width=6.5cm]{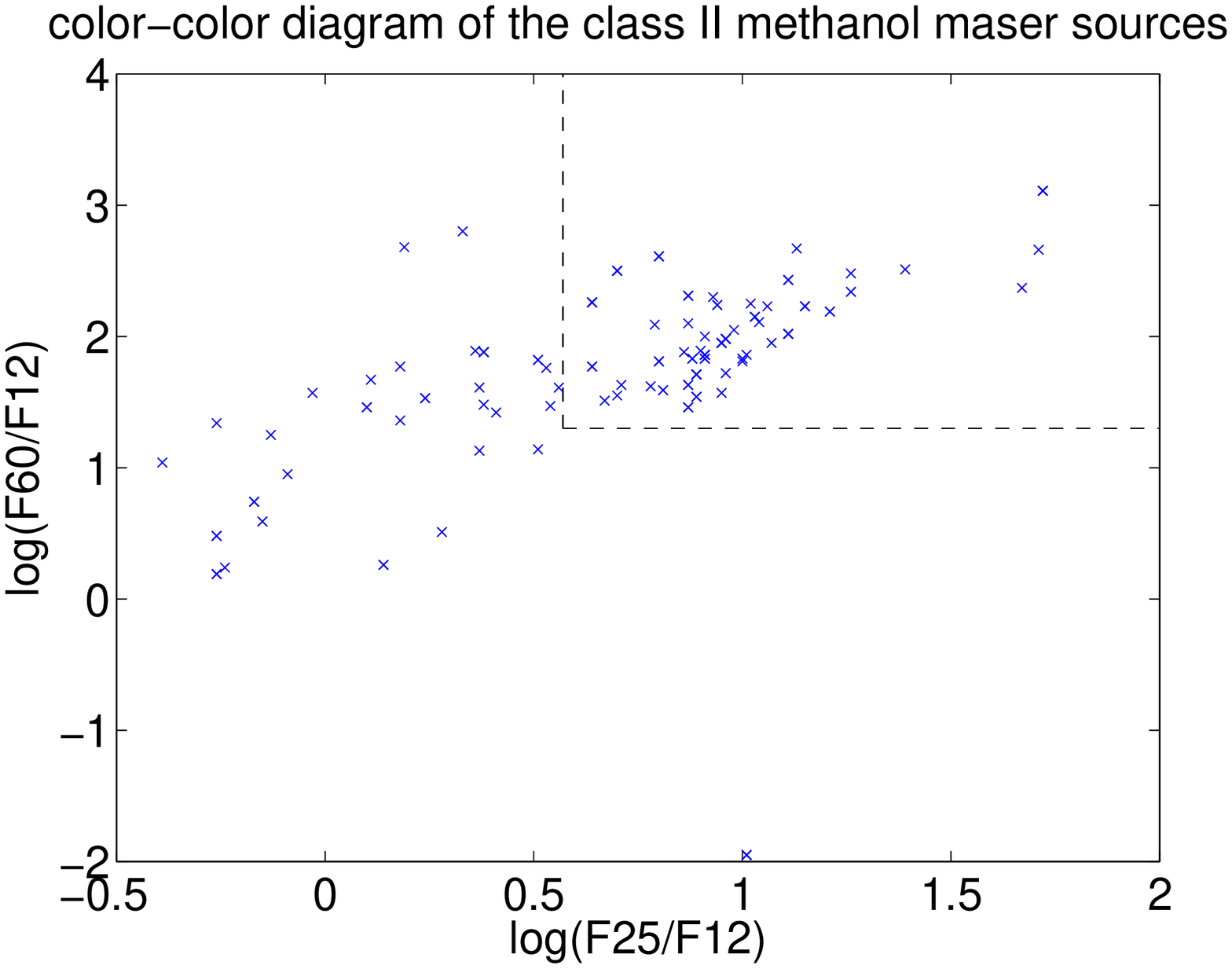}
\end{minipage}
\caption{The left Panel describes the color-color distribution of
the 29 Class I methanol maser sources and the right Panel is for the
69 Class $\rm II$ methanol maser sources. The color box of UC~H{\sc
ii} regions established by Wood \& Churchwell 1989 is indicated by
the dashed lines in both panels.}
\end{figure}

\begin{figure}
\begin{minipage}[c]{0.5\textwidth}
  \centering
  \includegraphics[width=30mm,height=65mm,angle=-90.0]{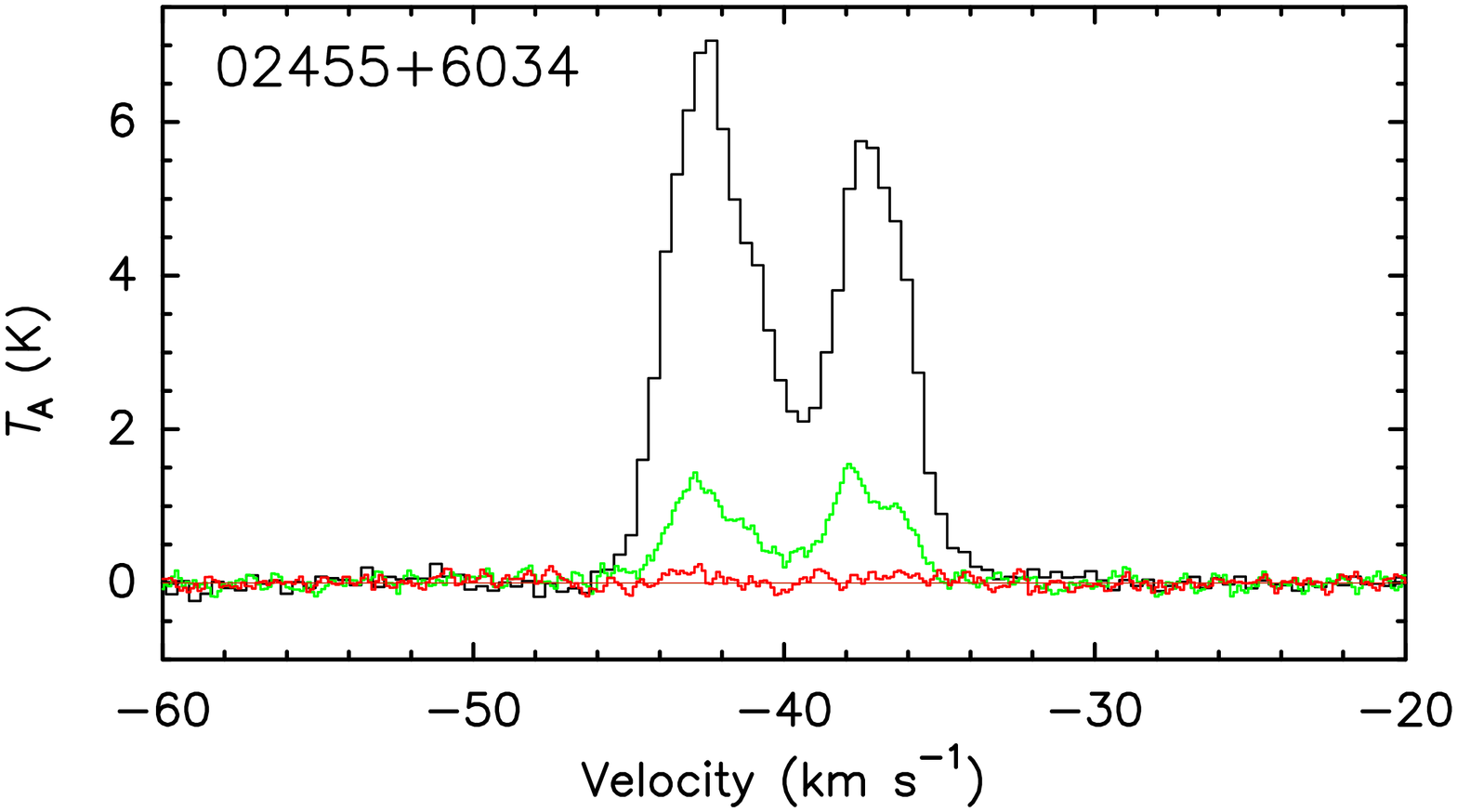}
\end{minipage}
\begin{minipage}[c]{0.5\textwidth}
  \centering
  \includegraphics[width=30mm,height=65mm,angle=-90.0]{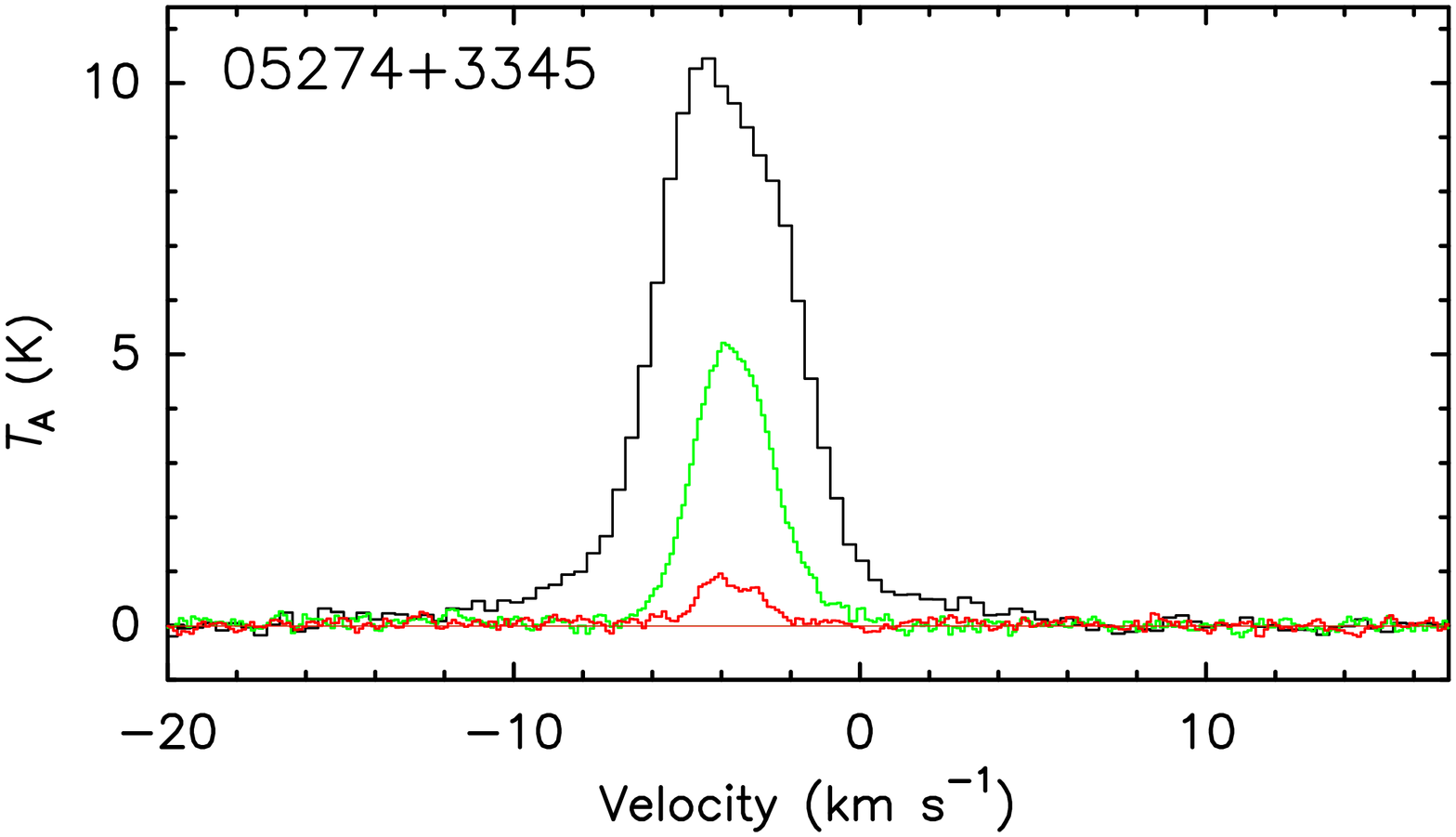}
\end{minipage}
\begin{minipage}[c]{0.5\textwidth}
  \centering
  \includegraphics[width=30mm,height=65mm,angle=-90.0]{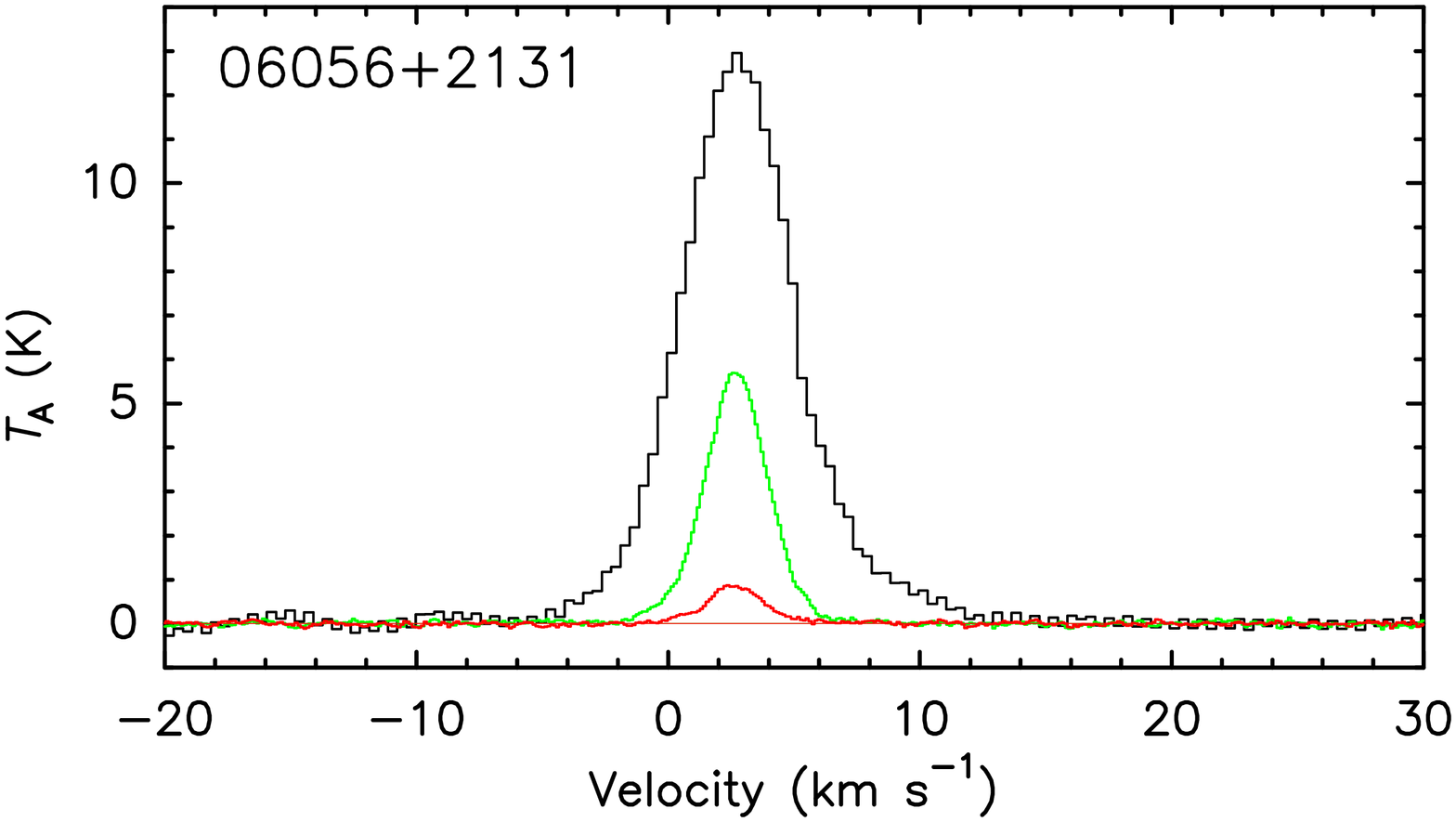}
\end{minipage}
\begin{minipage}[c]{0.5\textwidth}
  \centering
  \includegraphics[width=30mm,height=65mm,angle=-90.0]{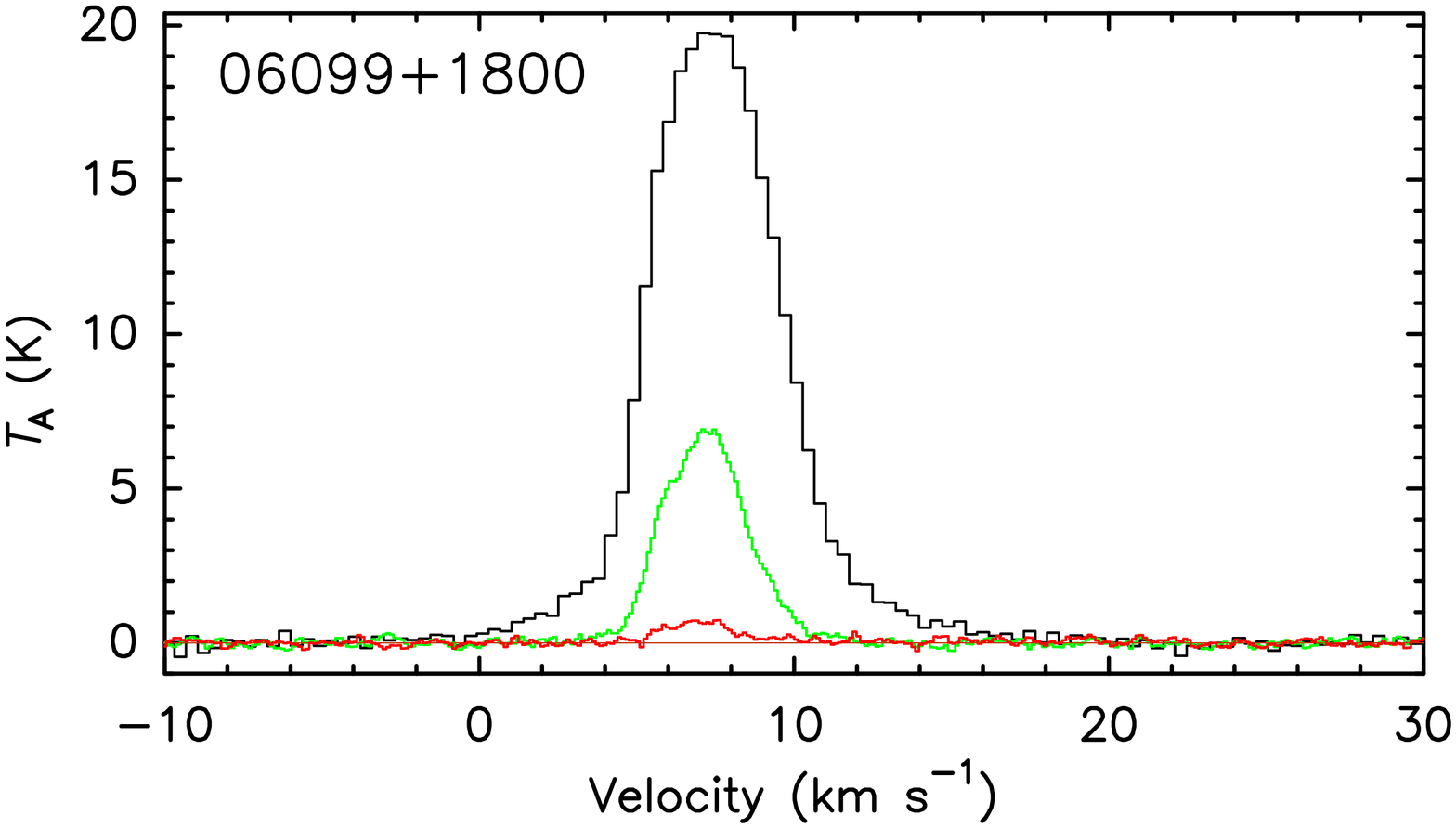}
\end{minipage}
\begin{minipage}[c]{0.5\textwidth}
  \centering
  \includegraphics[width=30mm,height=65mm,angle=-90.0]{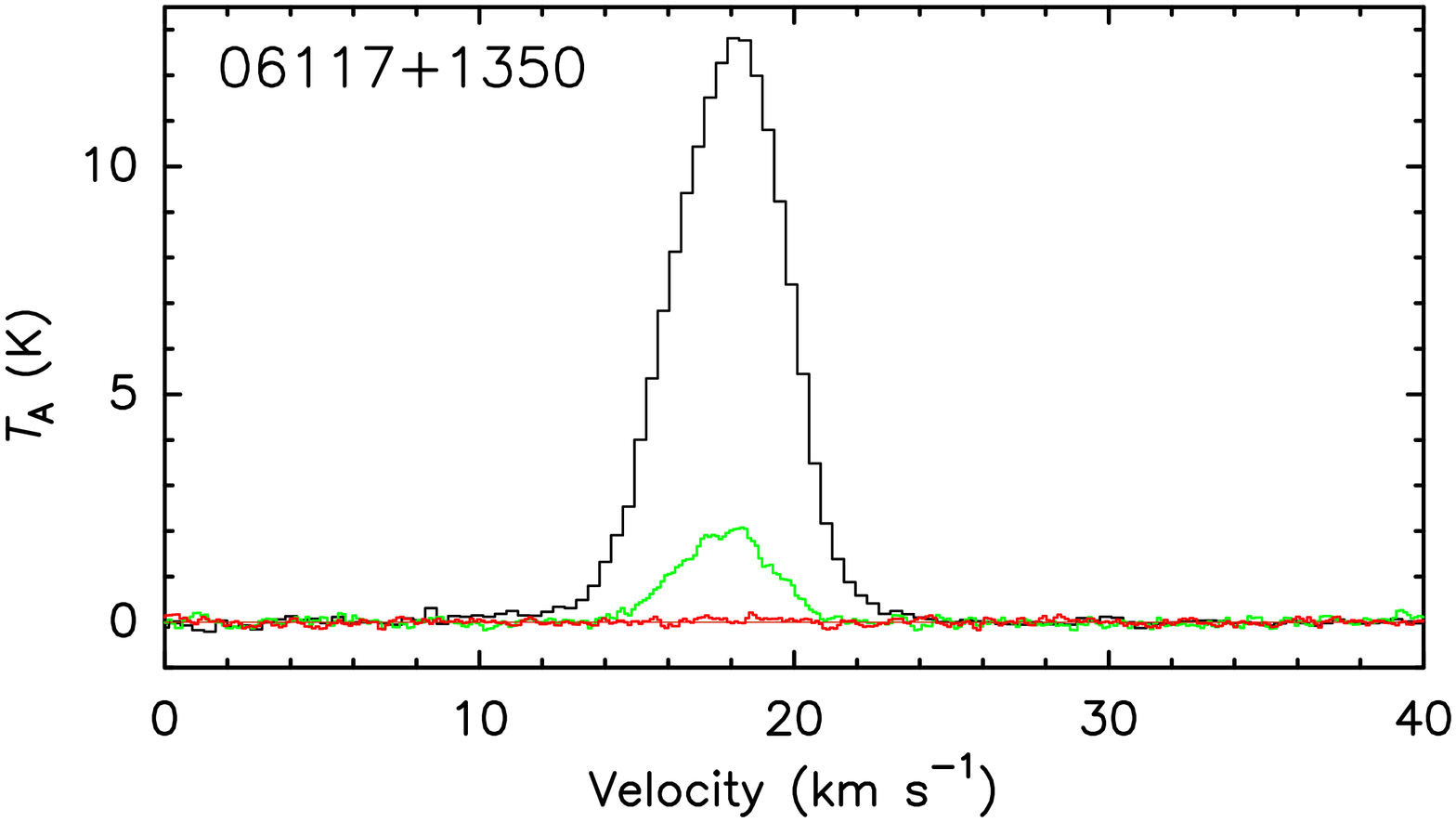}
\end{minipage}
\begin{minipage}[c]{0.5\textwidth}
  \centering
  \includegraphics[width=30mm,height=65mm,angle=-90.0]{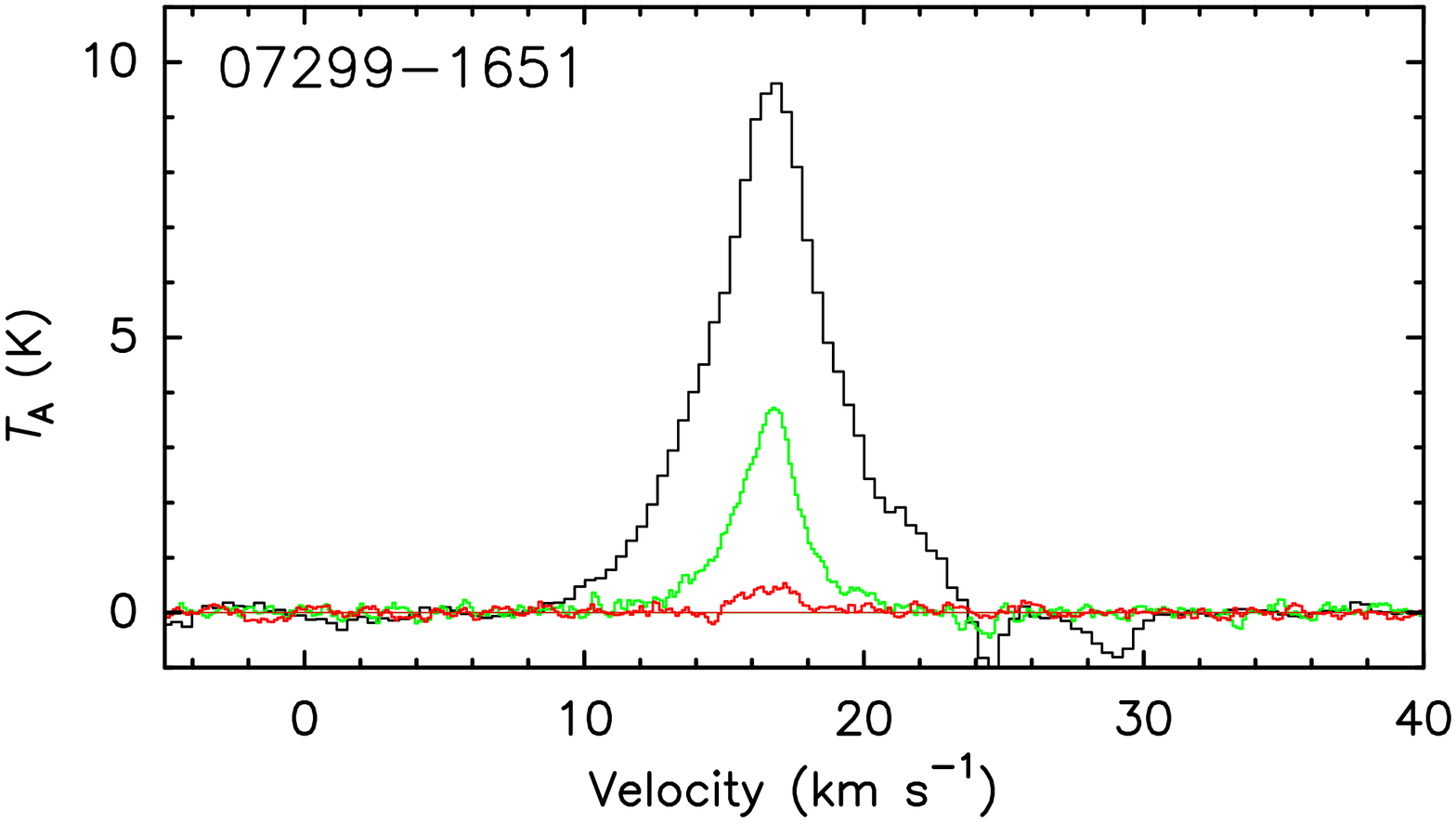}
\end{minipage}
\begin{minipage}[c]{0.5\textwidth}
  \centering
  \includegraphics[width=30mm,height=65mm,angle=-90.0]{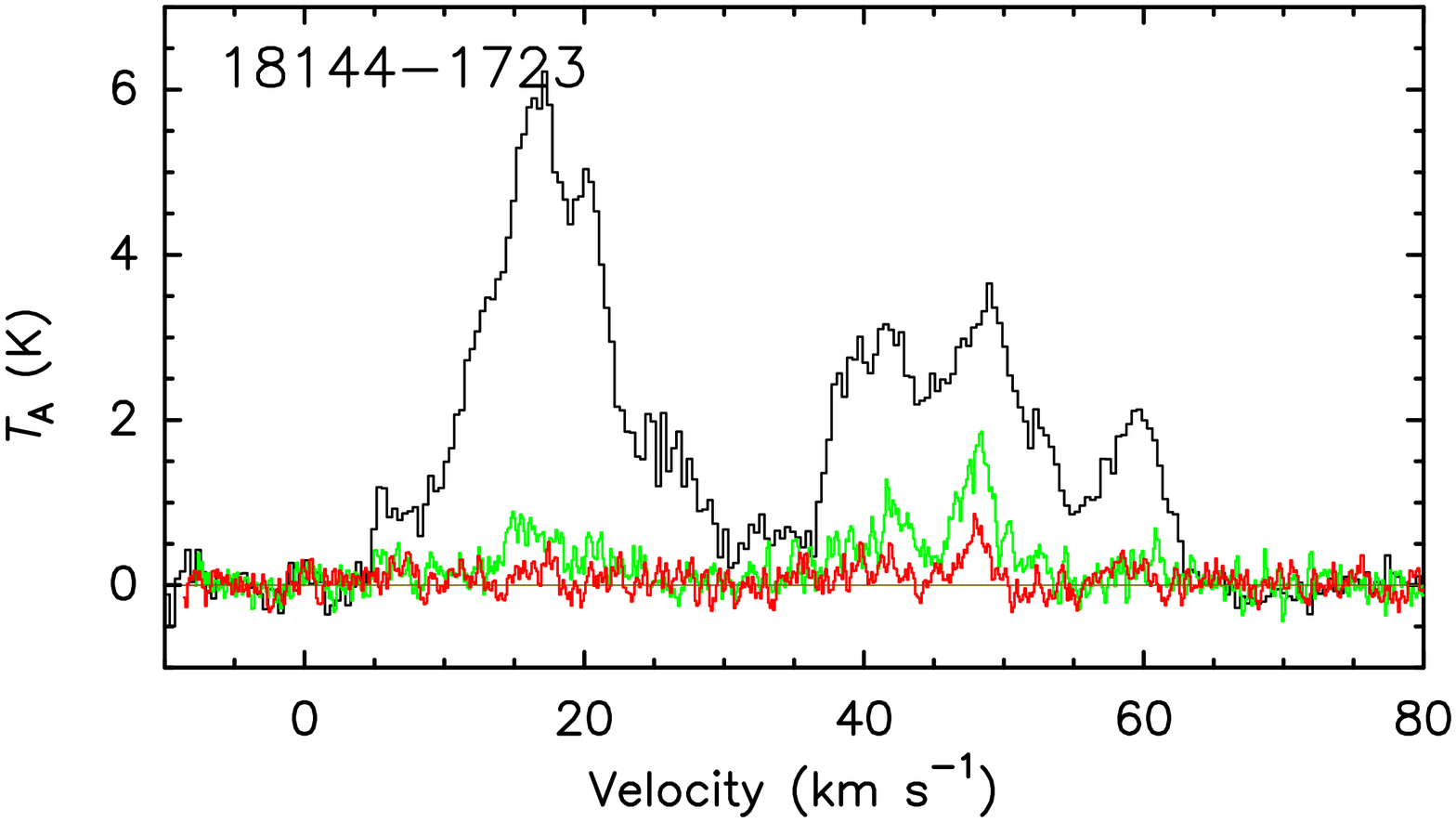}
\end{minipage}
\begin{minipage}[c]{0.5\textwidth}
  \centering
  \includegraphics[width=30mm,height=65mm,angle=-90.0]{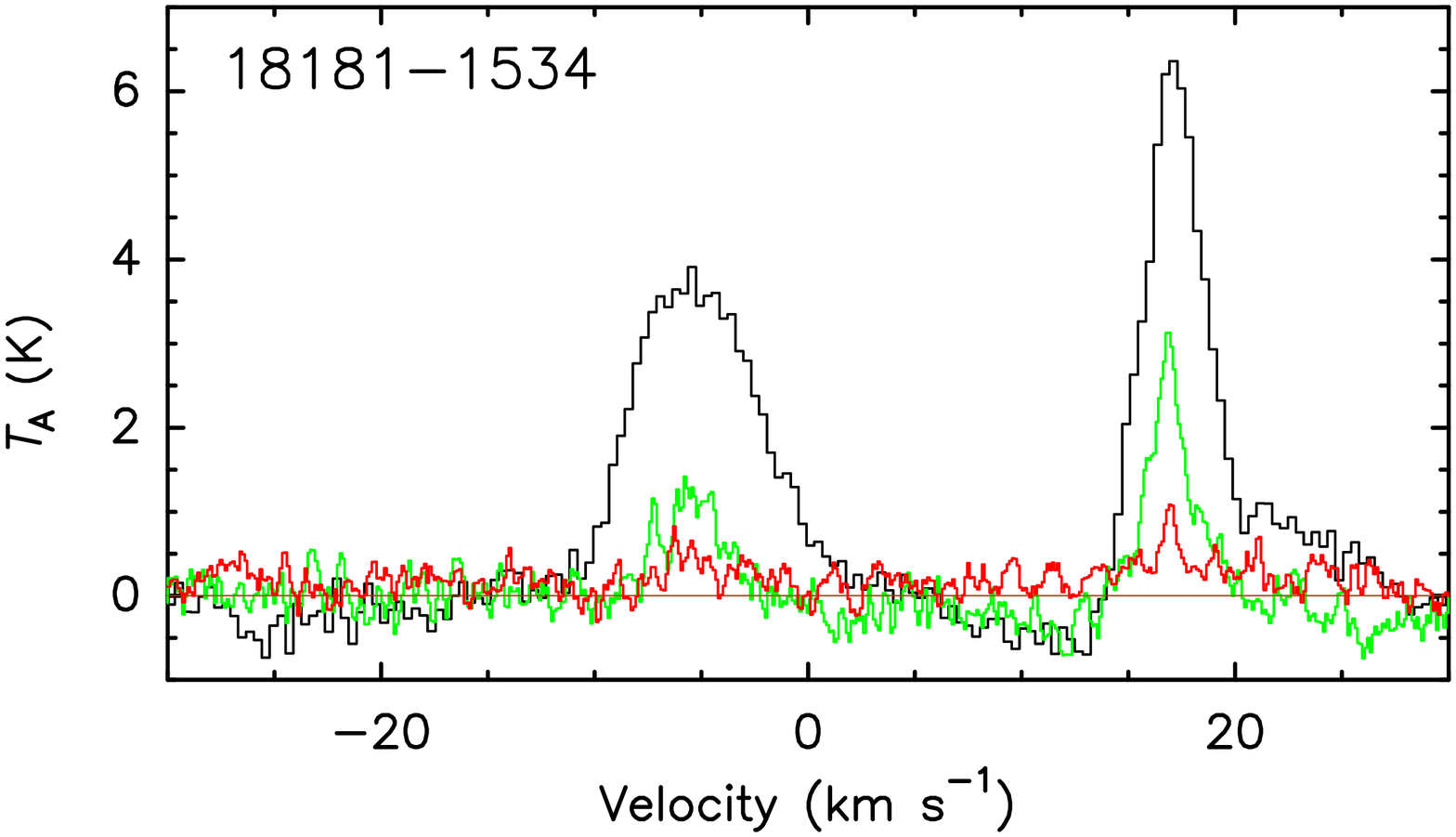}
  \end{minipage}
\begin{minipage}[c]{0.5\textwidth}
  \centering
  \includegraphics[width=30mm,height=65mm,angle=-90.0]{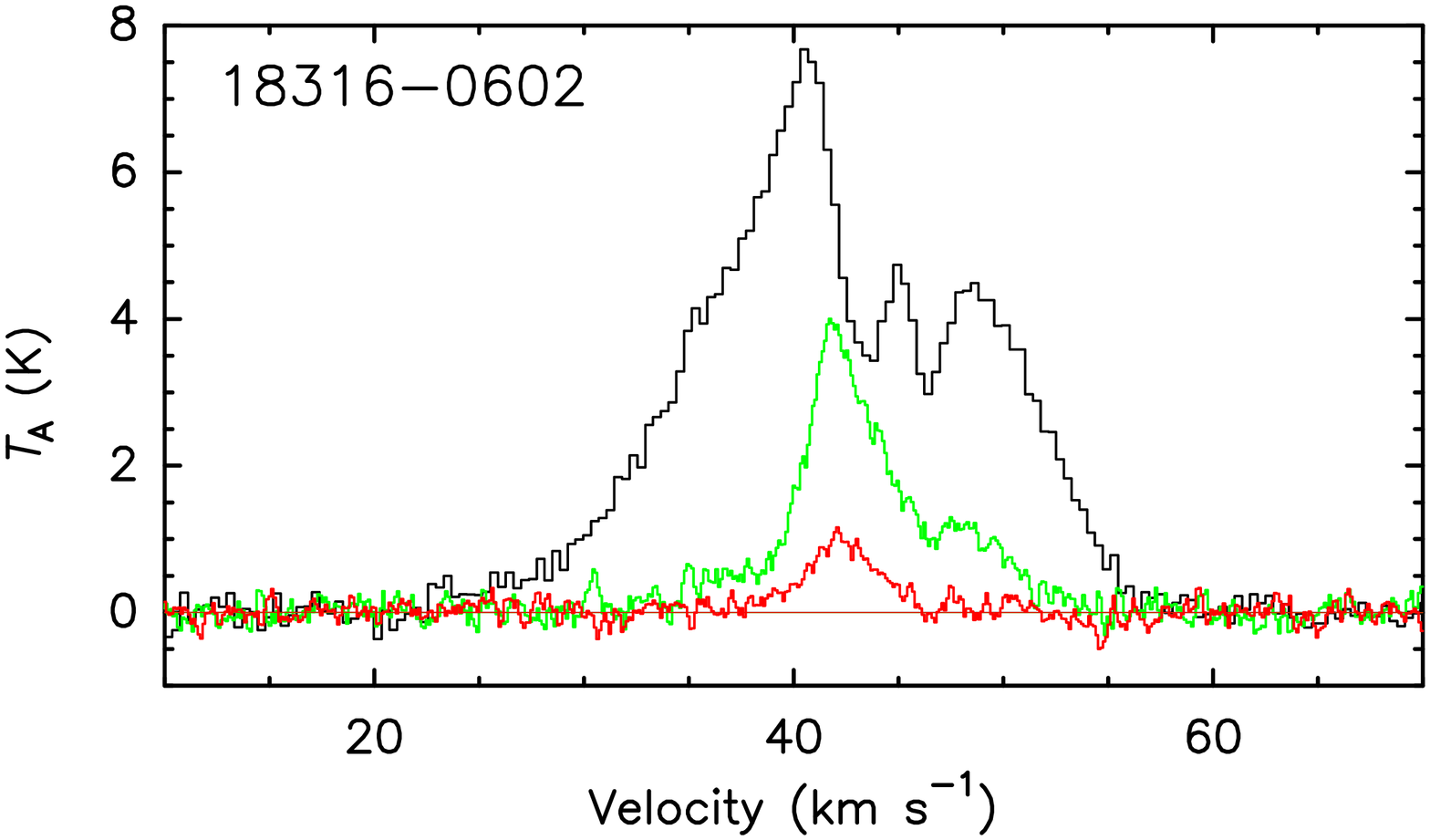}
\end{minipage}
\begin{minipage}[c]{0.5\textwidth}
  \centering
  \includegraphics[width=30mm,height=65mm,angle=-90.0]{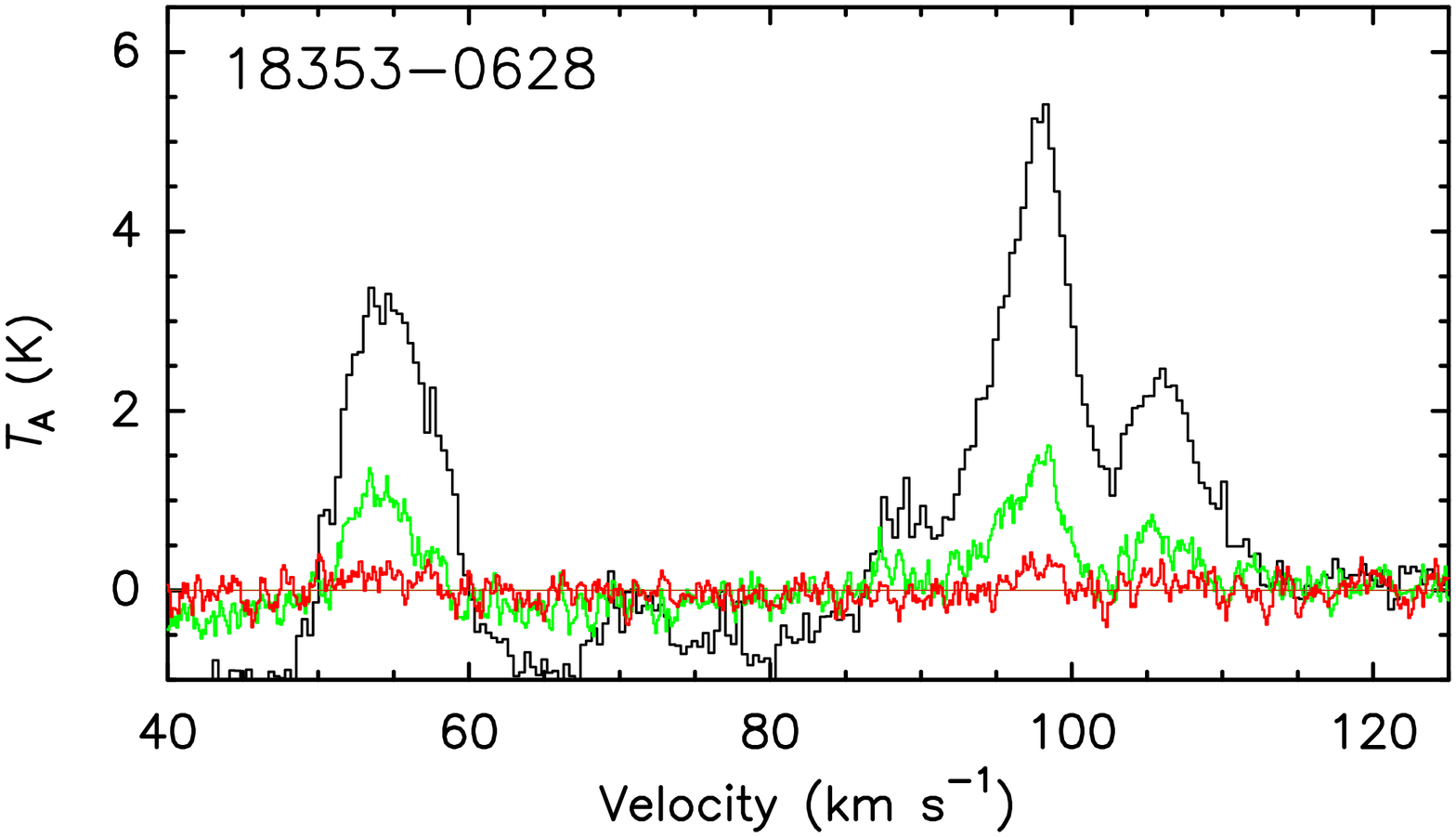}
\end{minipage}
\caption{Spectra of emission line samples. The grey, green and red
lines represent $^{12}$CO, $^{13}$CO and C$^{18}$O J=1-0 ,
respectively. The source names are drawn on the upper-left corners
of each panel. The $^{12}$CO spectra of IRAS 18414-1723, IRAS
18353-0628, G32.74-0.07 and G39.10+0.48 are blended, but the respect
$^{13}$CO emission spectra can be well distinguished. Source
G29.86-0.05 is a sample with blended $^{13}$CO emission lines.
Properties of these $^{13}$CO emission line profiles can be found in
column 12 of Table 2. }
\end{figure}

\setcounter{figure}{2}

\begin{figure}
\begin{minipage}[c]{0.5\textwidth}
  \centering
  \includegraphics[width=30mm,height=65mm,angle=-90.0]{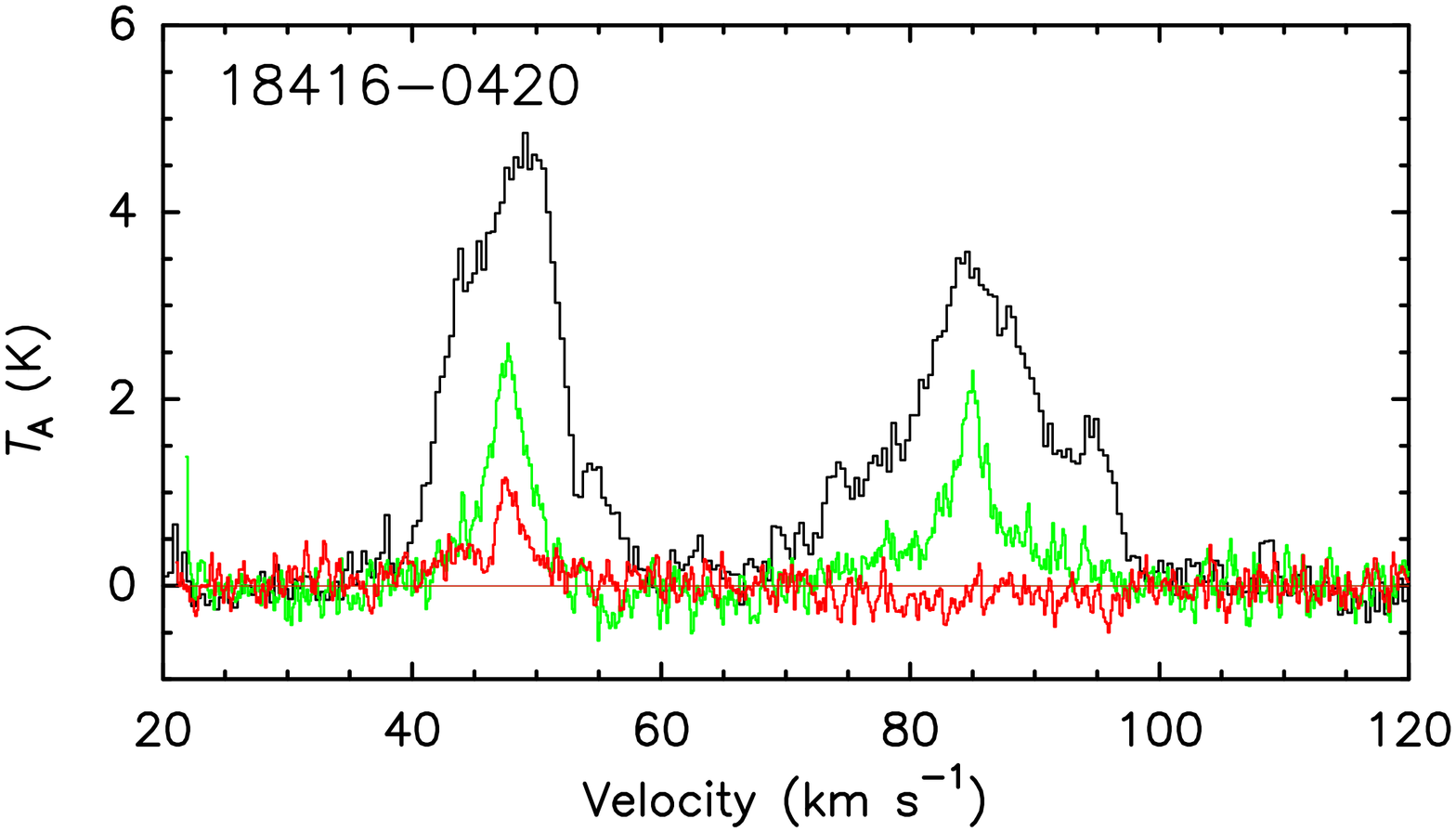}
\end{minipage}
\begin{minipage}[c]{0.5\textwidth}
  \centering
  \includegraphics[width=30mm,height=65mm,angle=-90.0]{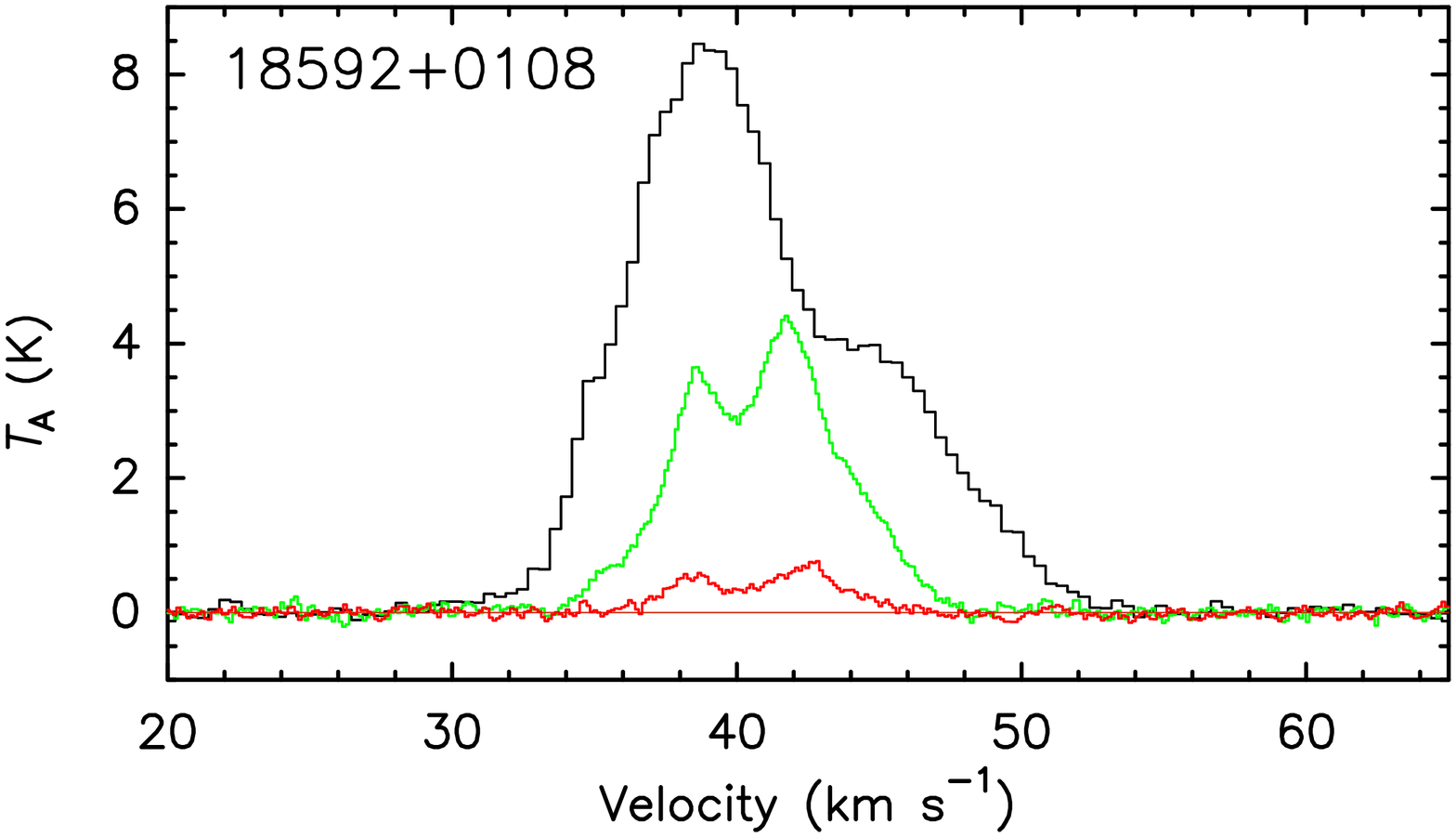}
\end{minipage}
\begin{minipage}[c]{0.5\textwidth}
  \centering
  \includegraphics[width=30mm,height=65mm,angle=-90.0]{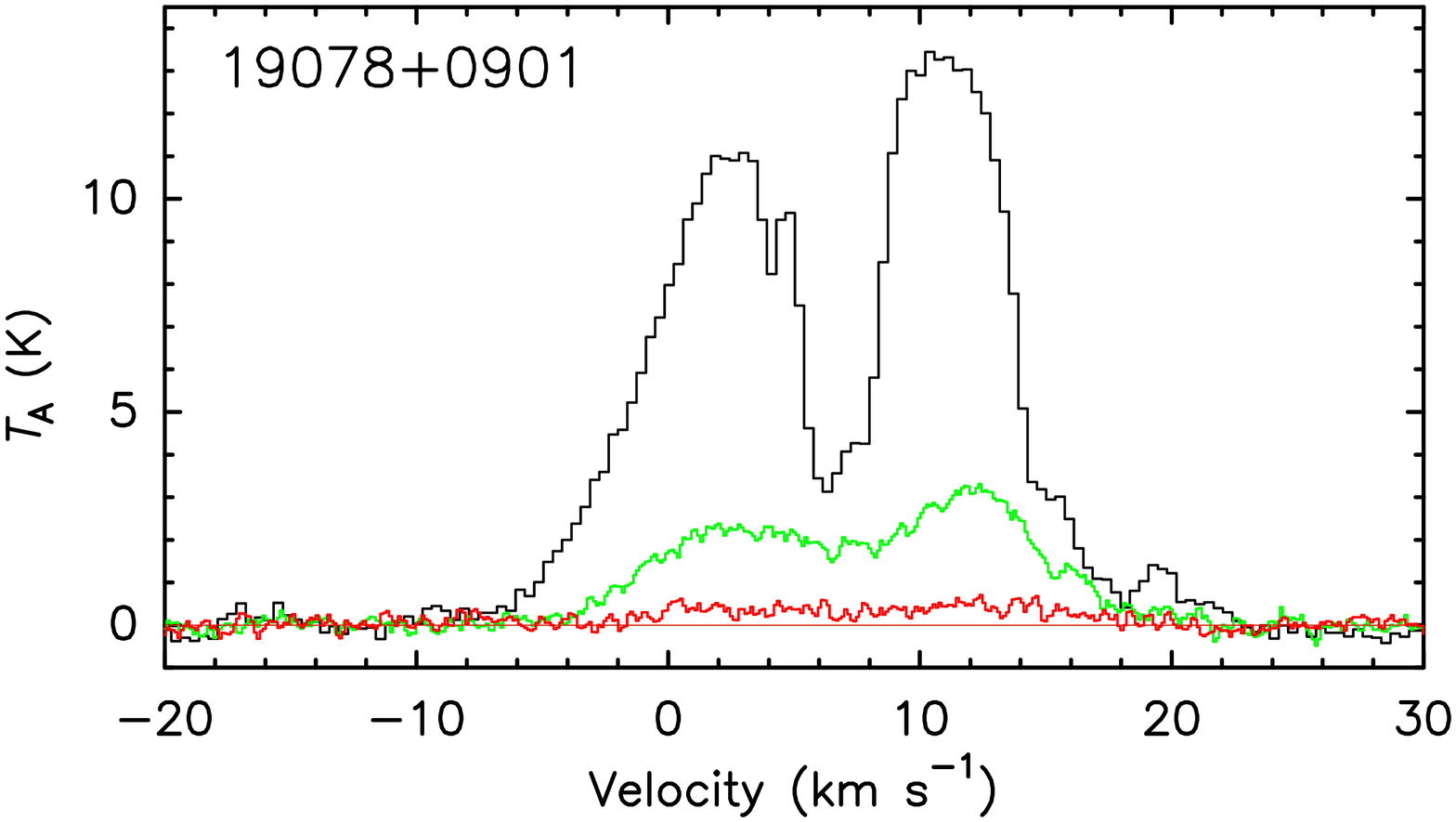}
\end{minipage}
\begin{minipage}[c]{0.5\textwidth}
  \centering
  \includegraphics[width=30mm,height=65mm,angle=-90.0]{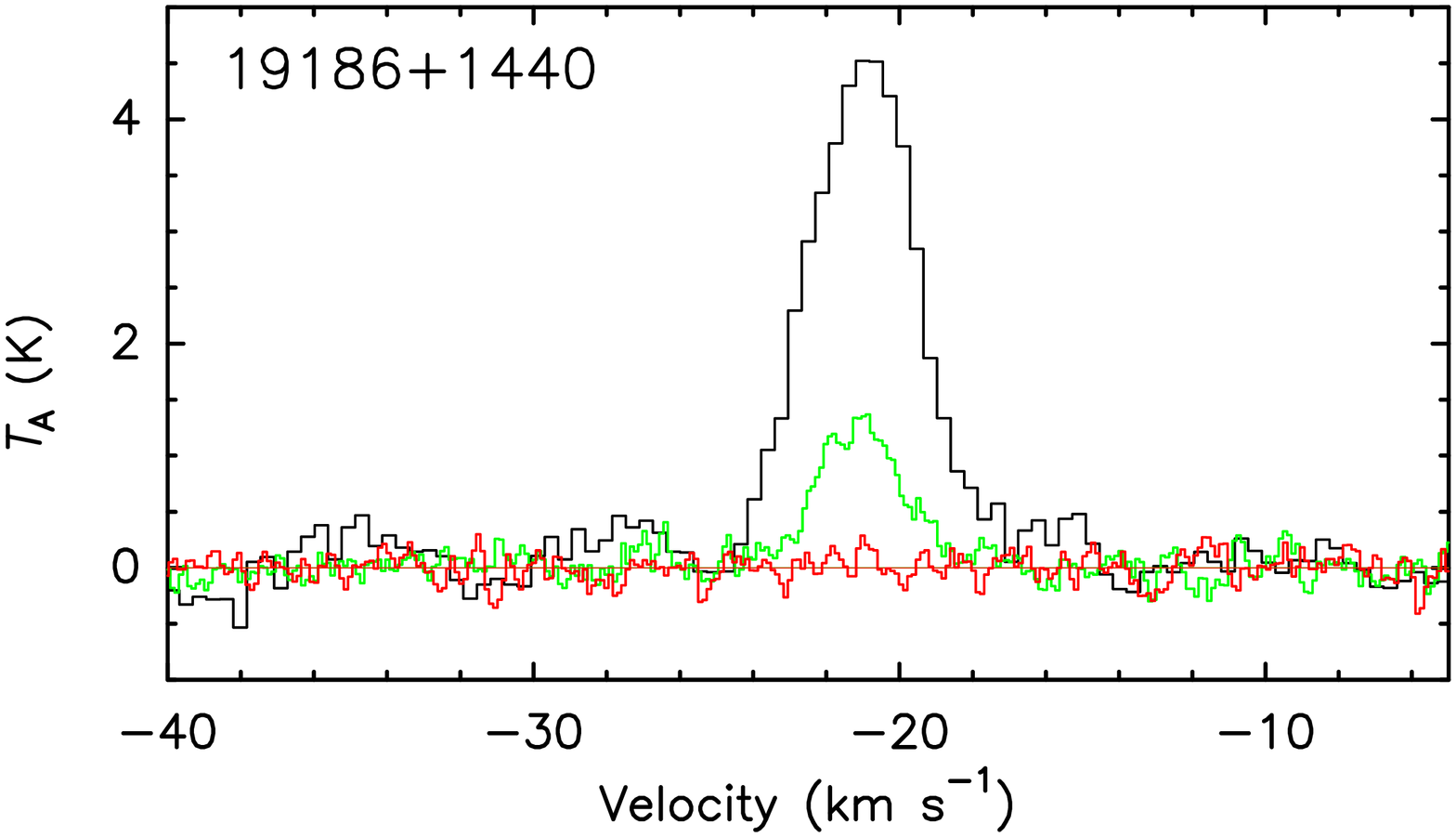}
\end{minipage}
\begin{minipage}[c]{0.5\textwidth}
  \centering
  \includegraphics[width=30mm,height=65mm,angle=-90.0]{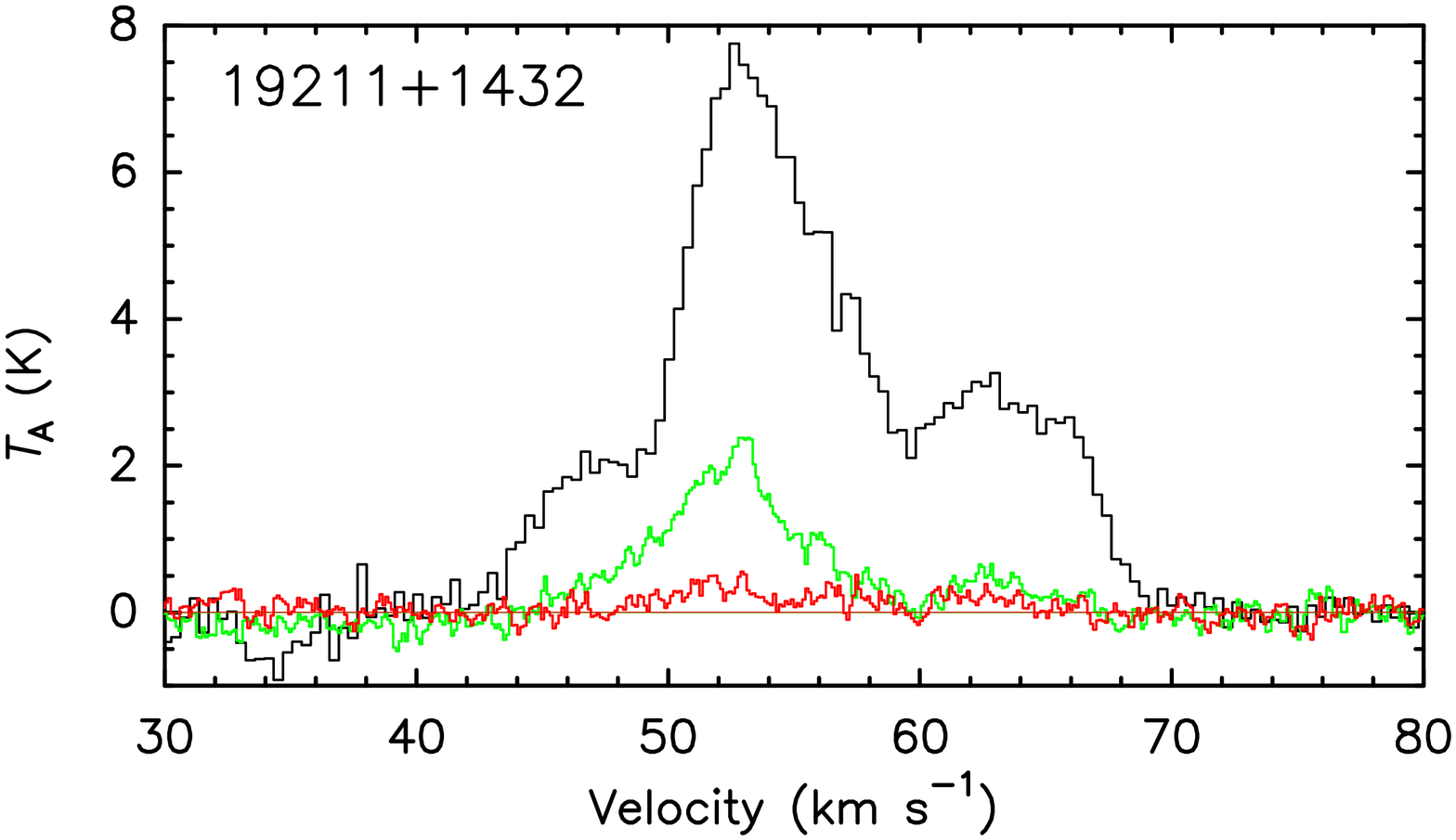}
\end{minipage}
\begin{minipage}[c]{0.5\textwidth}
  \centering
  \includegraphics[width=30mm,height=65mm,angle=-90.0]{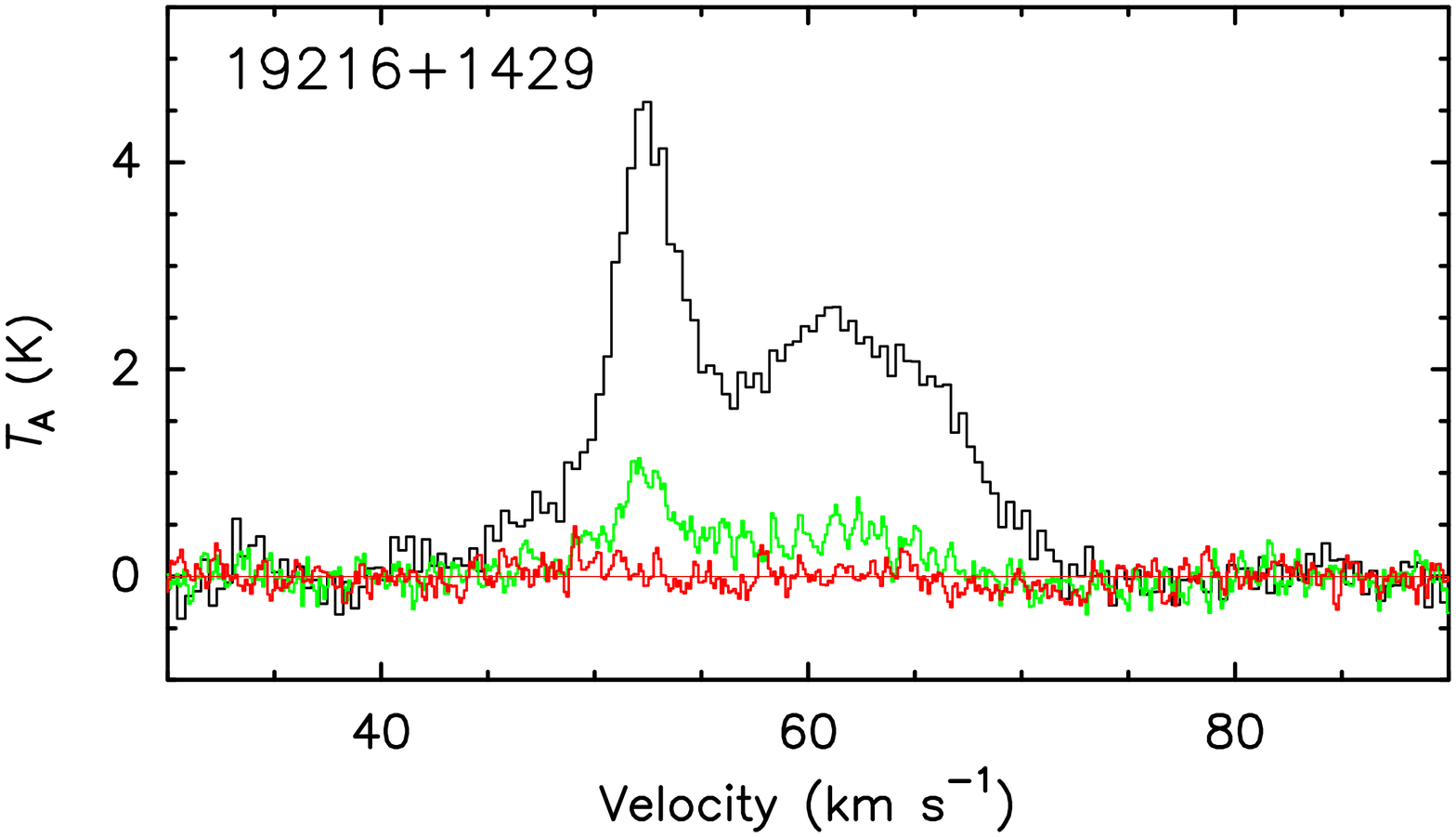}
  \end{minipage}
\begin{minipage}[c]{0.5\textwidth}
  \centering
  \includegraphics[width=30mm,height=65mm,angle=-90.0]{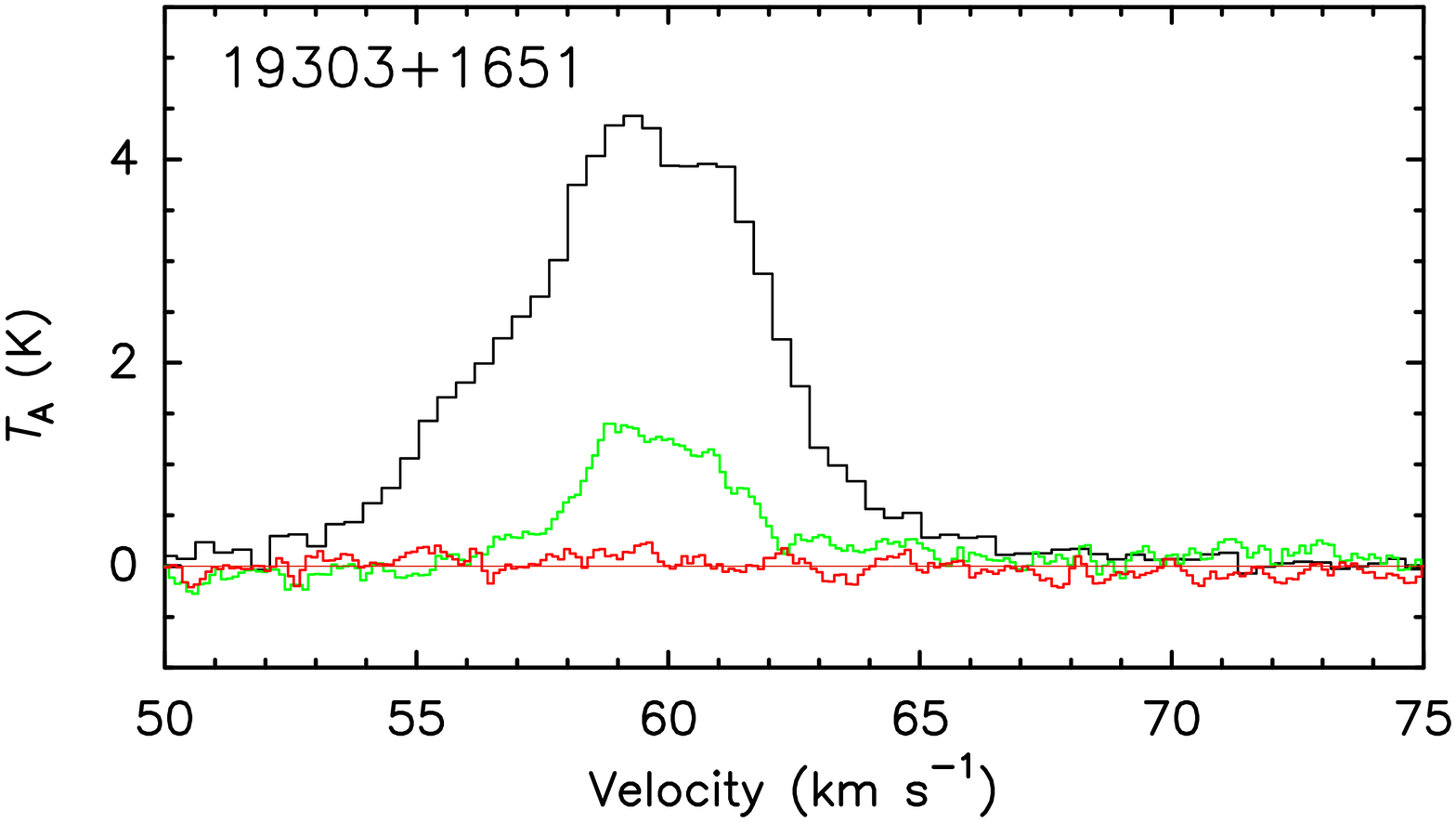}
\end{minipage}
\begin{minipage}[c]{0.5\textwidth}
  \centering
  \includegraphics[width=30mm,height=65mm,angle=-90.0]{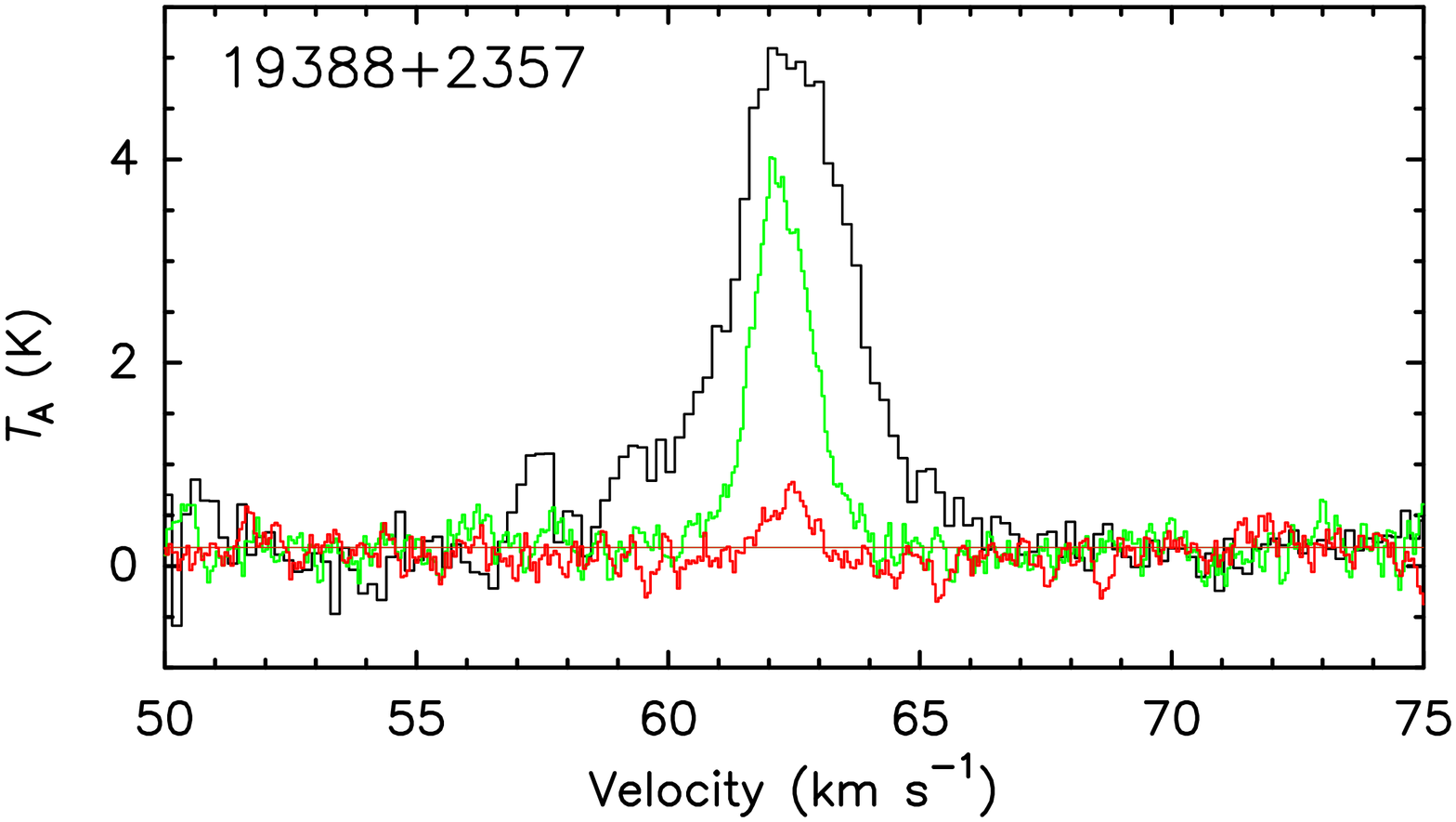}
\end{minipage}
\begin{minipage}[c]{0.5\textwidth}
  \centering
  \includegraphics[width=30mm,height=65mm,angle=-90.0]{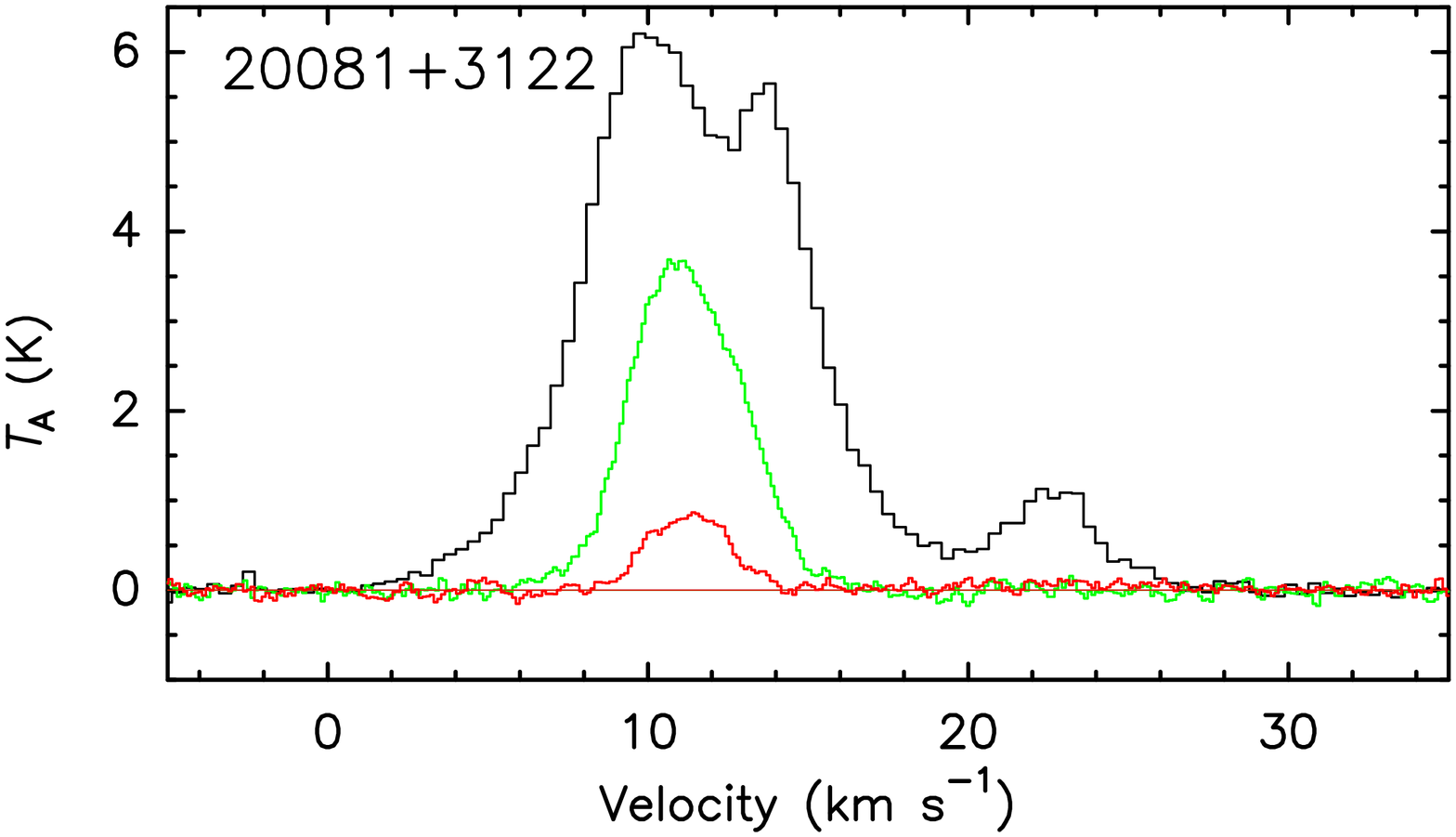}
\end{minipage}
\begin{minipage}[c]{0.5\textwidth}
  \centering
  \includegraphics[width=30mm,height=65mm,angle=-90.0]{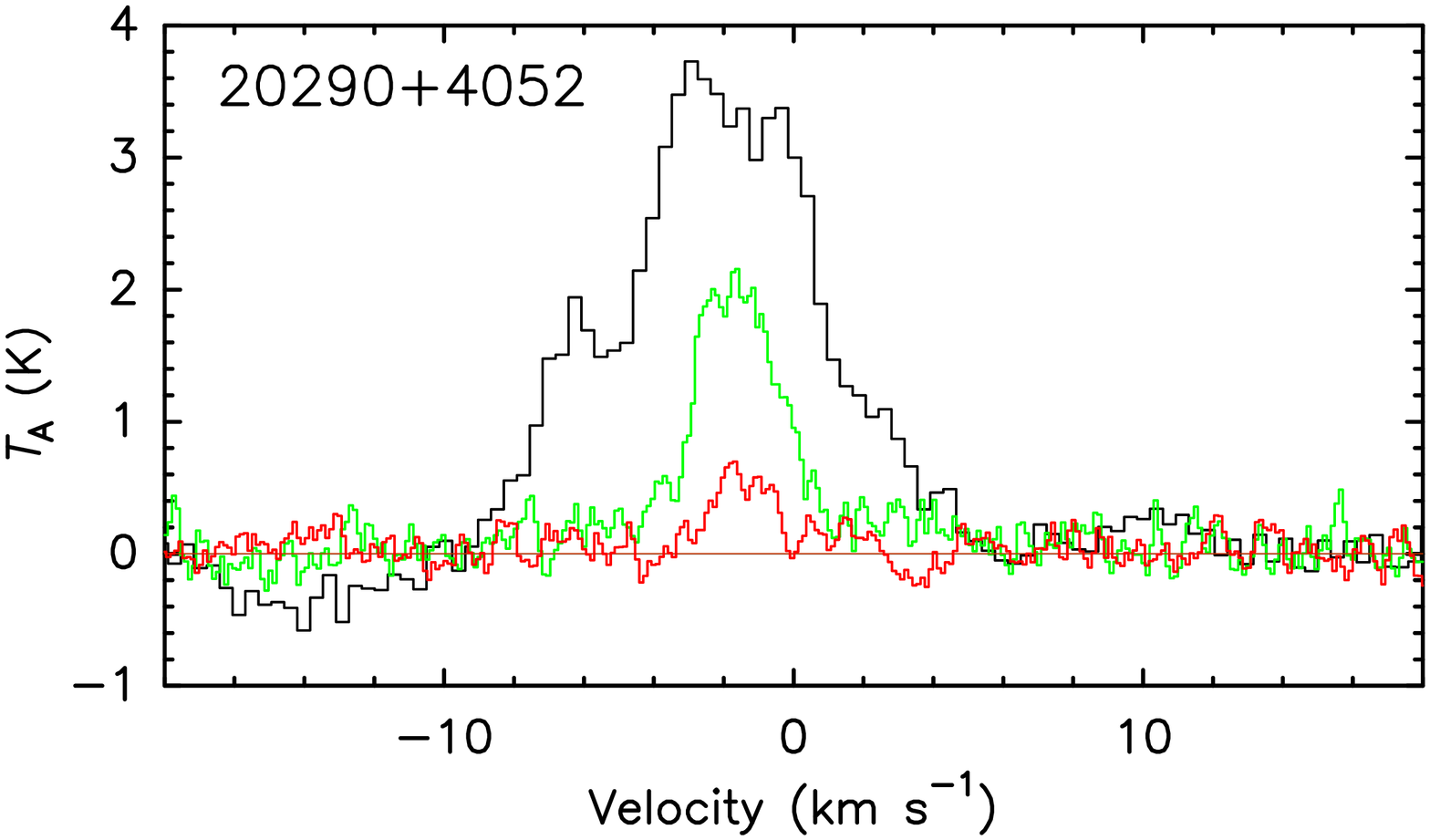}
\end{minipage}
\begin{minipage}[c]{0.5\textwidth}
  \centering
  \includegraphics[width=30mm,height=65mm,angle=-90.0]{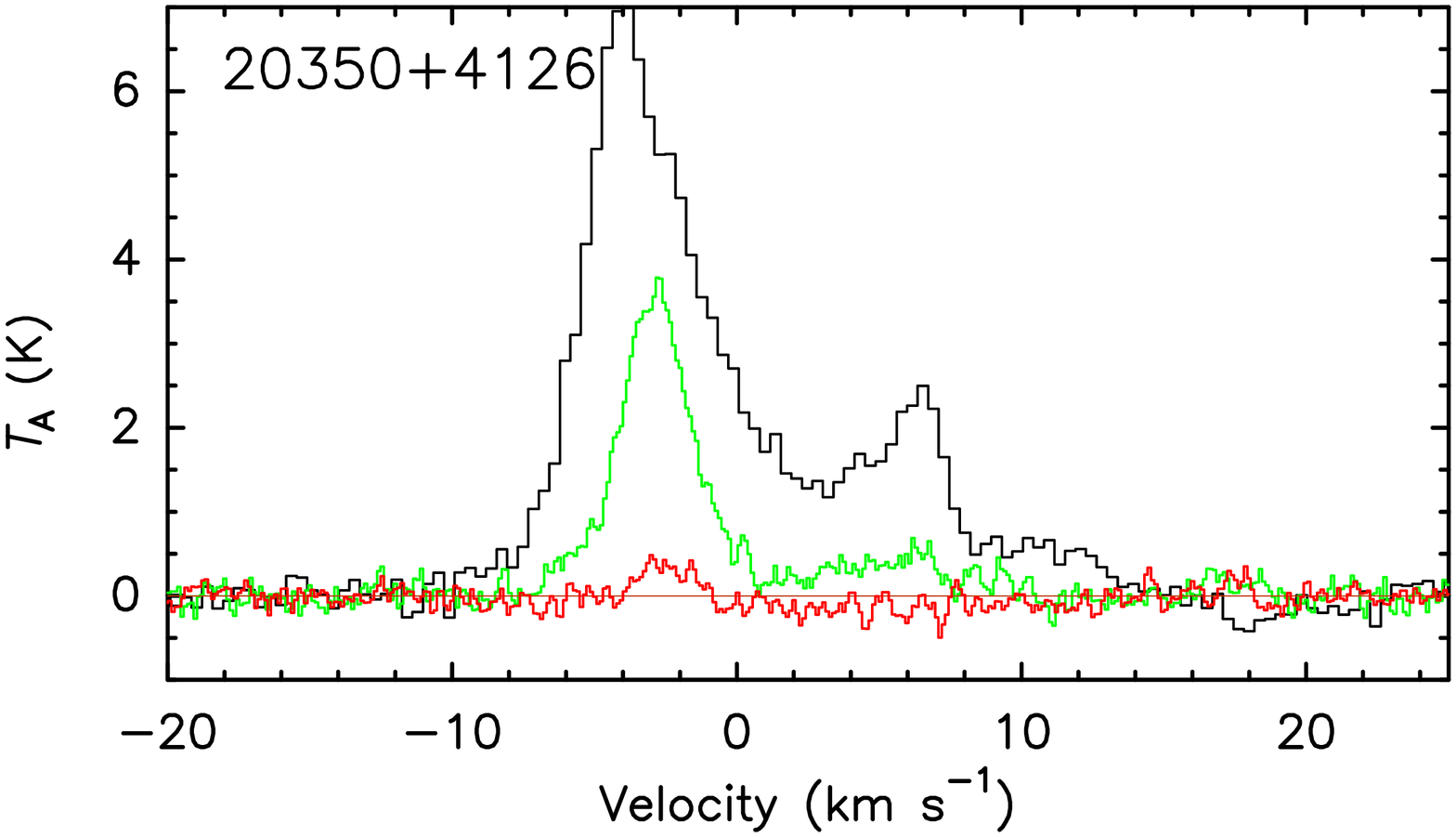}
\end{minipage}
\begin{minipage}[c]{0.5\textwidth}
  \centering
  \includegraphics[width=30mm,height=65mm,angle=-90.0]{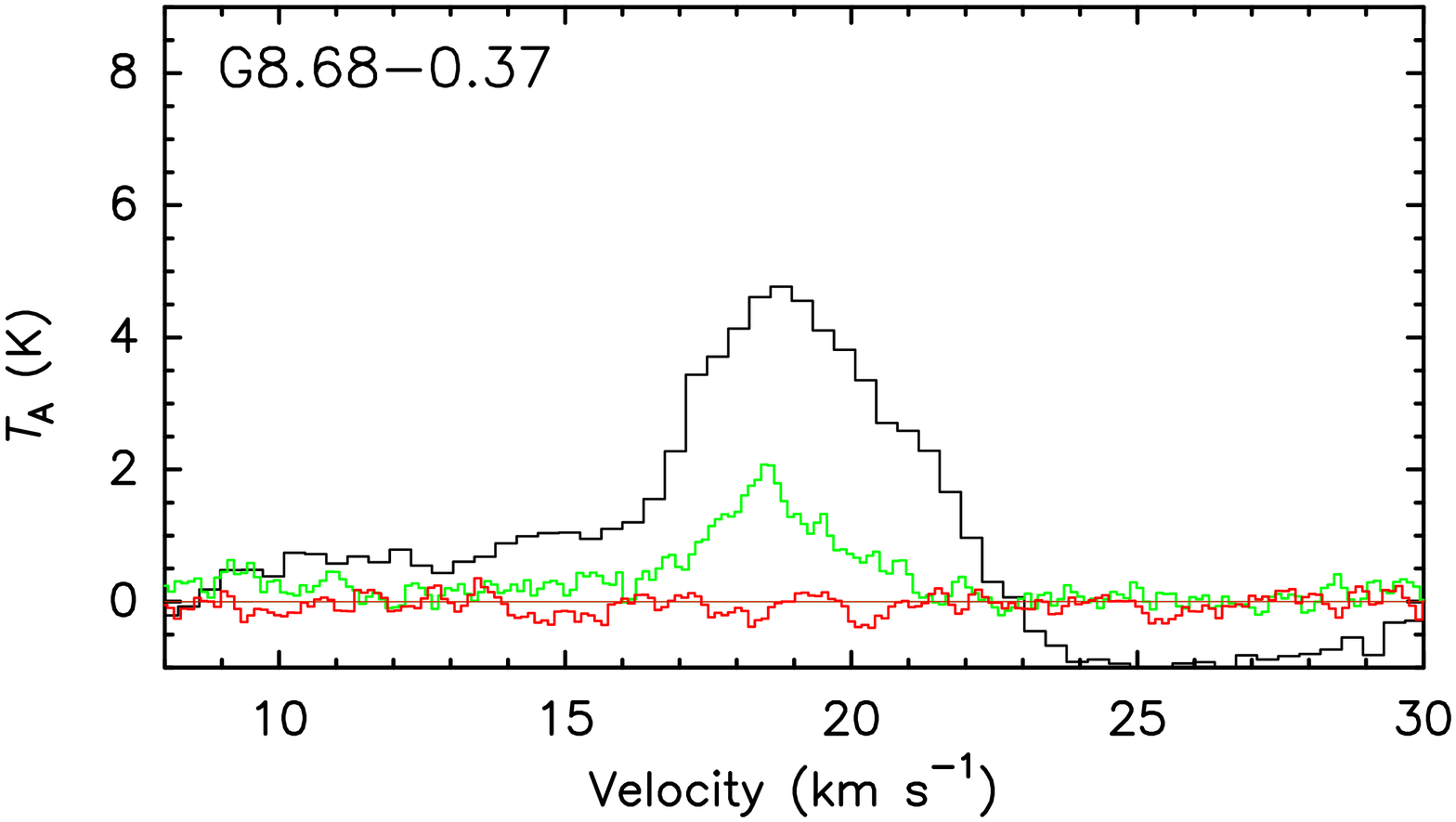}
\end{minipage}
\caption{Continued }
\end{figure}

\setcounter{figure}{2}

\begin{figure}
\begin{minipage}[c]{0.5\textwidth}
  \centering
  \includegraphics[width=30mm,height=65mm,angle=-90.0]{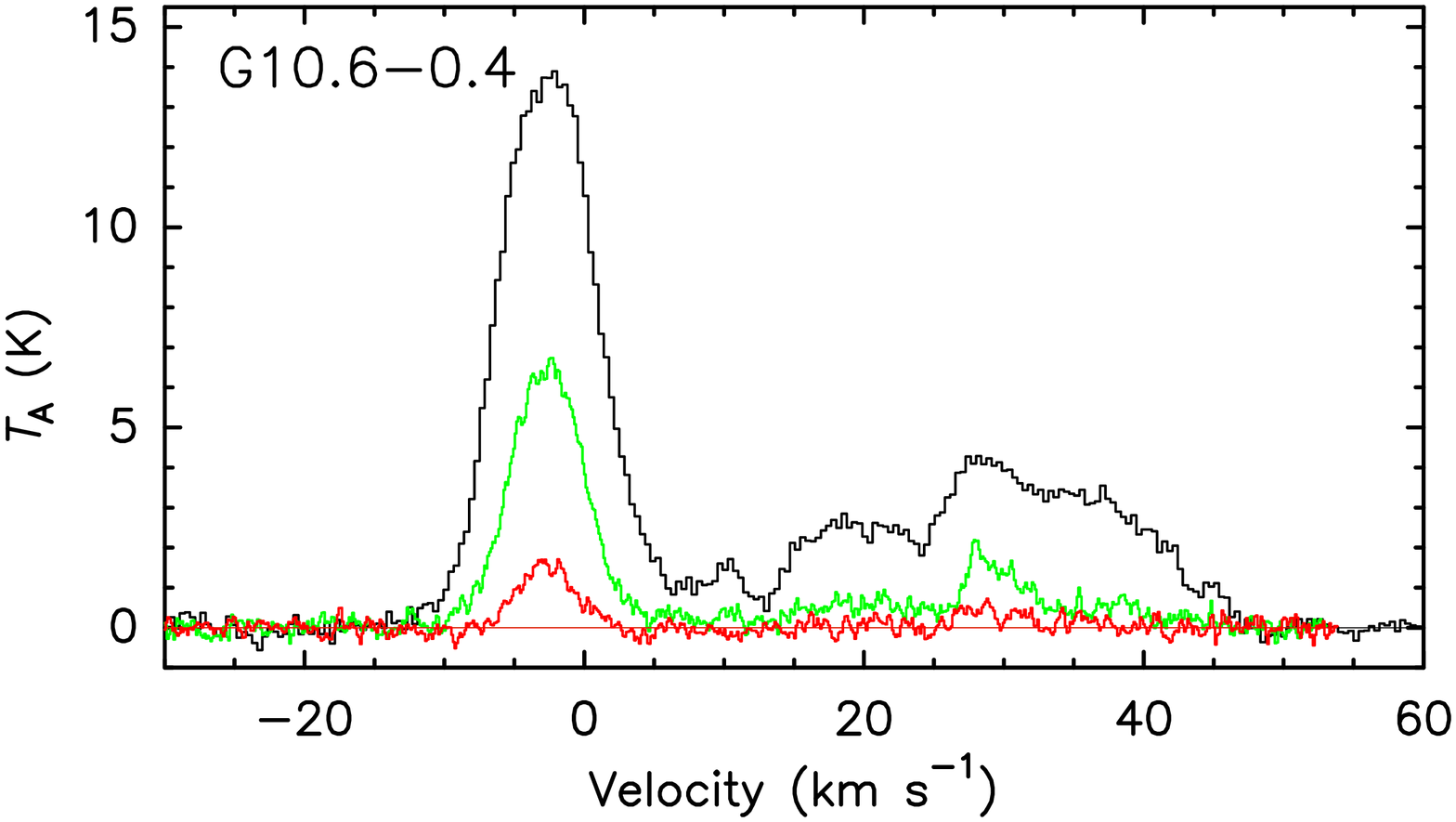}
\end{minipage}
\begin{minipage}[c]{0.5\textwidth}
  \centering
  \includegraphics[width=30mm,height=65mm,angle=-90.0]{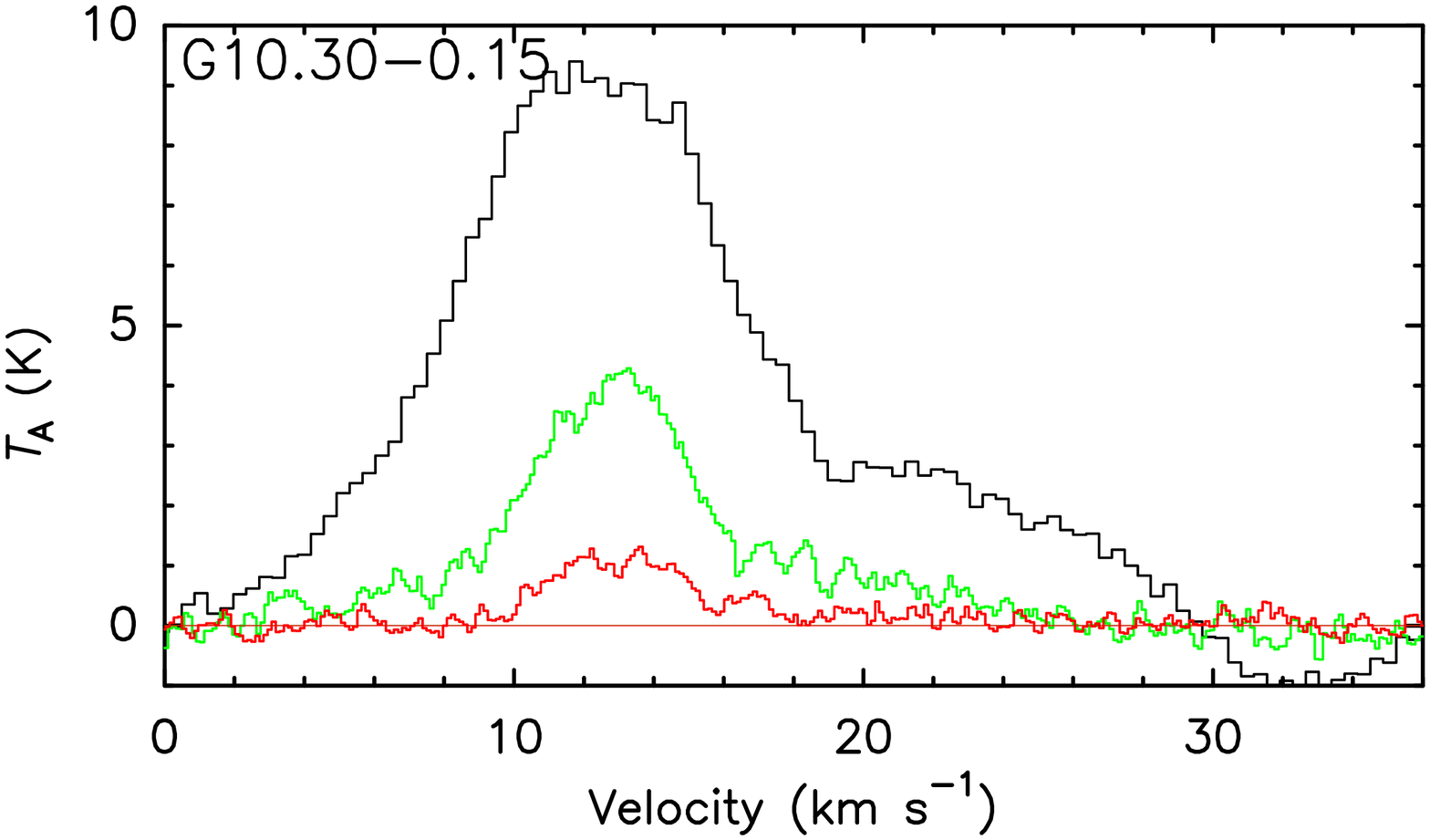}
\end{minipage}
\begin{minipage}[c]{0.5\textwidth}
  \centering
  \includegraphics[width=30mm,height=65mm,angle=-90.0]{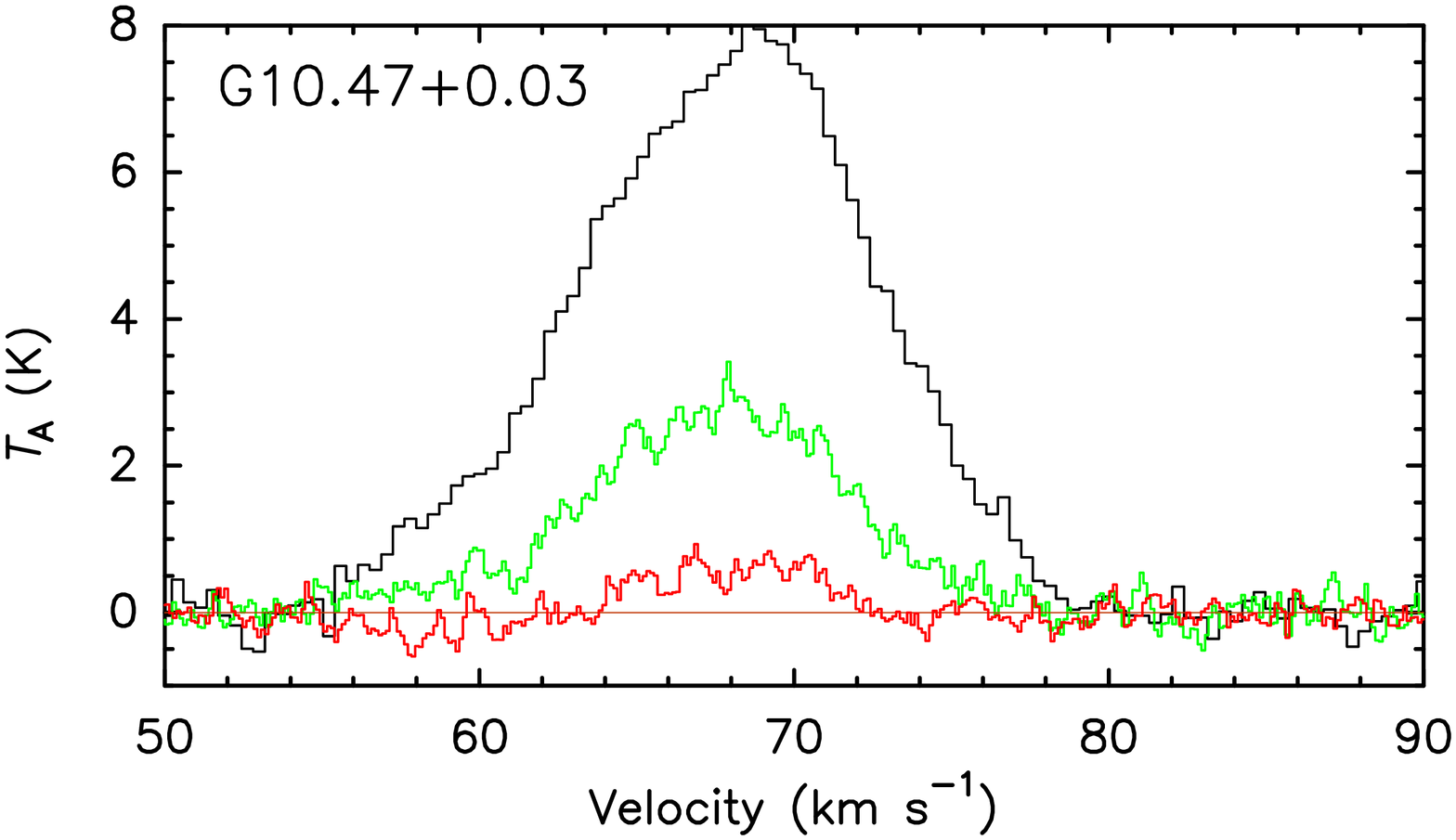}
\end{minipage}
\begin{minipage}[c]{0.5\textwidth}
  \centering
  \includegraphics[width=30mm,height=65mm,angle=-90.0]{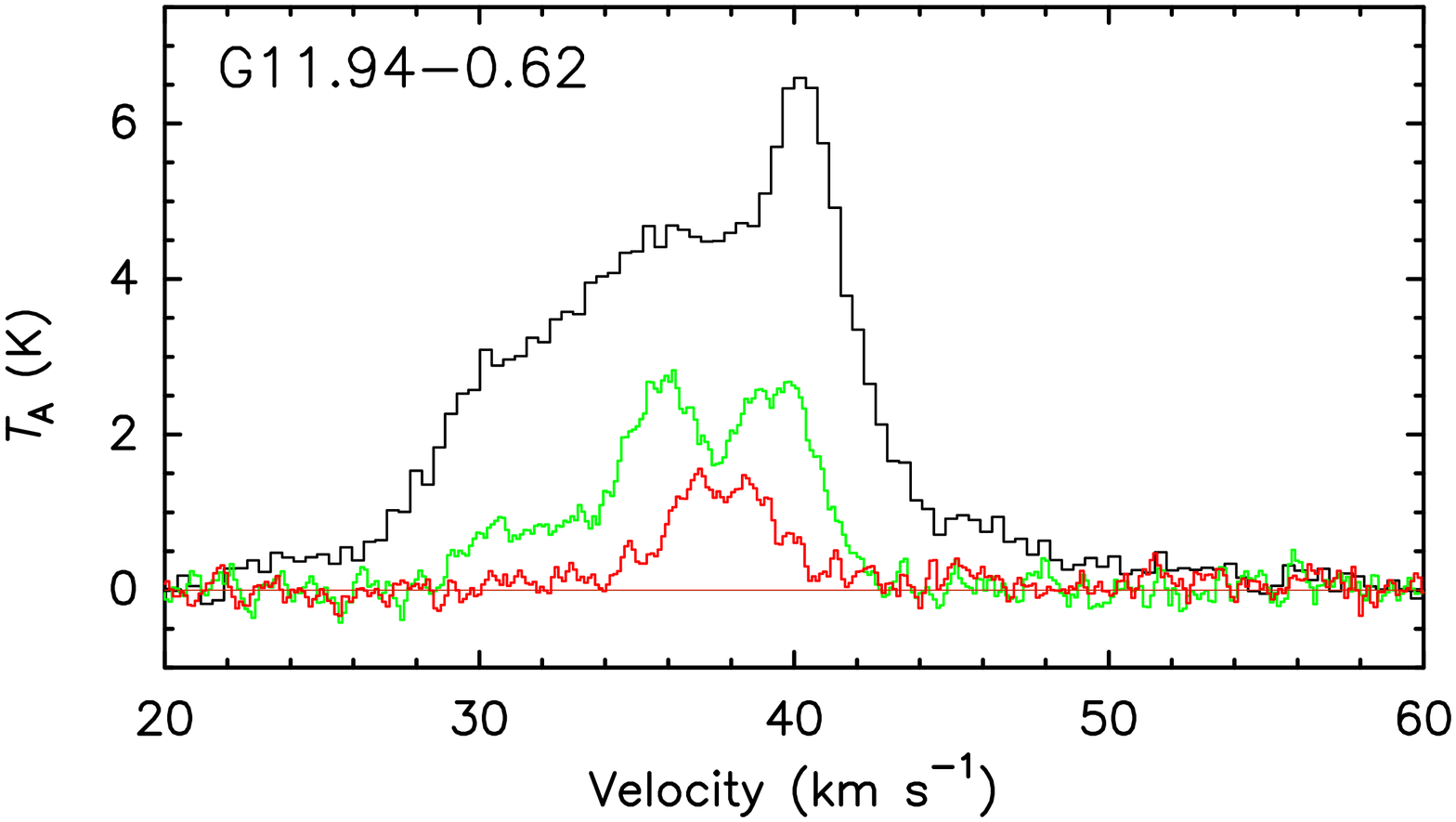}
\end{minipage}
\begin{minipage}[c]{0.5\textwidth}
  \centering
  \includegraphics[width=30mm,height=65mm,angle=-90.0]{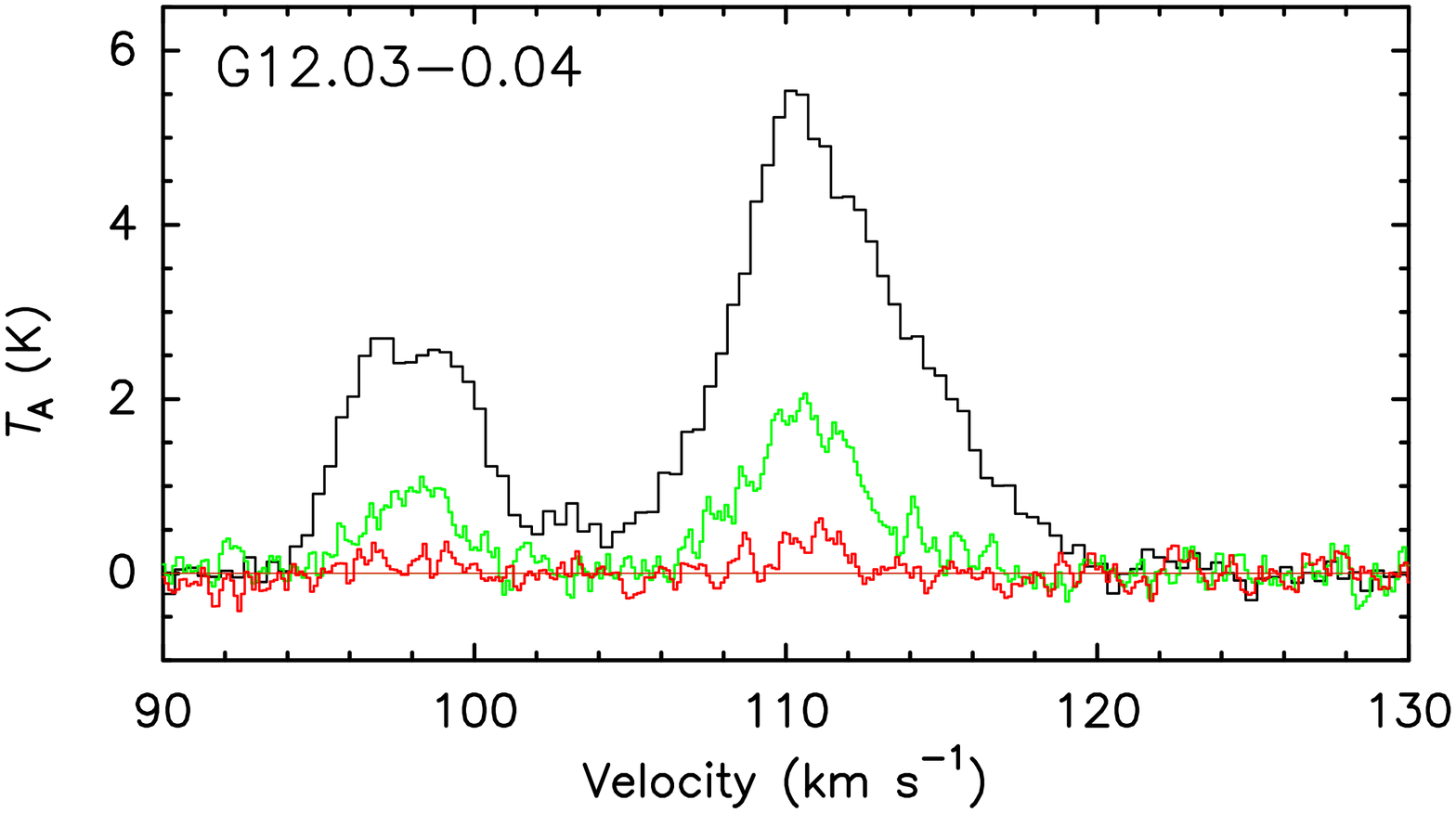}
\end{minipage}
\begin{minipage}[c]{0.5\textwidth}
  \centering
  \includegraphics[width=30mm,height=65mm,angle=-90.0]{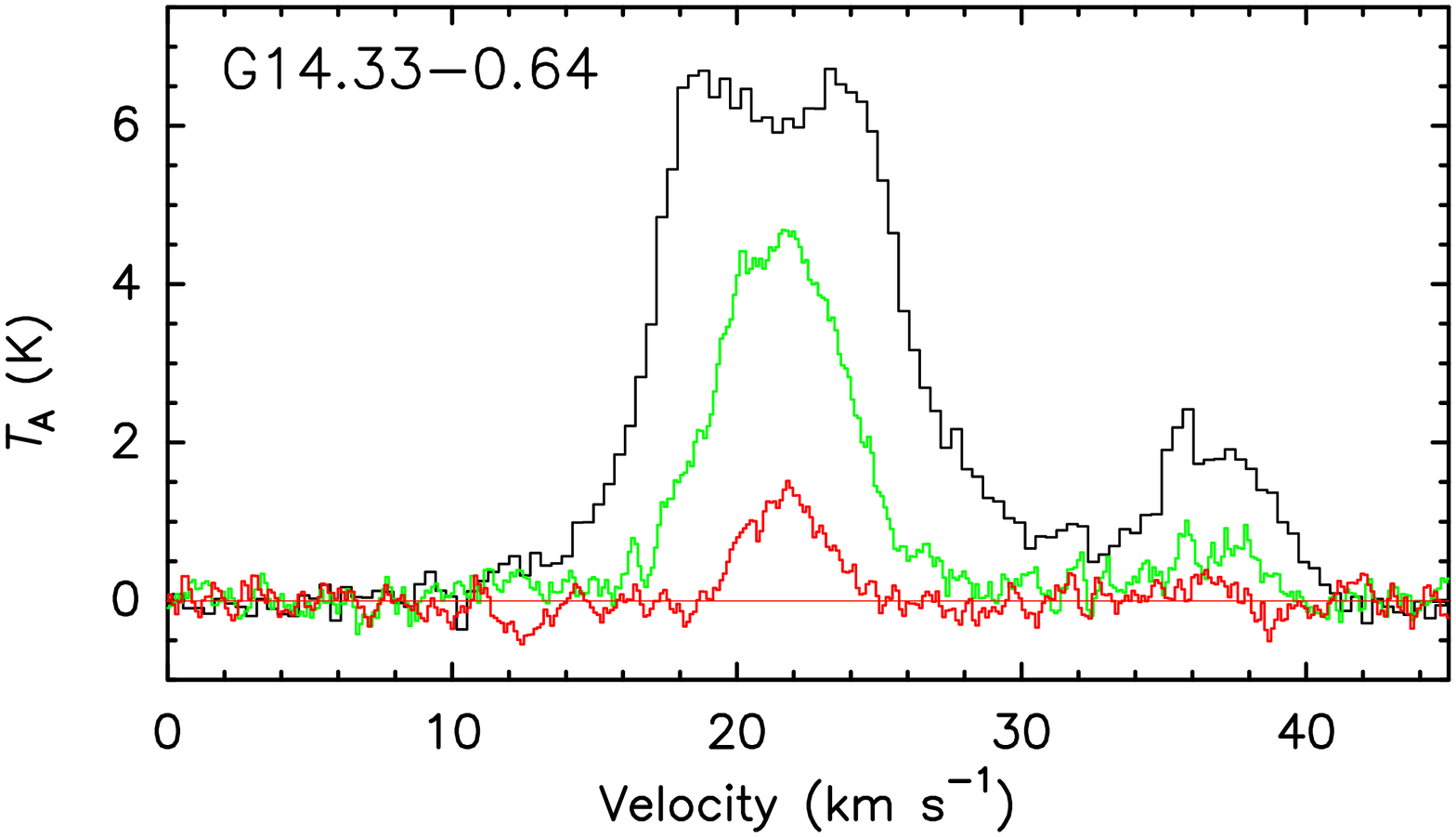}
\end{minipage}
\begin{minipage}[c]{0.5\textwidth}
  \centering
  \includegraphics[width=30mm,height=65mm,angle=-90.0]{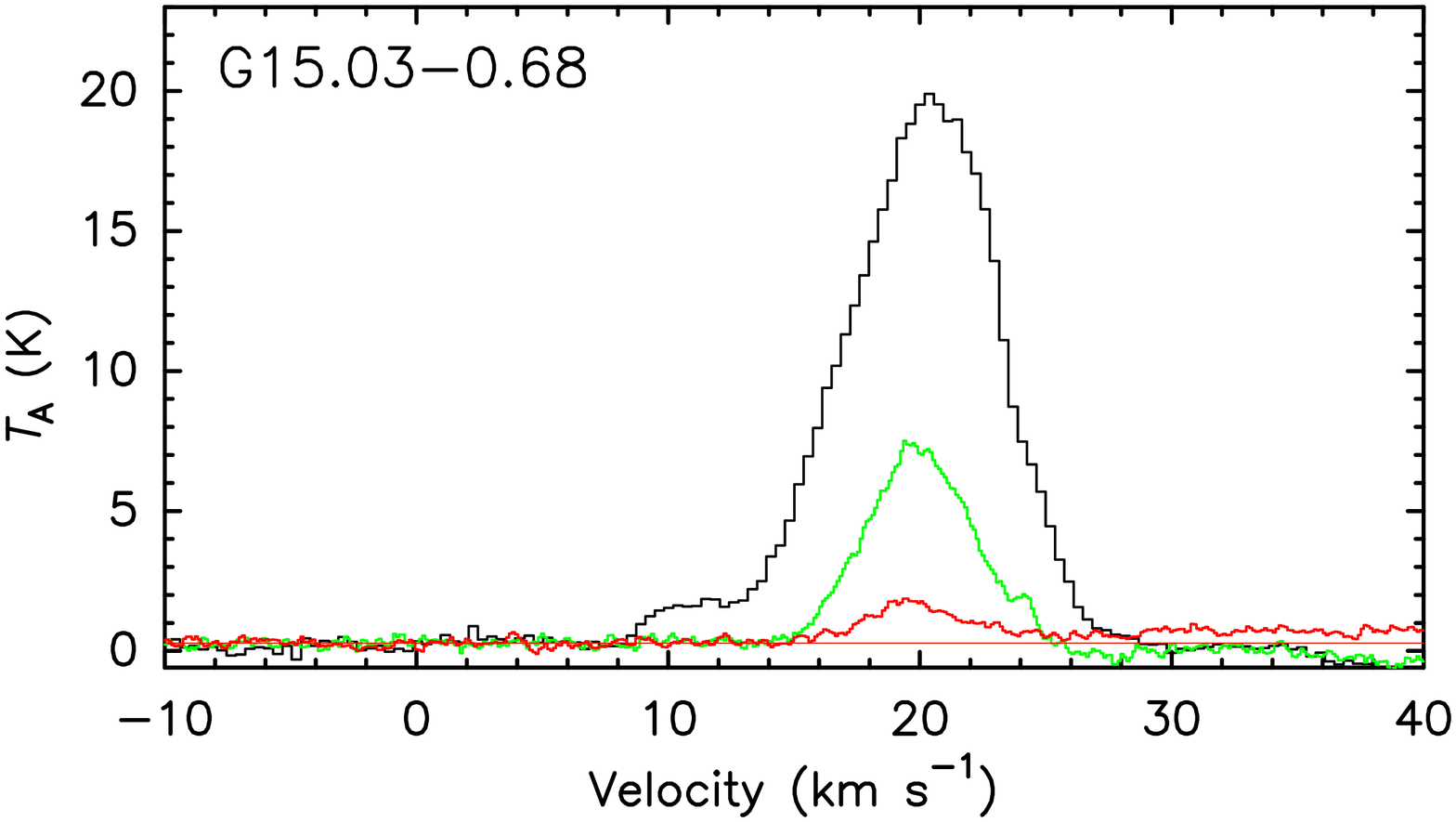}
\end{minipage}
\begin{minipage}[c]{0.5\textwidth}
  \centering
  \includegraphics[width=30mm,height=65mm,angle=-90.0]{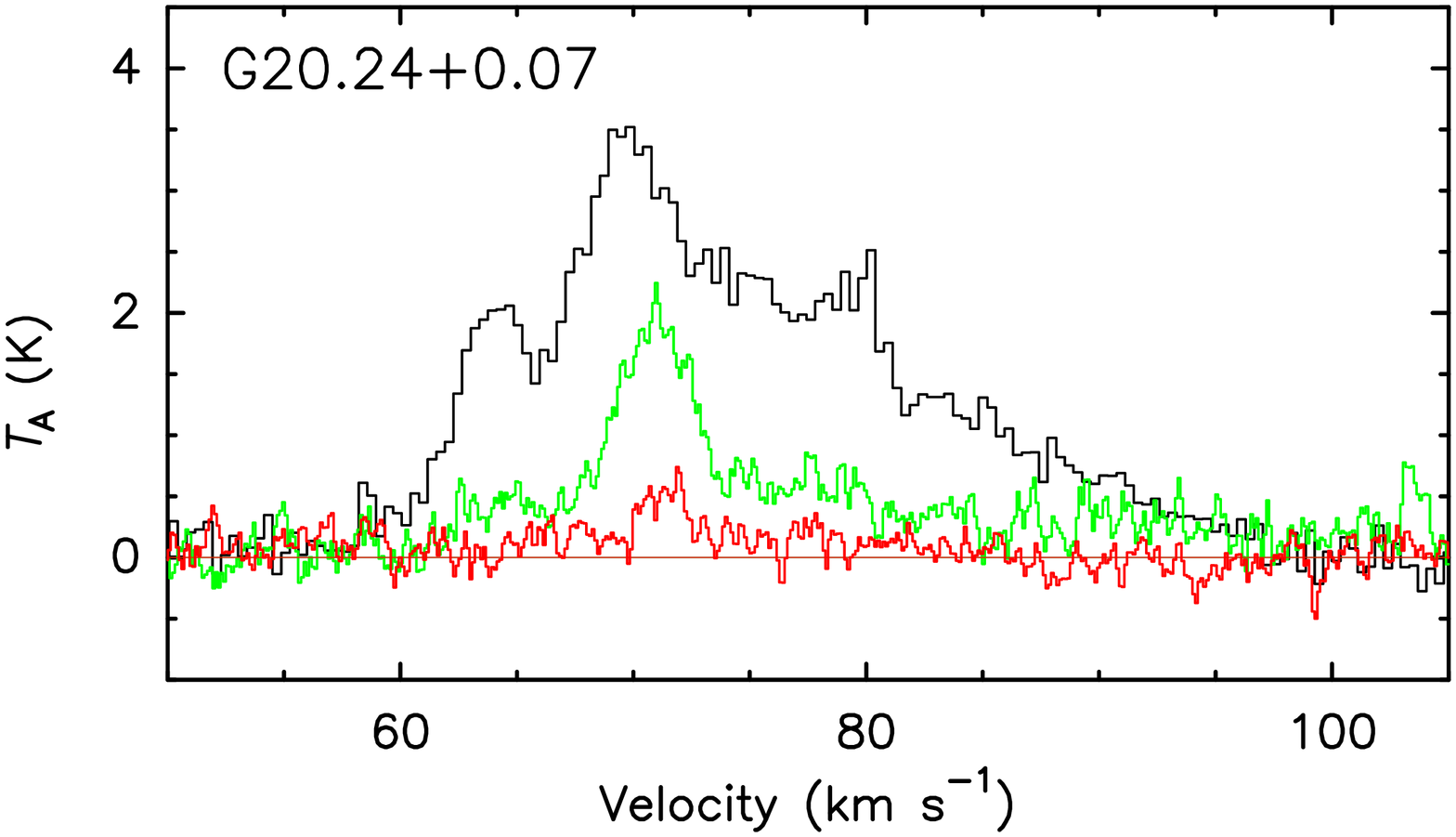}
\end{minipage}
\begin{minipage}[c]{0.5\textwidth}
  \centering
  \includegraphics[width=30mm,height=65mm,angle=-90.0]{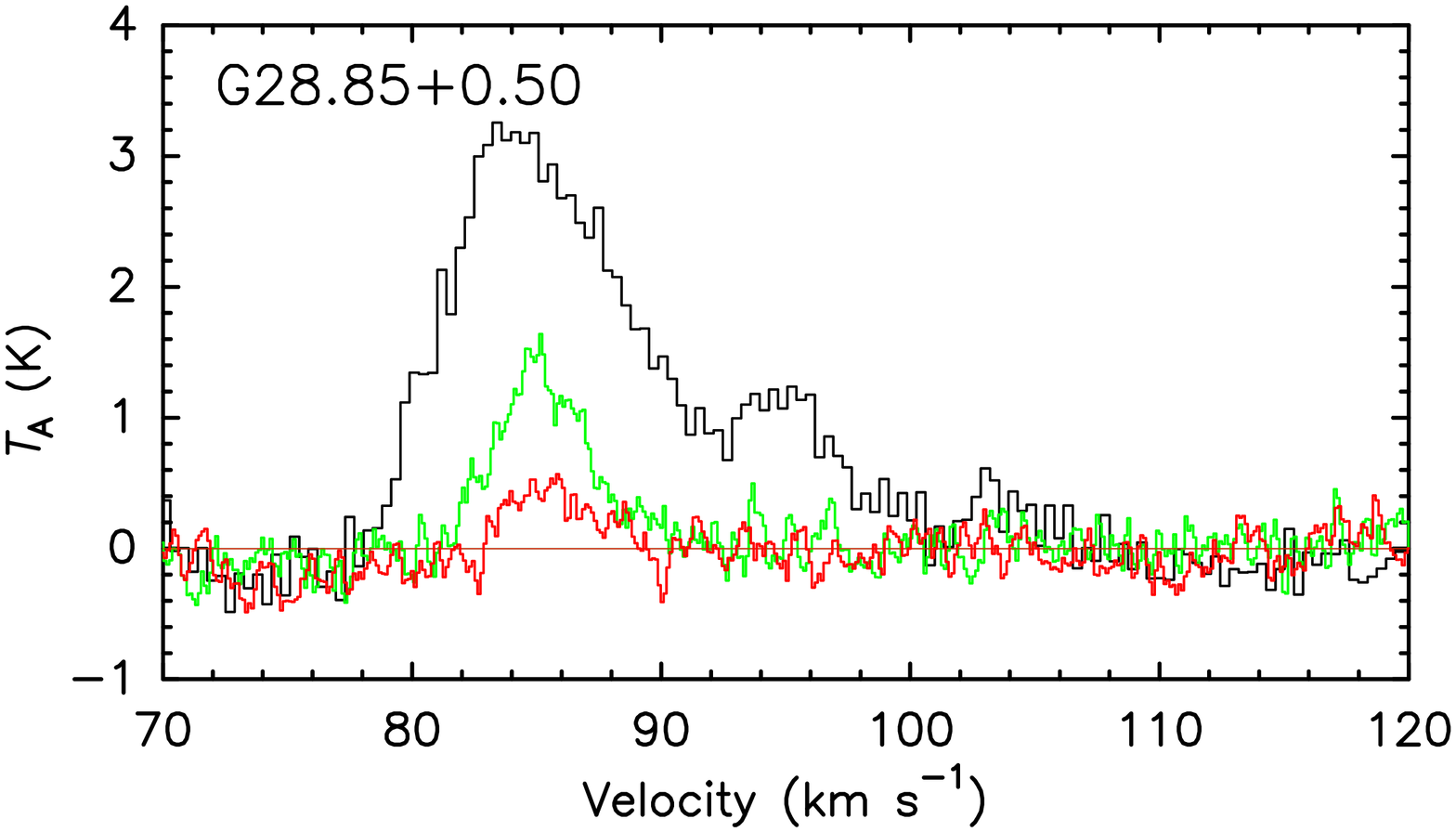}
\end{minipage}
\begin{minipage}[c]{0.5\textwidth}
  \centering
  \includegraphics[width=30mm,height=65mm,angle=-90.0]{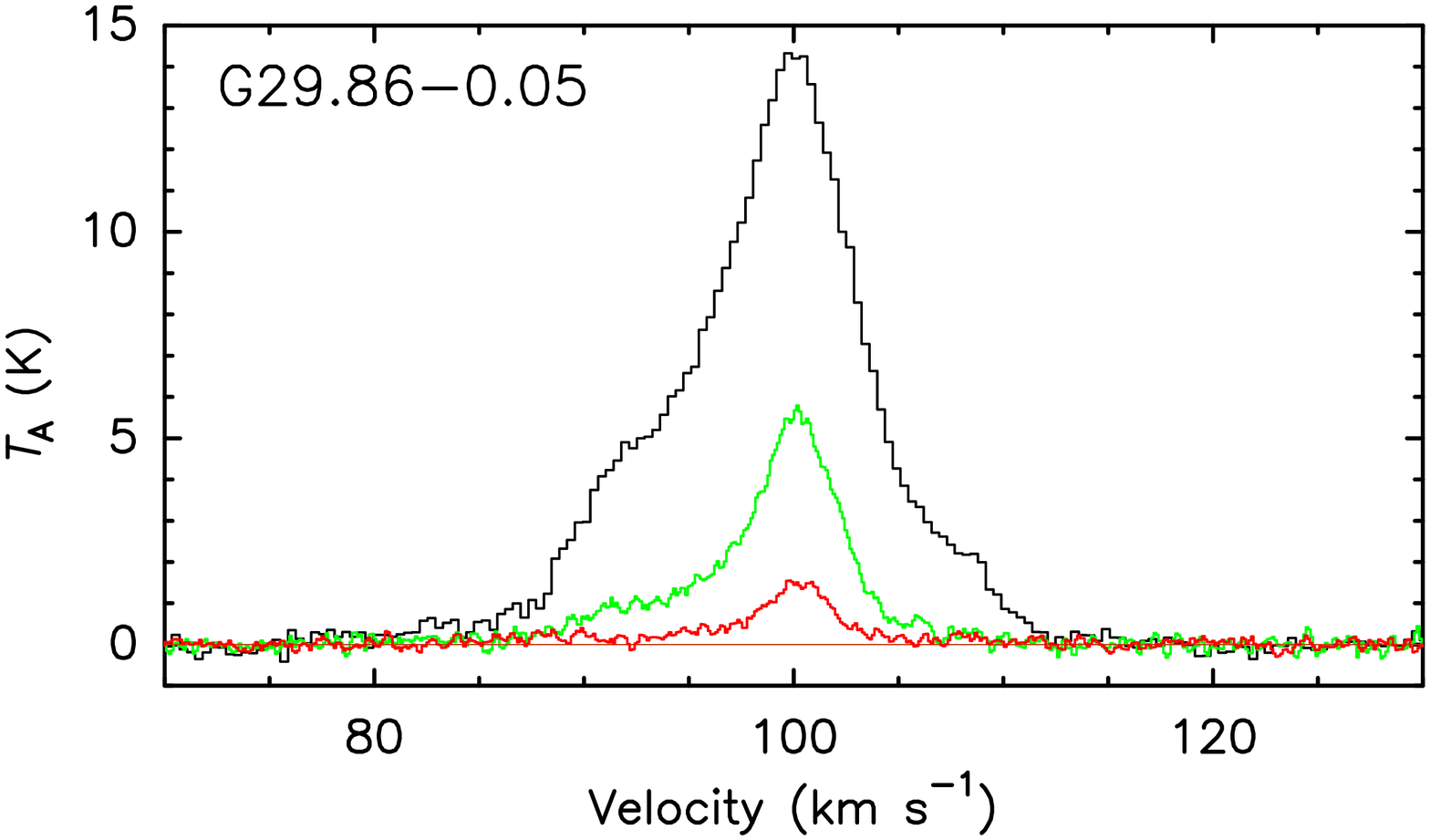}
  \end{minipage}
\begin{minipage}[c]{0.5\textwidth}
  \centering
  \includegraphics[width=30mm,height=65mm,angle=-90.0]{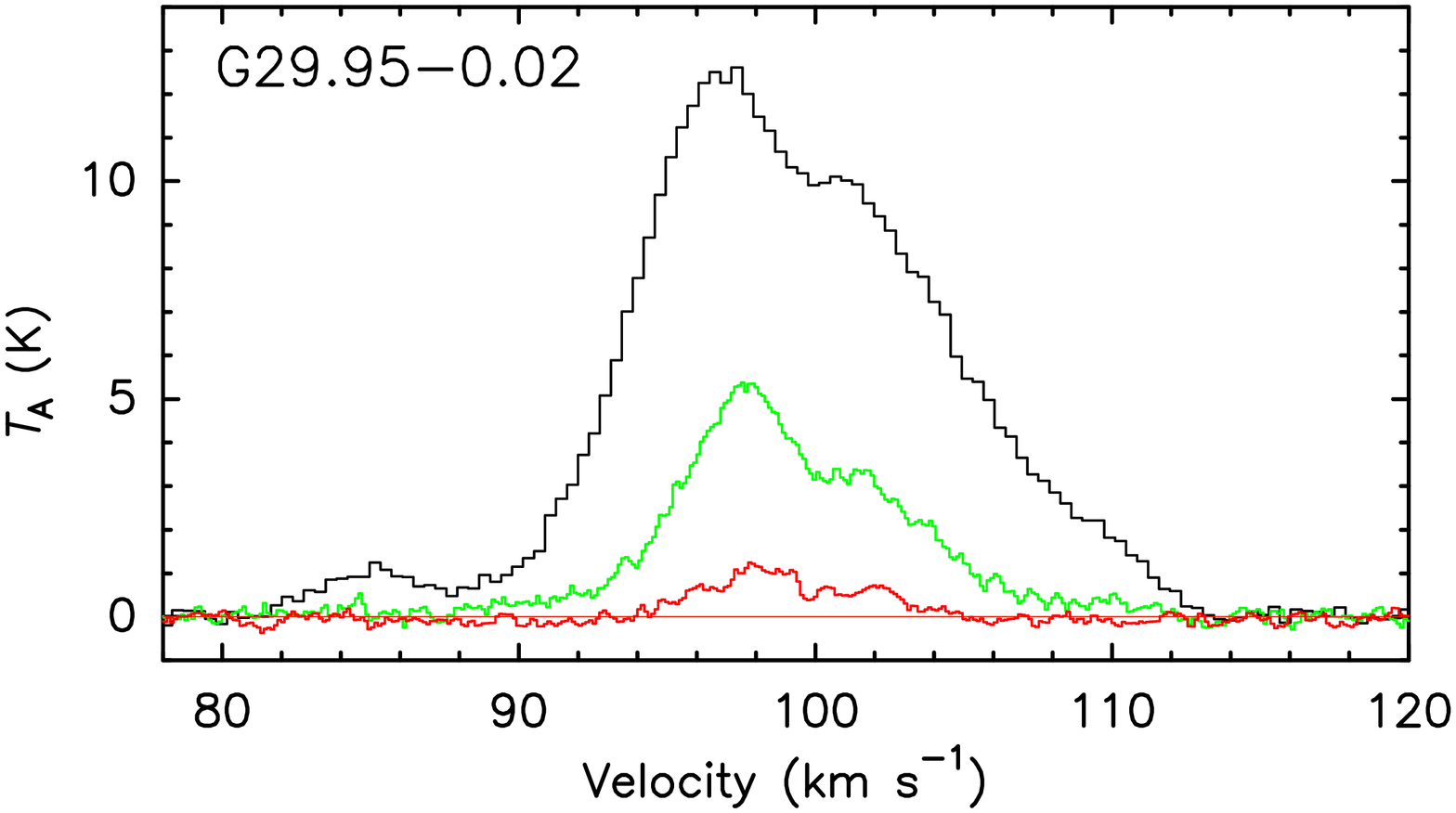}
\end{minipage}
\begin{minipage}[c]{0.5\textwidth}
  \centering
  \includegraphics[width=30mm,height=65mm,angle=-90.0]{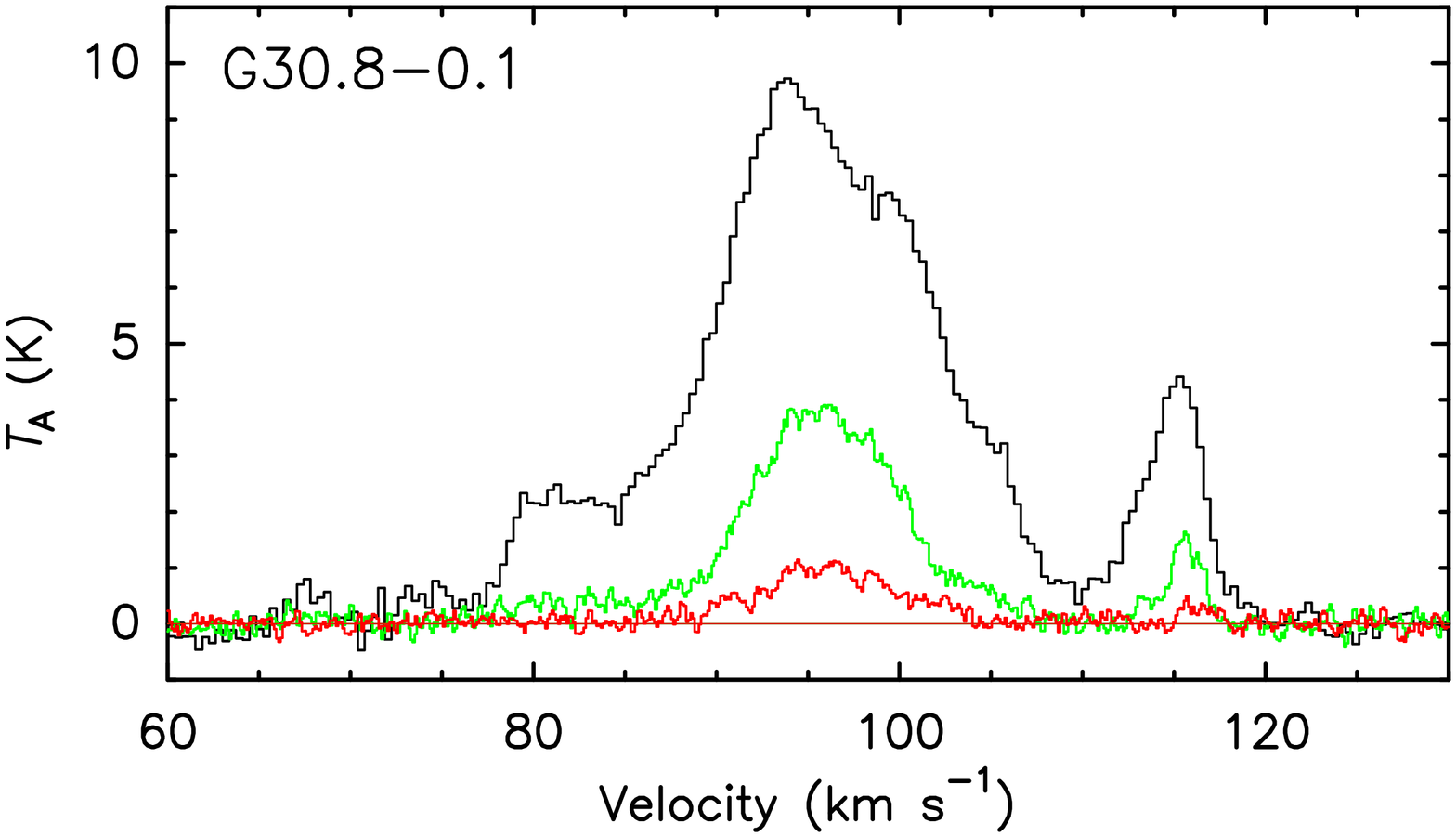}
\end{minipage}
\caption{Continued }
\end{figure}

\setcounter{figure}{2}

\begin{figure}

\begin{minipage}[c]{0.5\textwidth}
  \centering
  \includegraphics[width=30mm,height=65mm,angle=-90.0]{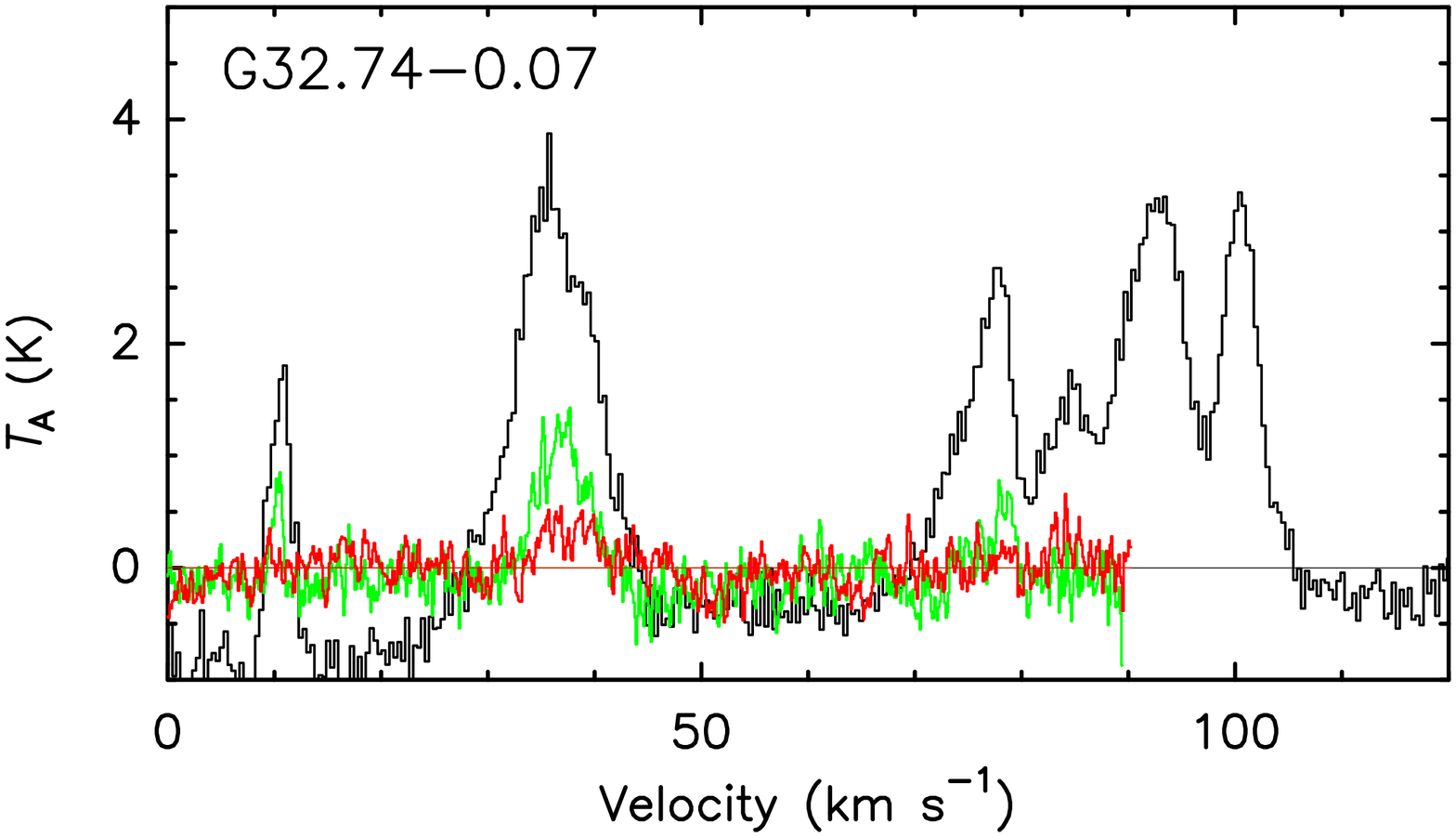}
\end{minipage}
\begin{minipage}[c]{0.5\textwidth}
  \centering
  \includegraphics[width=30mm,height=65mm,angle=-90.0]{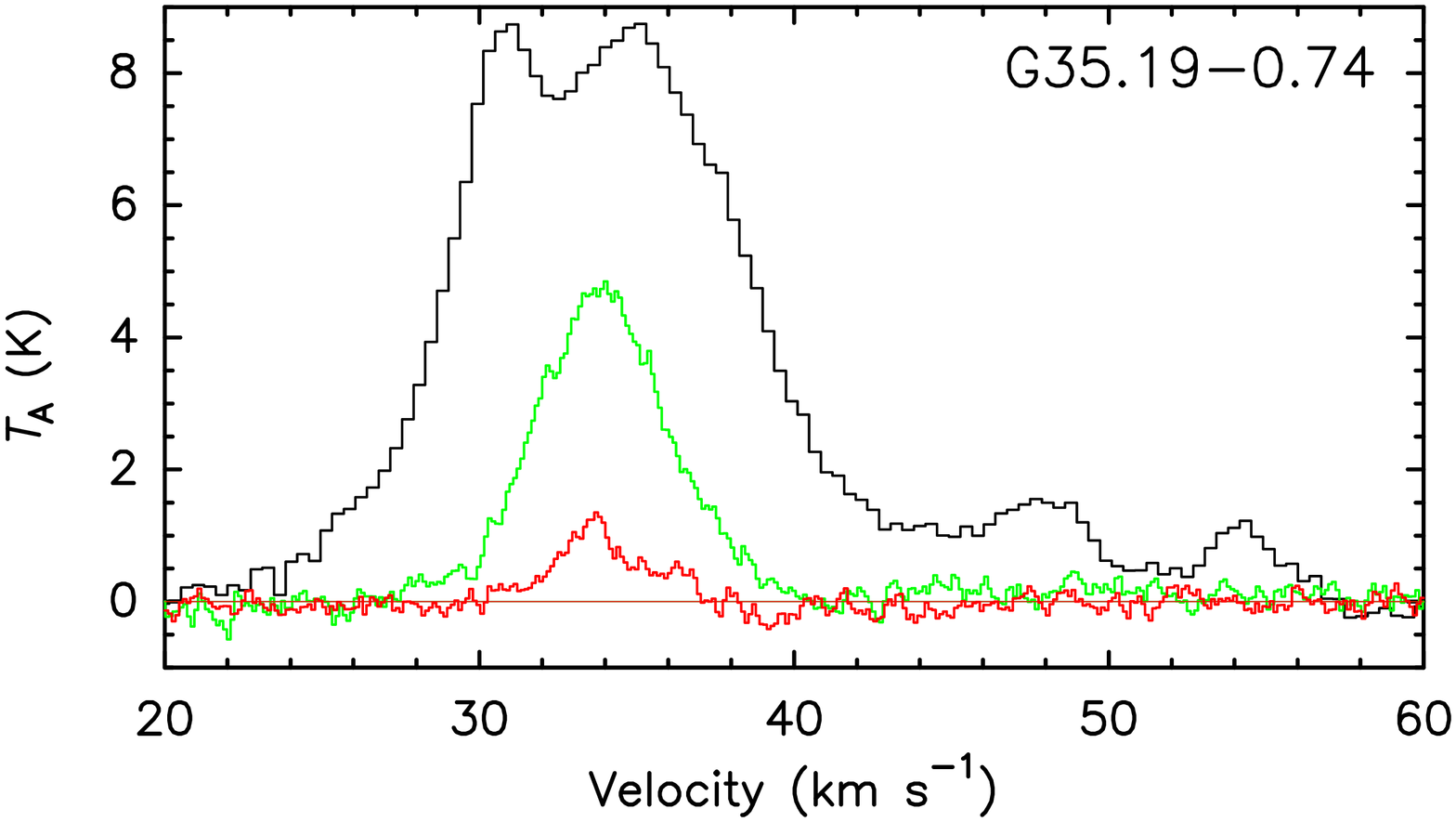}
\end{minipage}
\begin{minipage}[c]{0.5\textwidth}
  \centering
  \includegraphics[width=30mm,height=65mm,angle=-90.0]{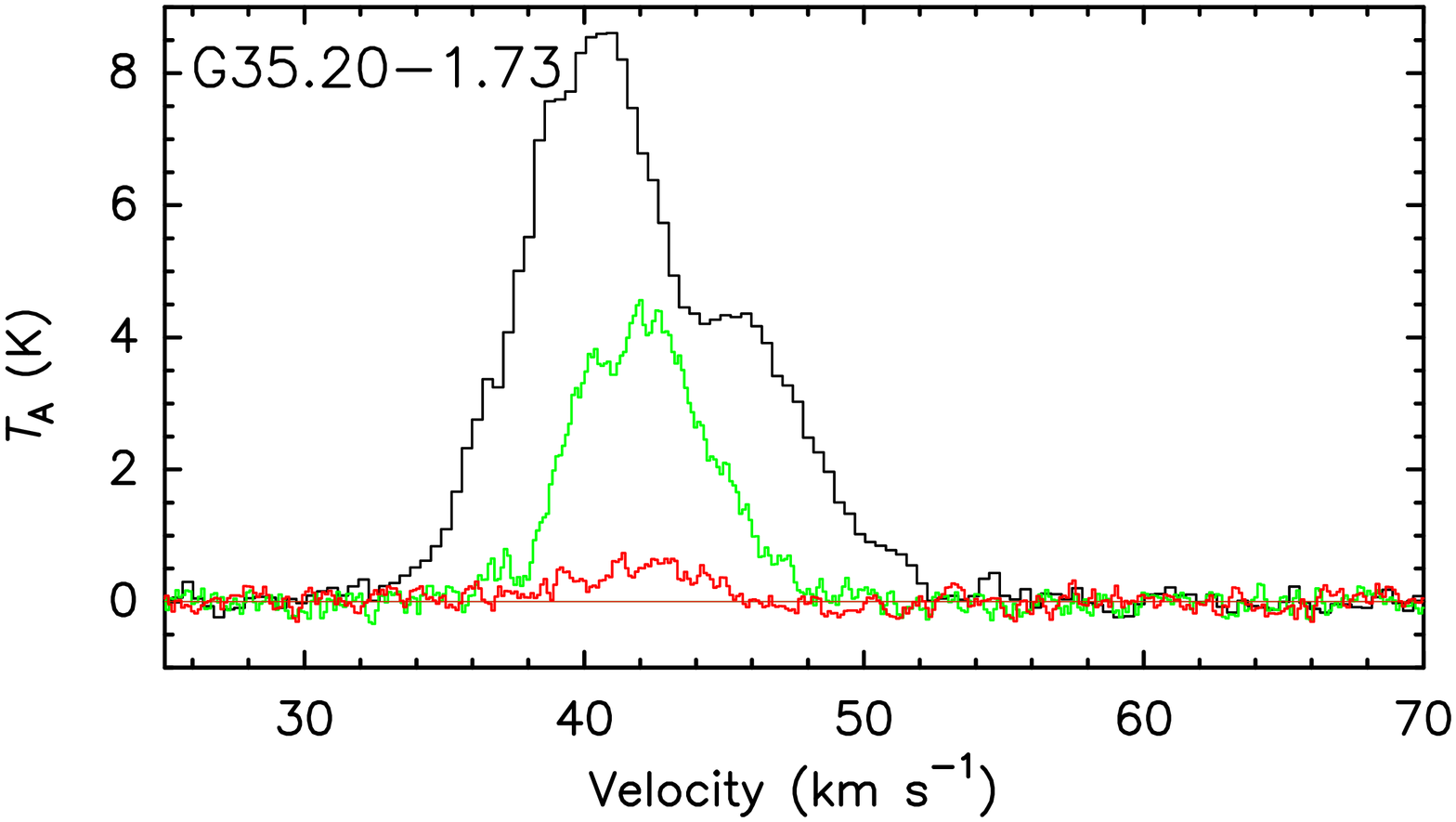}
\end{minipage}
\begin{minipage}[c]{0.5\textwidth}
  \centering
  \includegraphics[width=30mm,height=65mm,angle=-90.0]{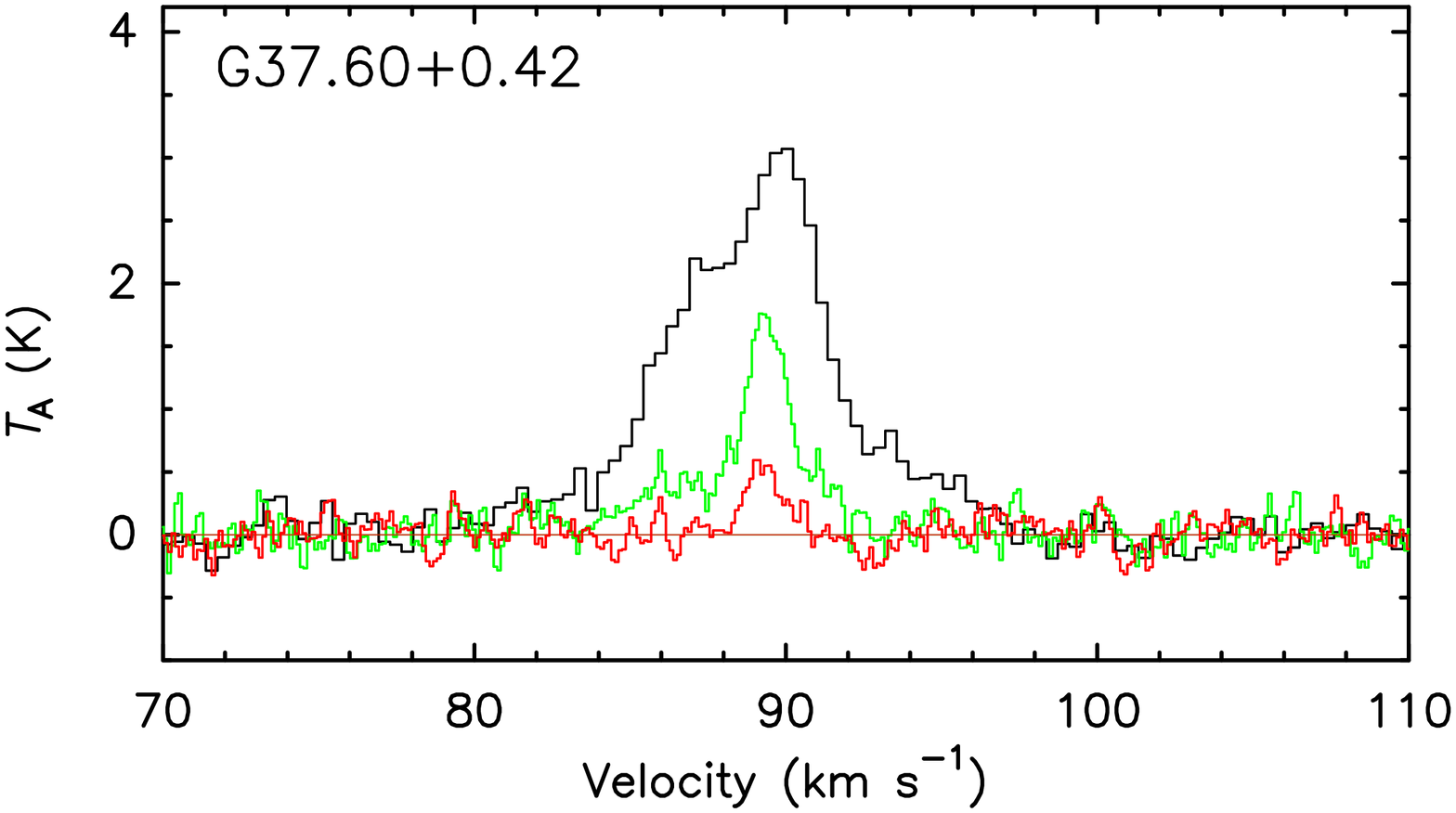}
\end{minipage}
\begin{minipage}[c]{0.5\textwidth}
  \centering
  \includegraphics[width=30mm,height=65mm,angle=-90.0]{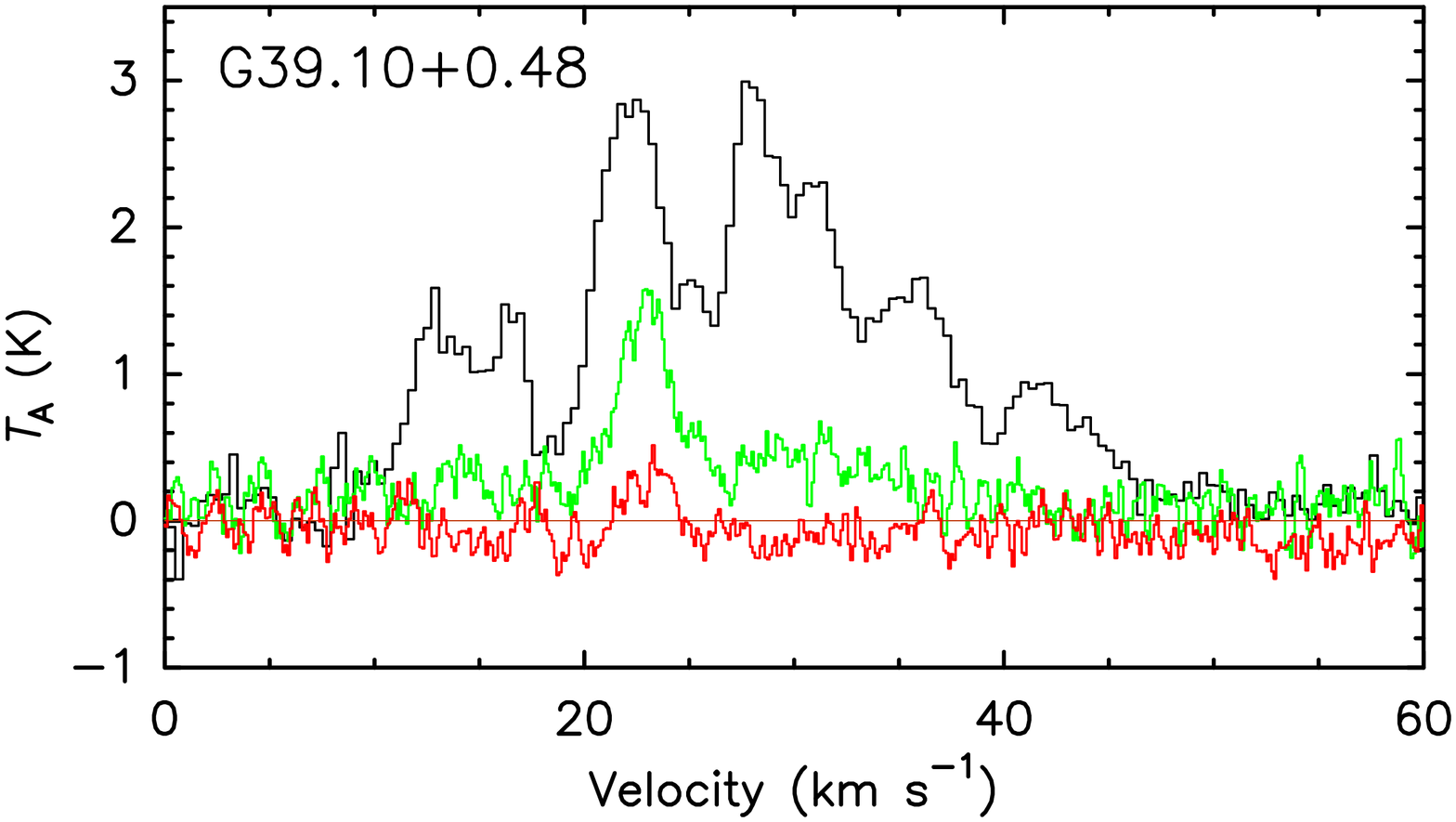}
\end{minipage}
\begin{minipage}[c]{0.5\textwidth}
  \centering
  \includegraphics[width=30mm,height=65mm,angle=-90.0]{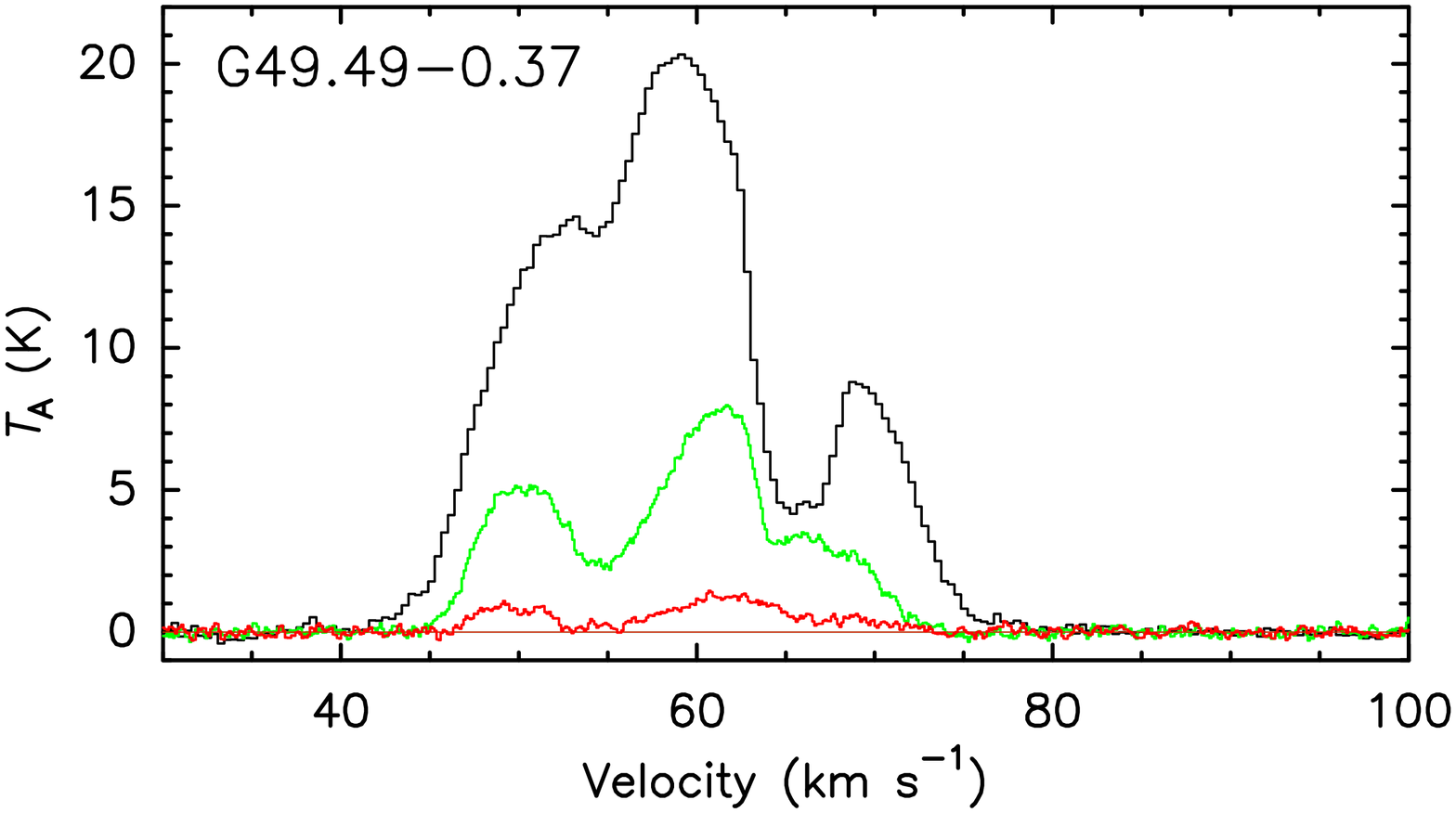}
\end{minipage}
\begin{minipage}[c]{0.5\textwidth}
  \centering
  \includegraphics[width=30mm,height=65mm,angle=-90.0]{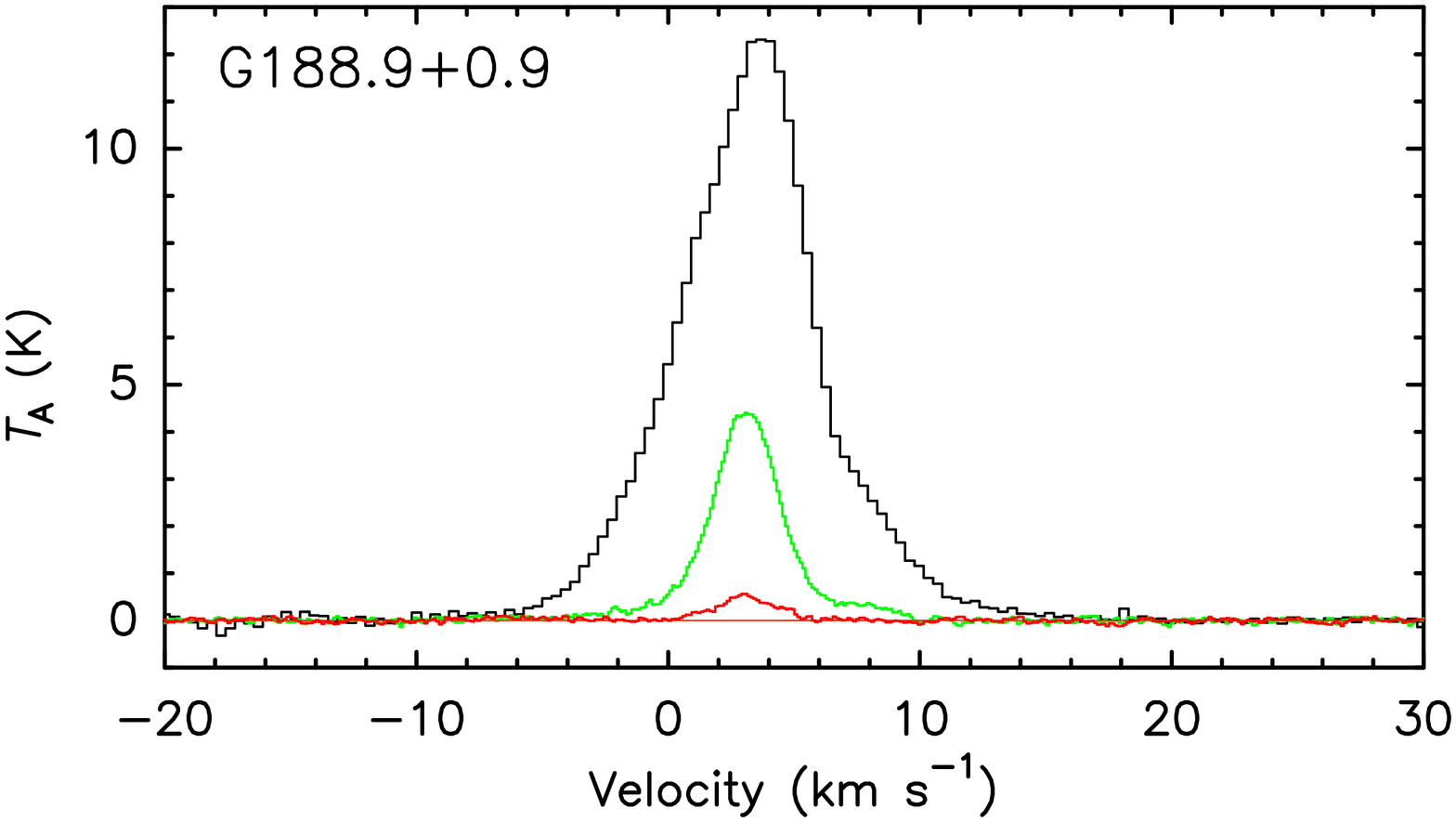}
\end{minipage}
\begin{minipage}[c]{0.5\textwidth}
  \centering
  \includegraphics[width=30mm,height=65mm,angle=-90.0]{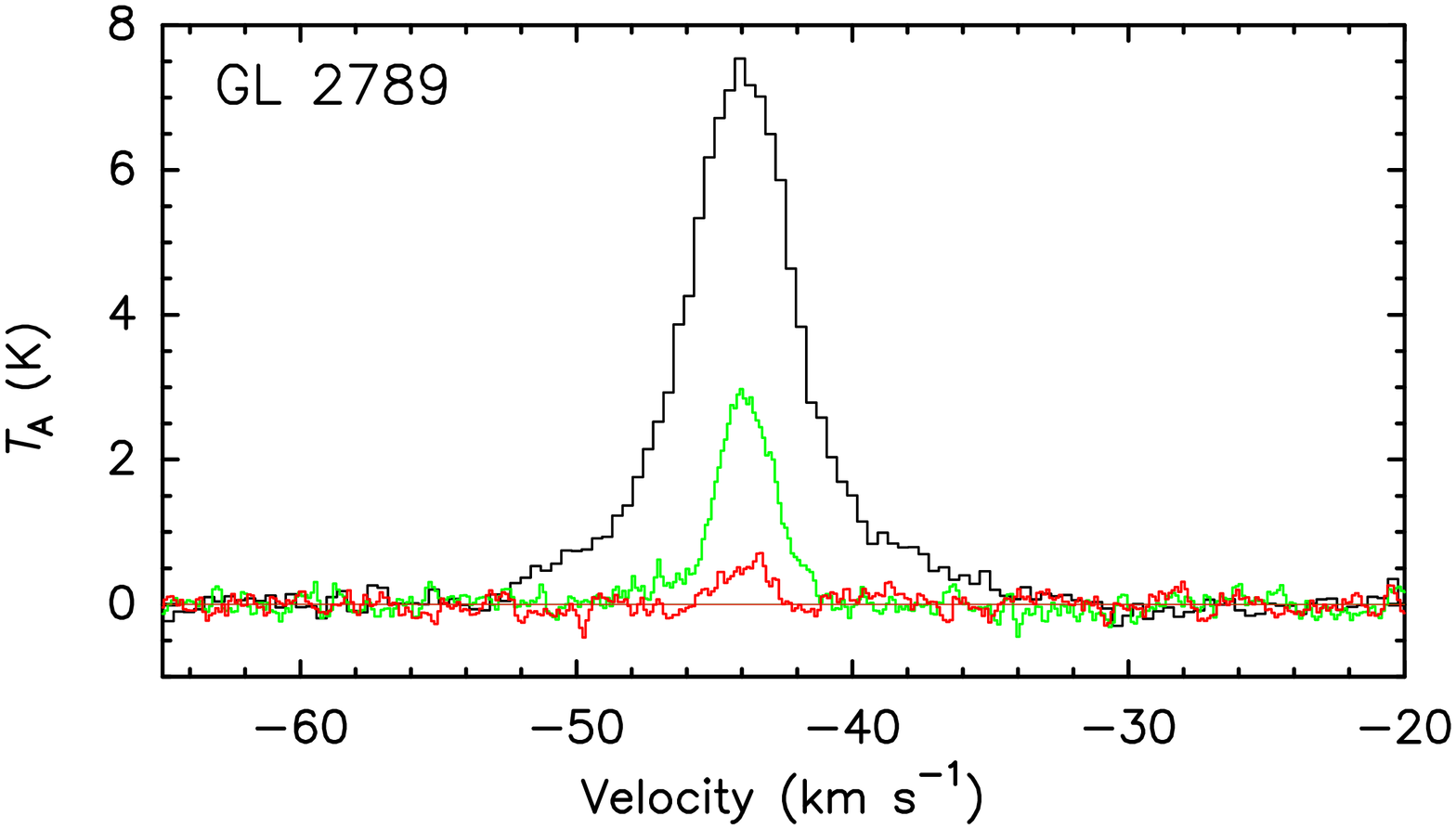}
\end{minipage}
\begin{minipage}[c]{0.5\textwidth}
  \centering
  \includegraphics[width=30mm,height=65mm,angle=-90.0]{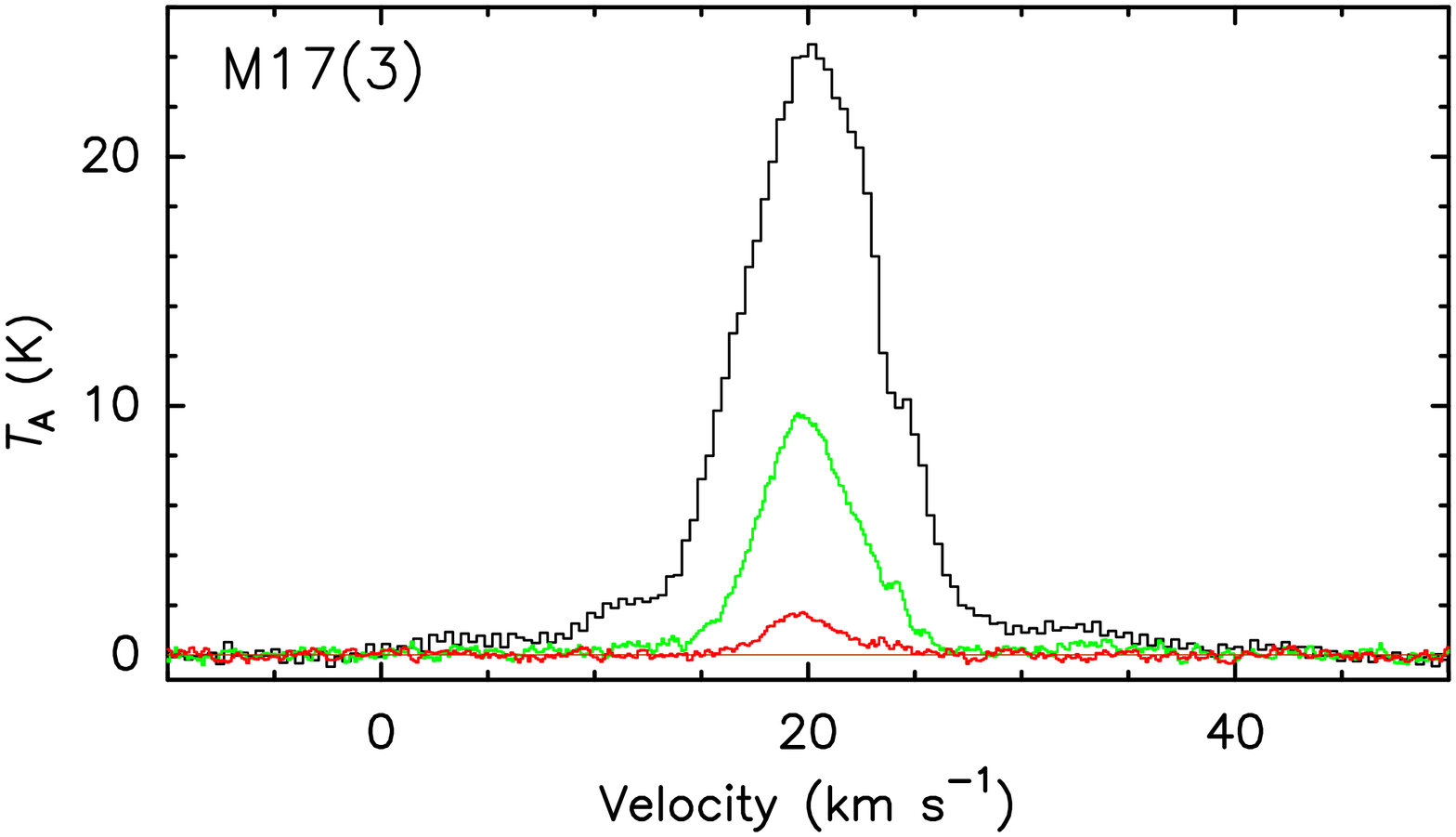}
\end{minipage}
\begin{minipage}[c]{0.5\textwidth}
  \centering
  \includegraphics[width=30mm,height=65mm,angle=-90.0]{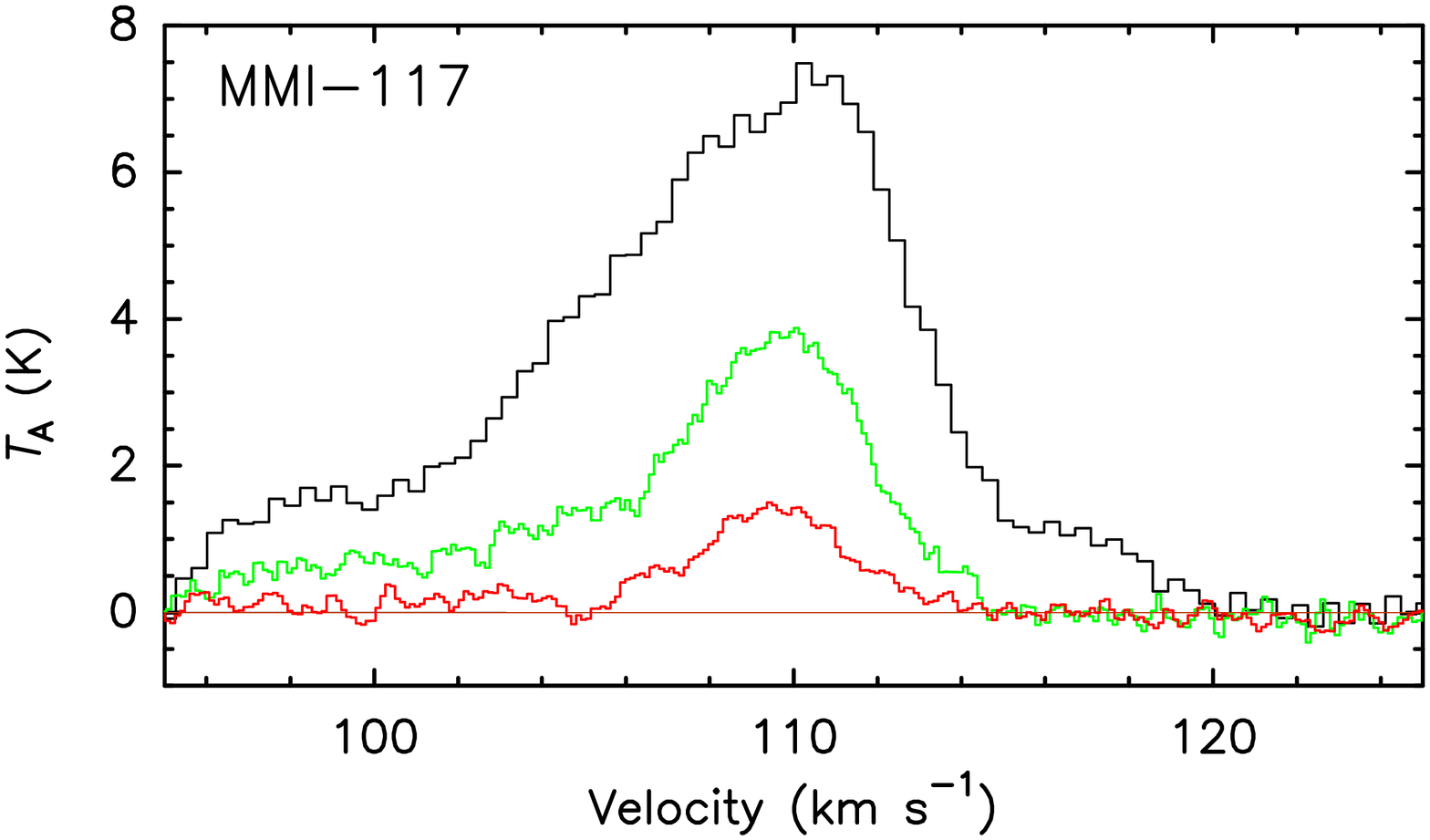}
\end{minipage}
\begin{minipage}[c]{0.5\textwidth}
  \centering
  \includegraphics[width=30mm,height=65mm,angle=-90.0]{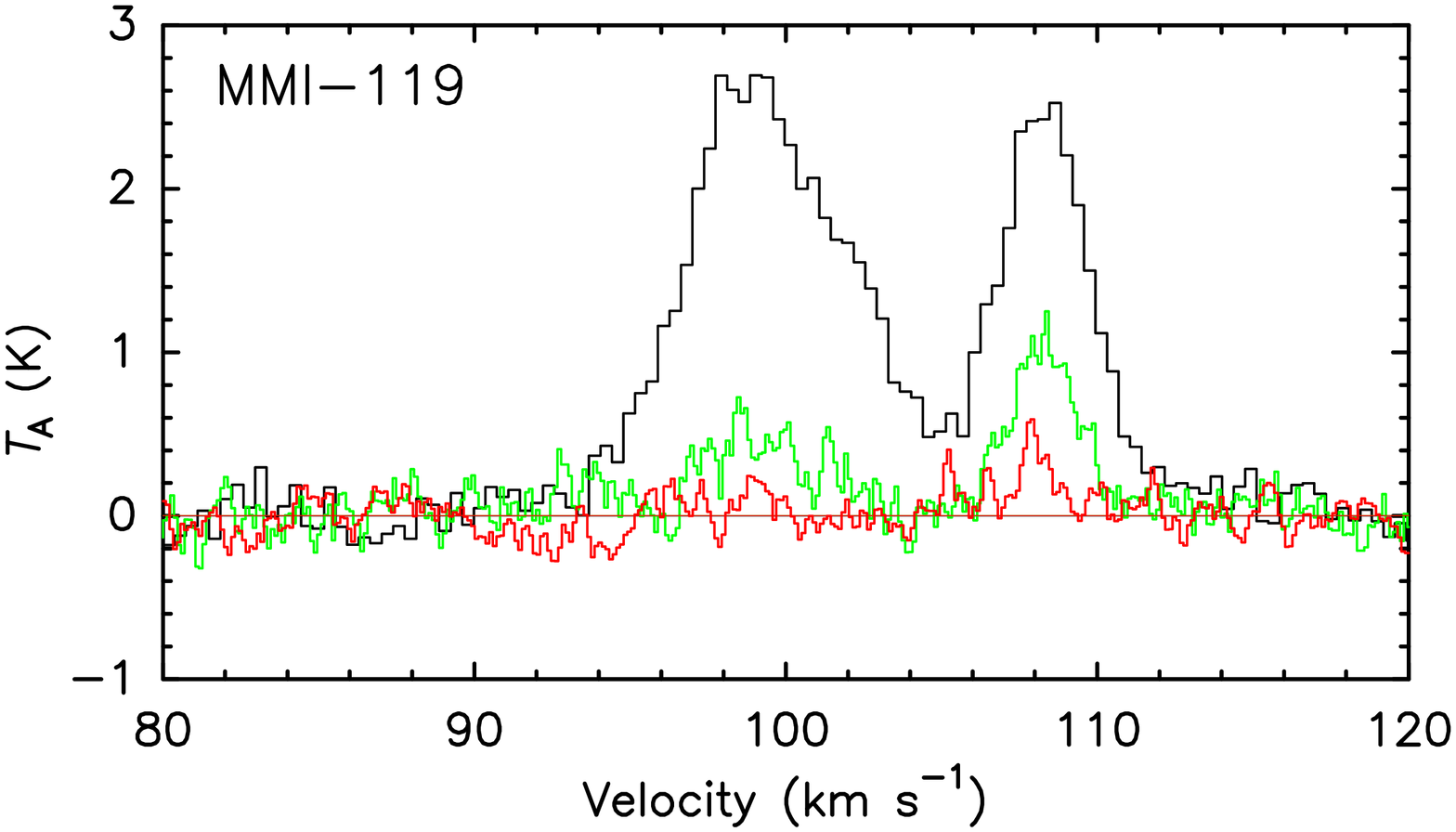}
\end{minipage}
\begin{minipage}[c]{0.5\textwidth}
  \centering
  \includegraphics[width=30mm,height=65mm,angle=-90.0]{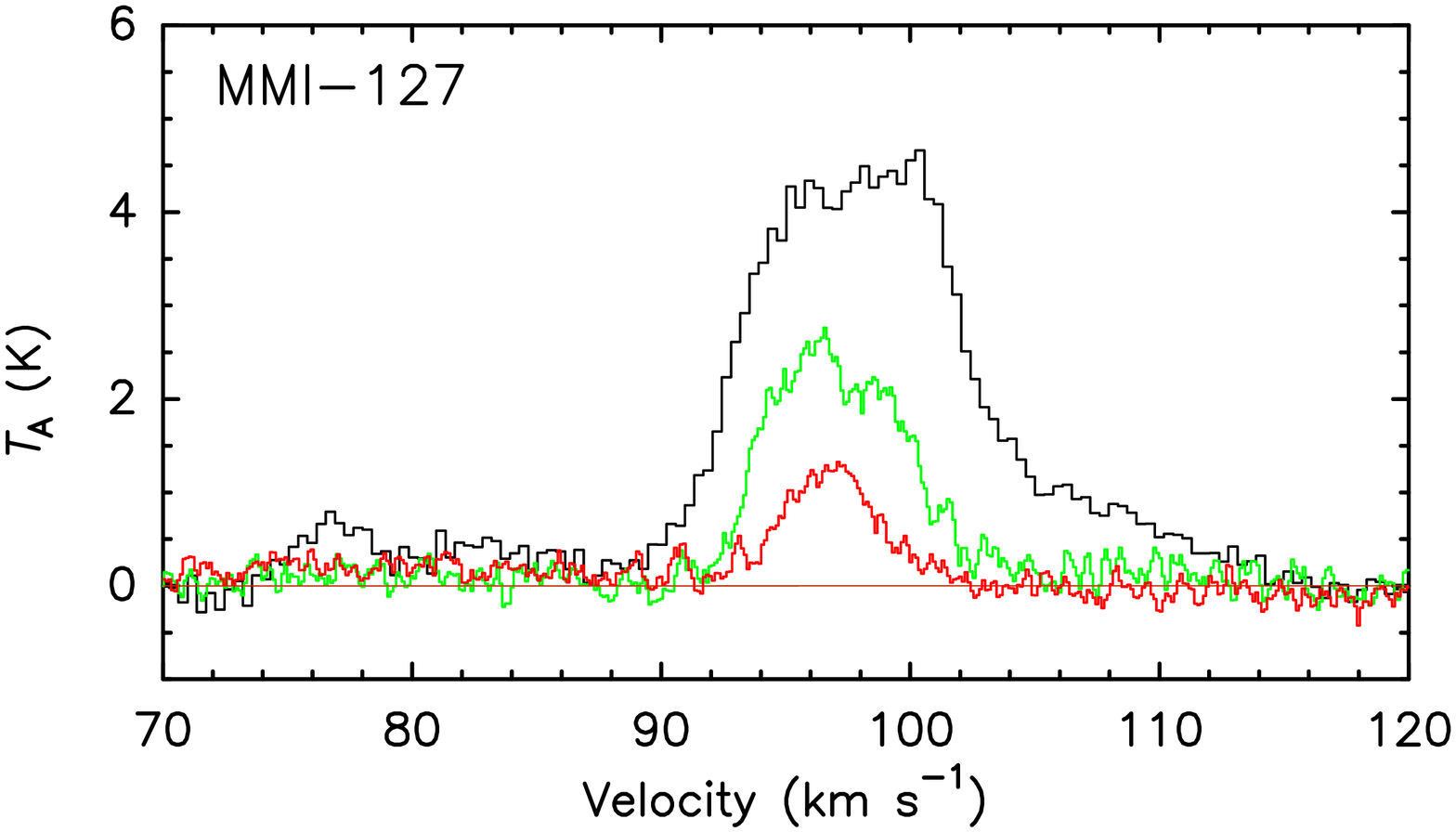}
\end{minipage}
\caption{Continued }
\end{figure}

\setcounter{figure}{2}

\begin{figure}
\begin{minipage}[c]{0.5\textwidth}
  \centering
  \includegraphics[width=30mm,height=65mm,angle=-90.0]{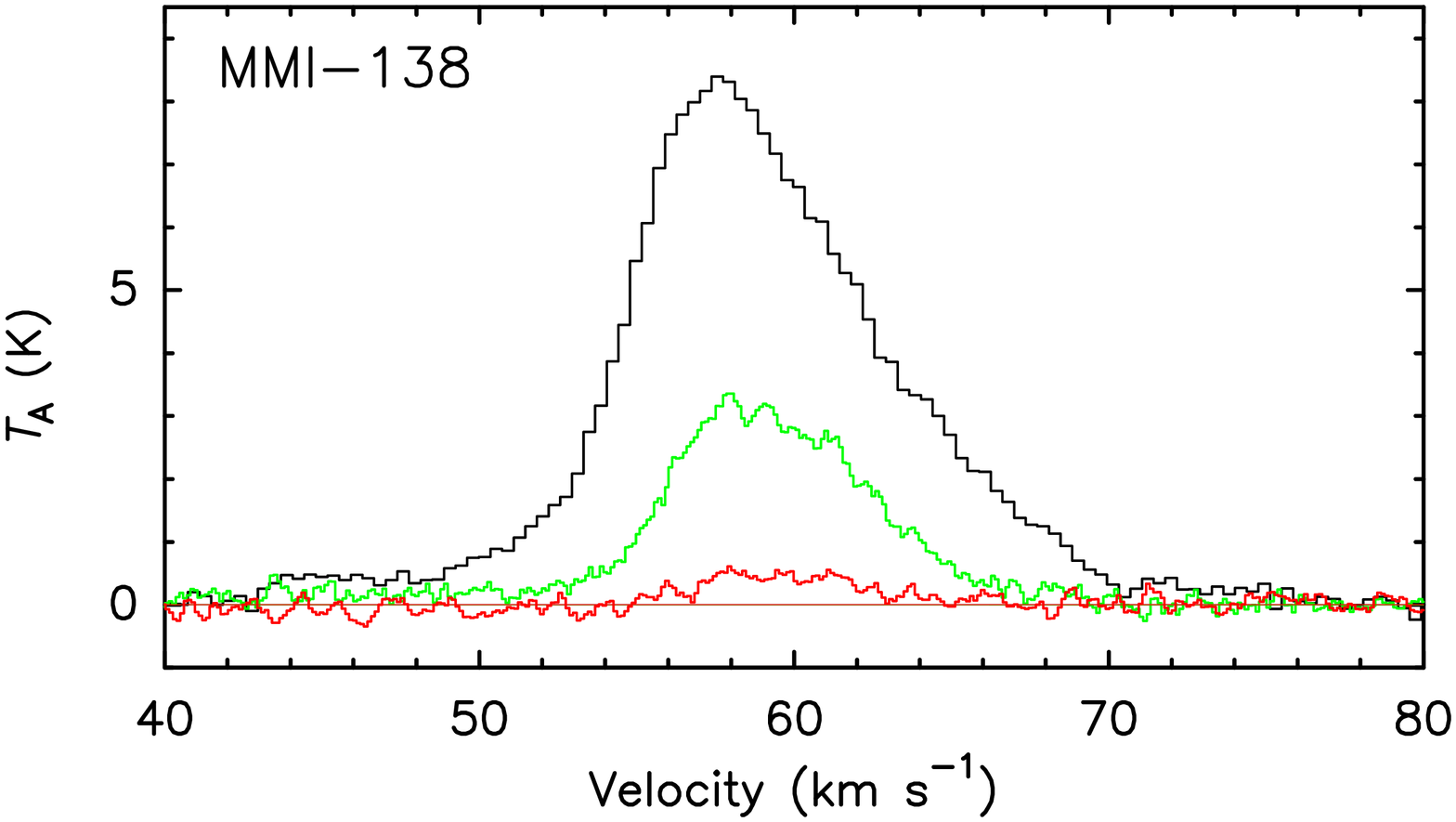}
\end{minipage}
\begin{minipage}[c]{0.5\textwidth}
  \centering
  \includegraphics[width=30mm,height=65mm,angle=-90.0]{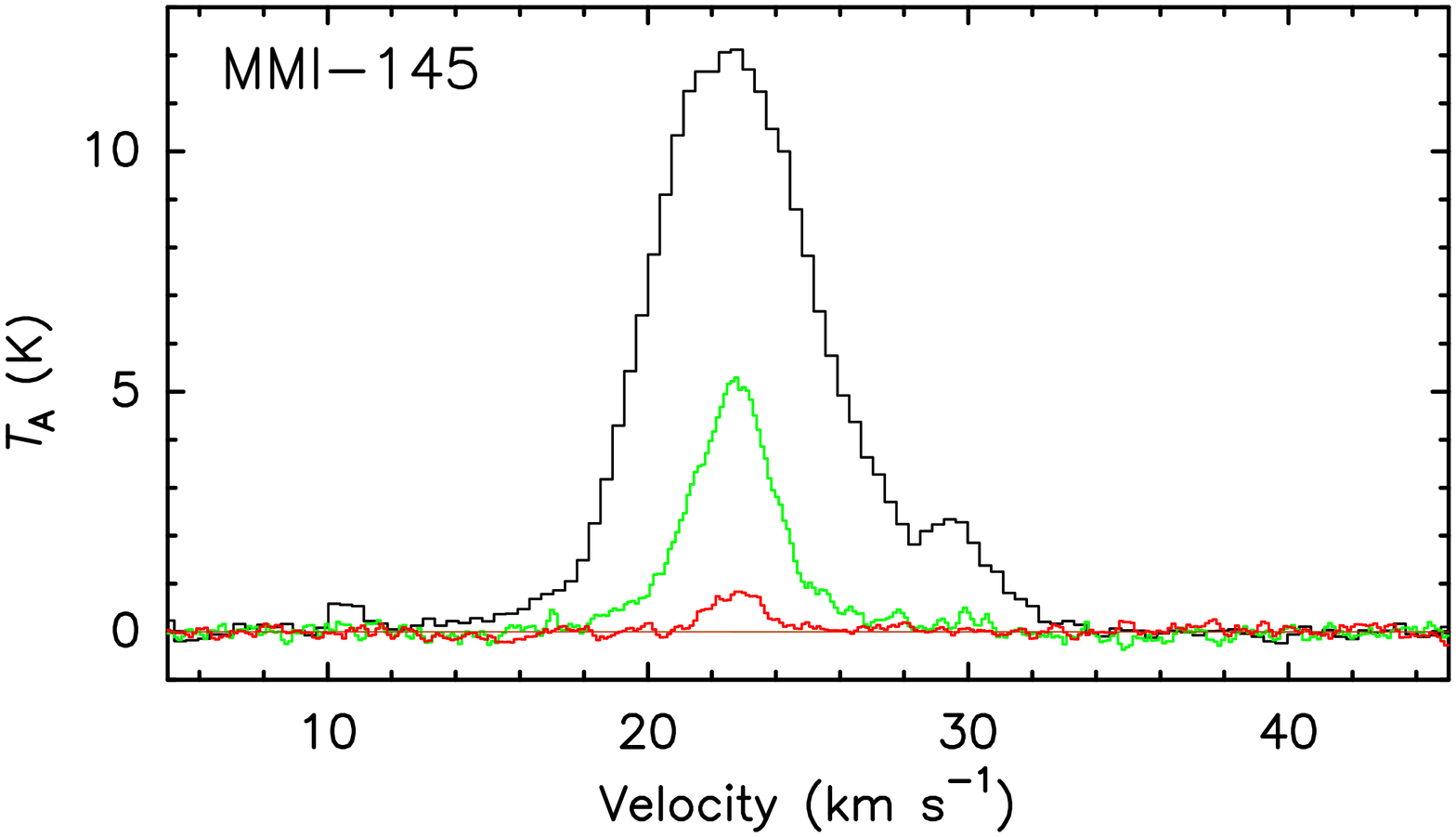}
\end{minipage}
\begin{minipage}[c]{0.5\textwidth}
  \centering
  \includegraphics[width=30mm,height=65mm,angle=-90.0]{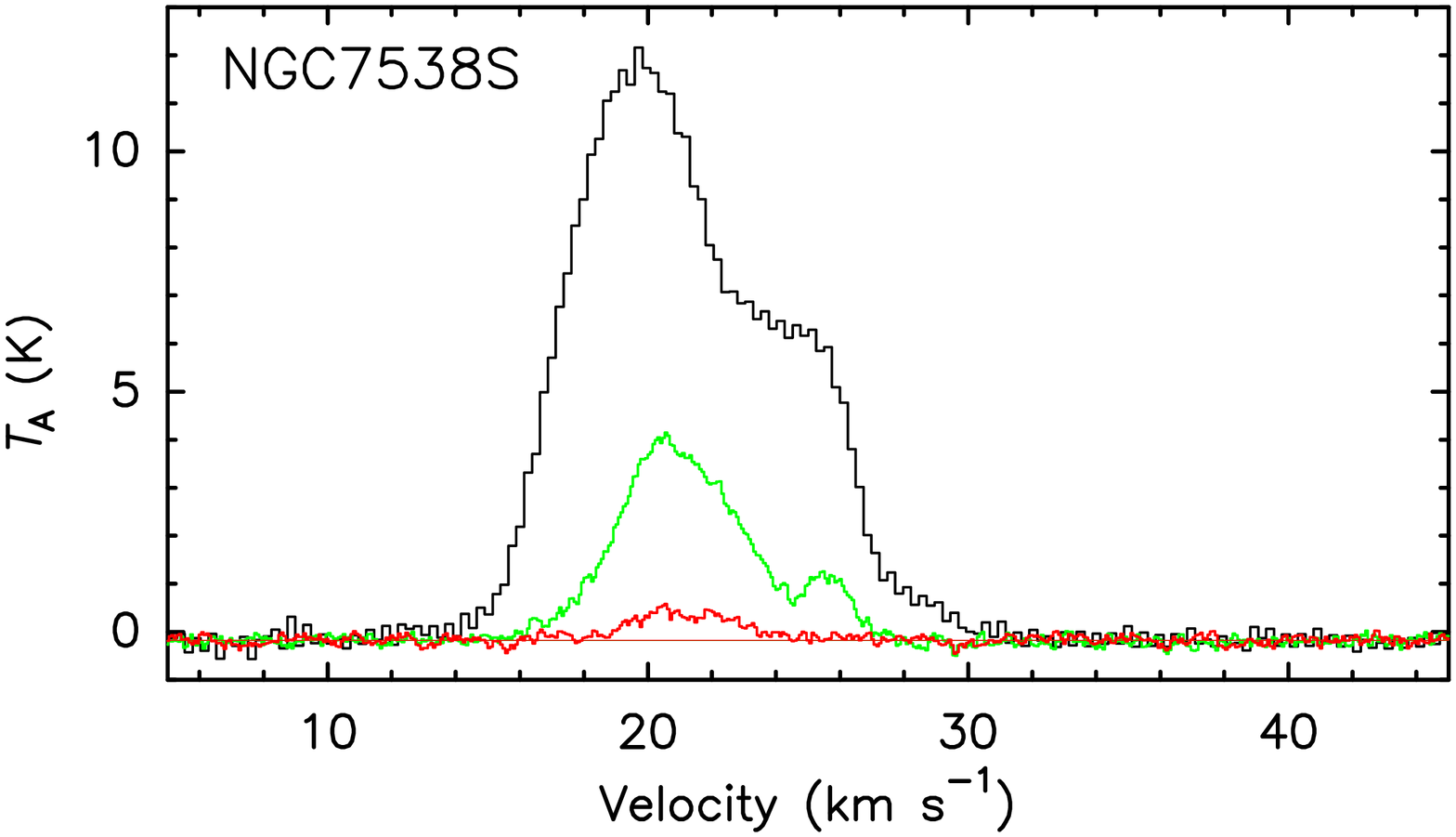}
\end{minipage}
\begin{minipage}[c]{0.5\textwidth}
  \centering
  \includegraphics[width=30mm,height=58mm,angle=-90.0]{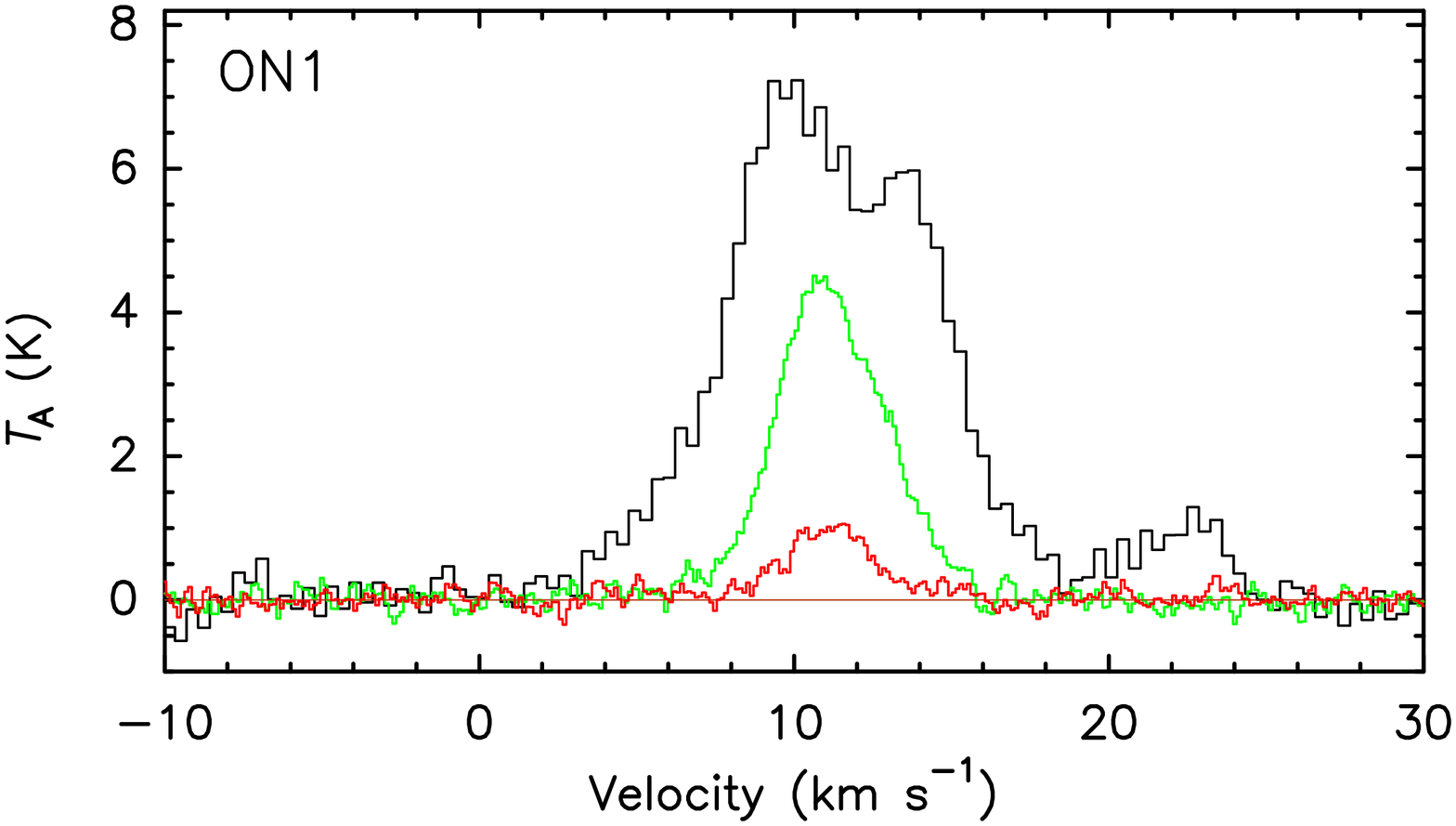}
\end{minipage}
\begin{minipage}[c]{0.5\textwidth}
  \centering
  \includegraphics[width=30mm,height=65mm,angle=-90.0]{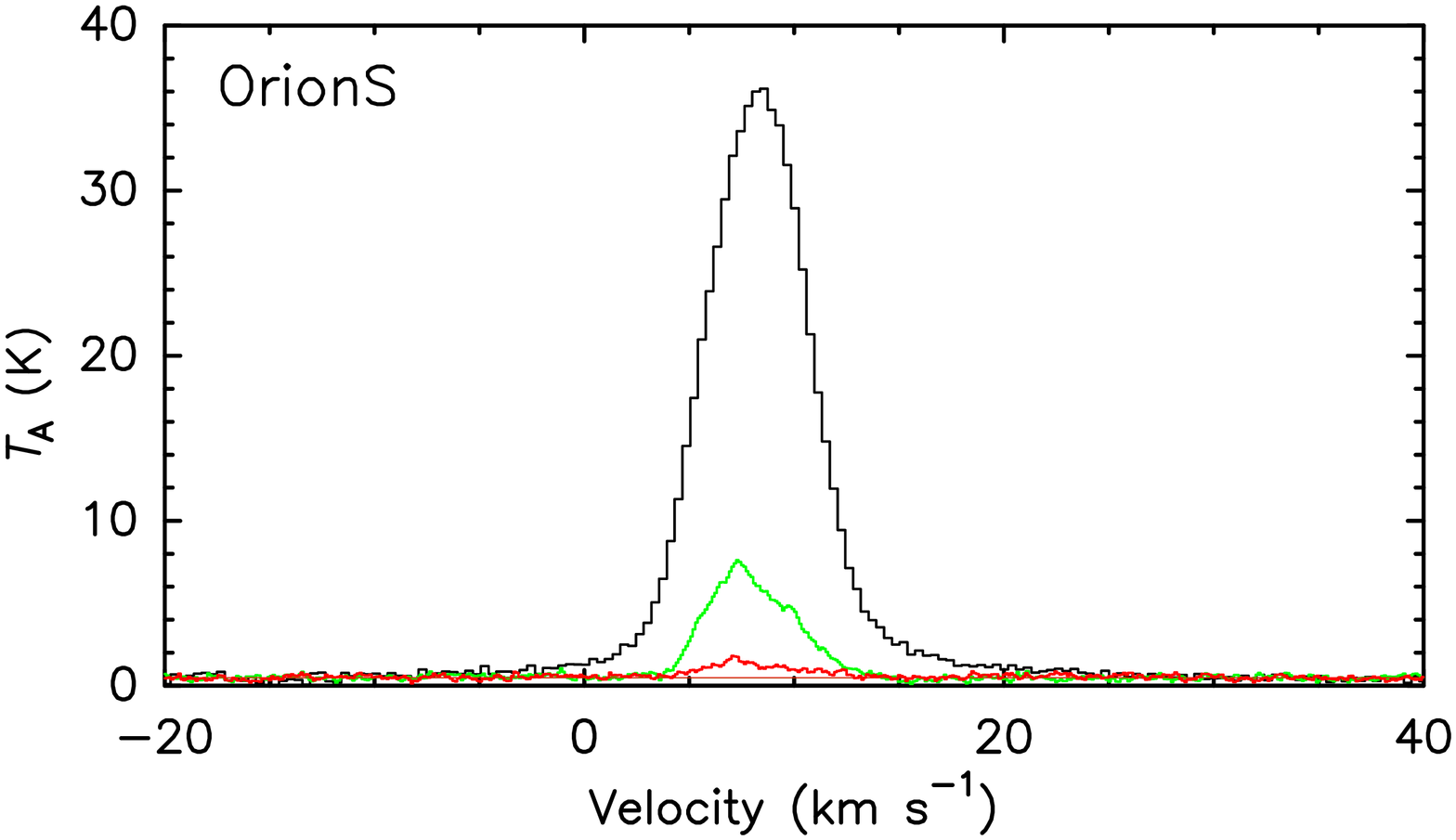}
\end{minipage}
\begin{minipage}[c]{0.5\textwidth}
  \centering
  \includegraphics[width=30mm,height=65mm,angle=-90.0]{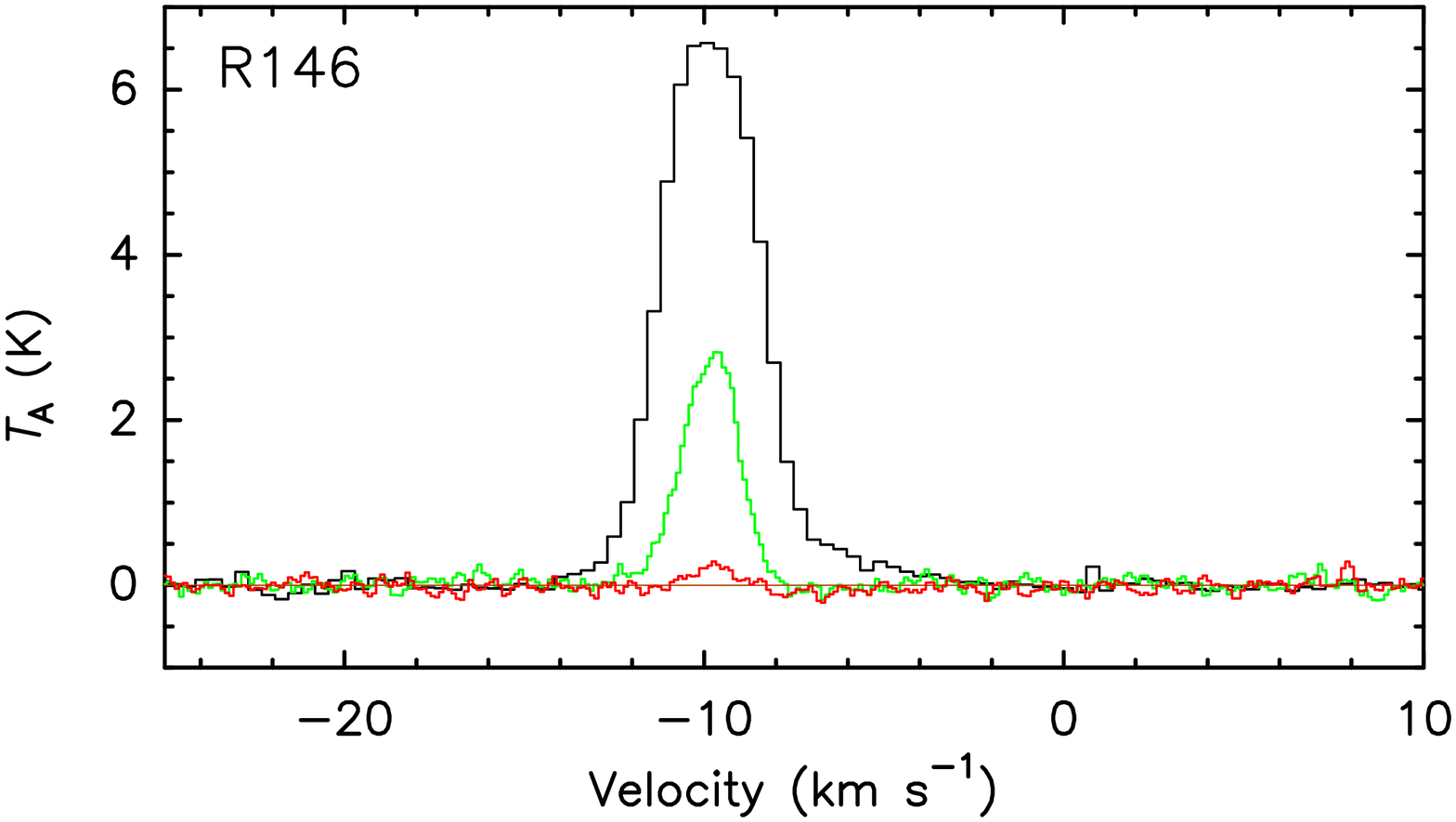}
\end{minipage}
\begin{minipage}[c]{0.5\textwidth}
  \centering
  \includegraphics[width=30mm,height=65mm,angle=-90.0]{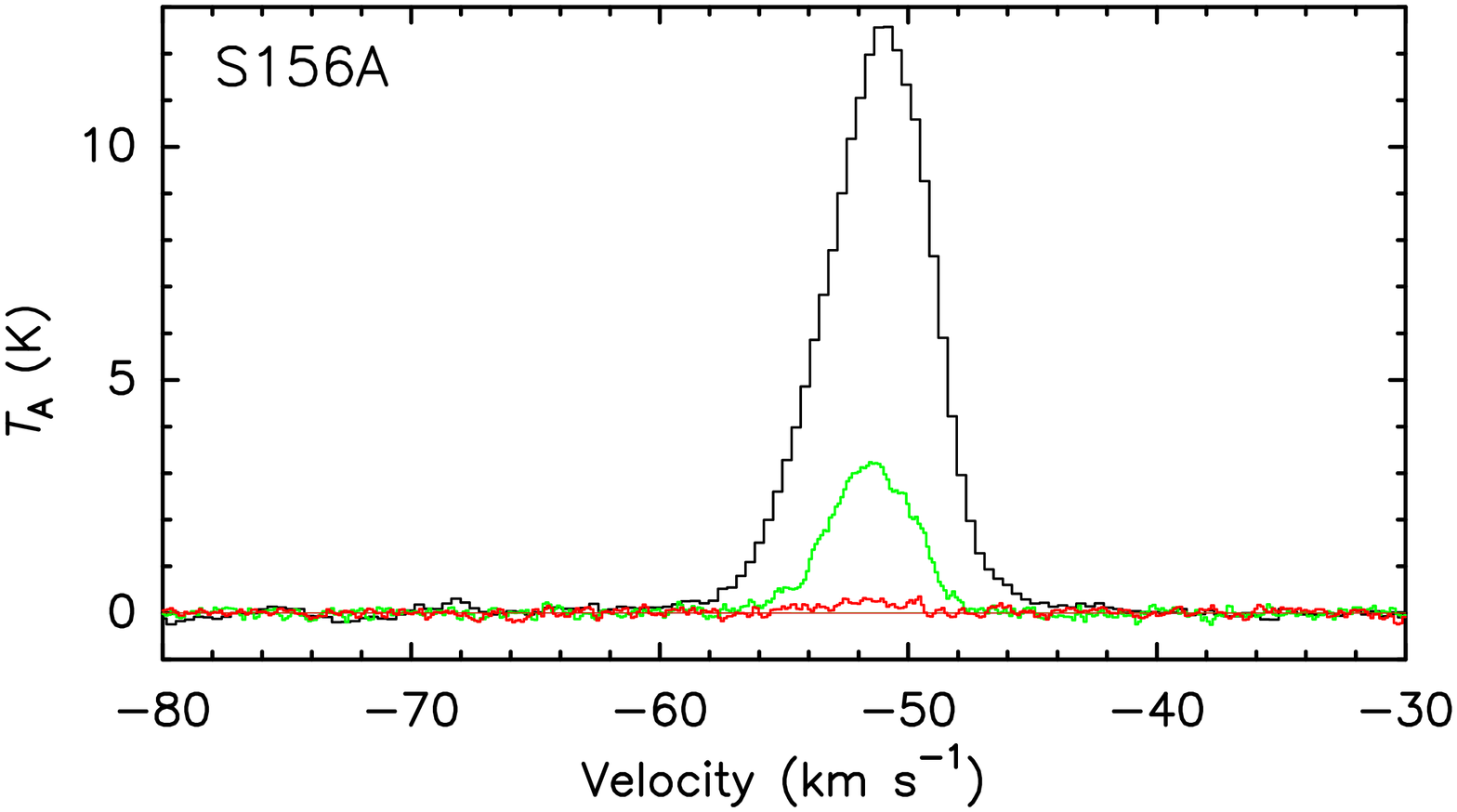}
\end{minipage}
\begin{minipage}[c]{0.5\textwidth}
  \centering
  \includegraphics[width=30mm,height=65mm,angle=-90.0]{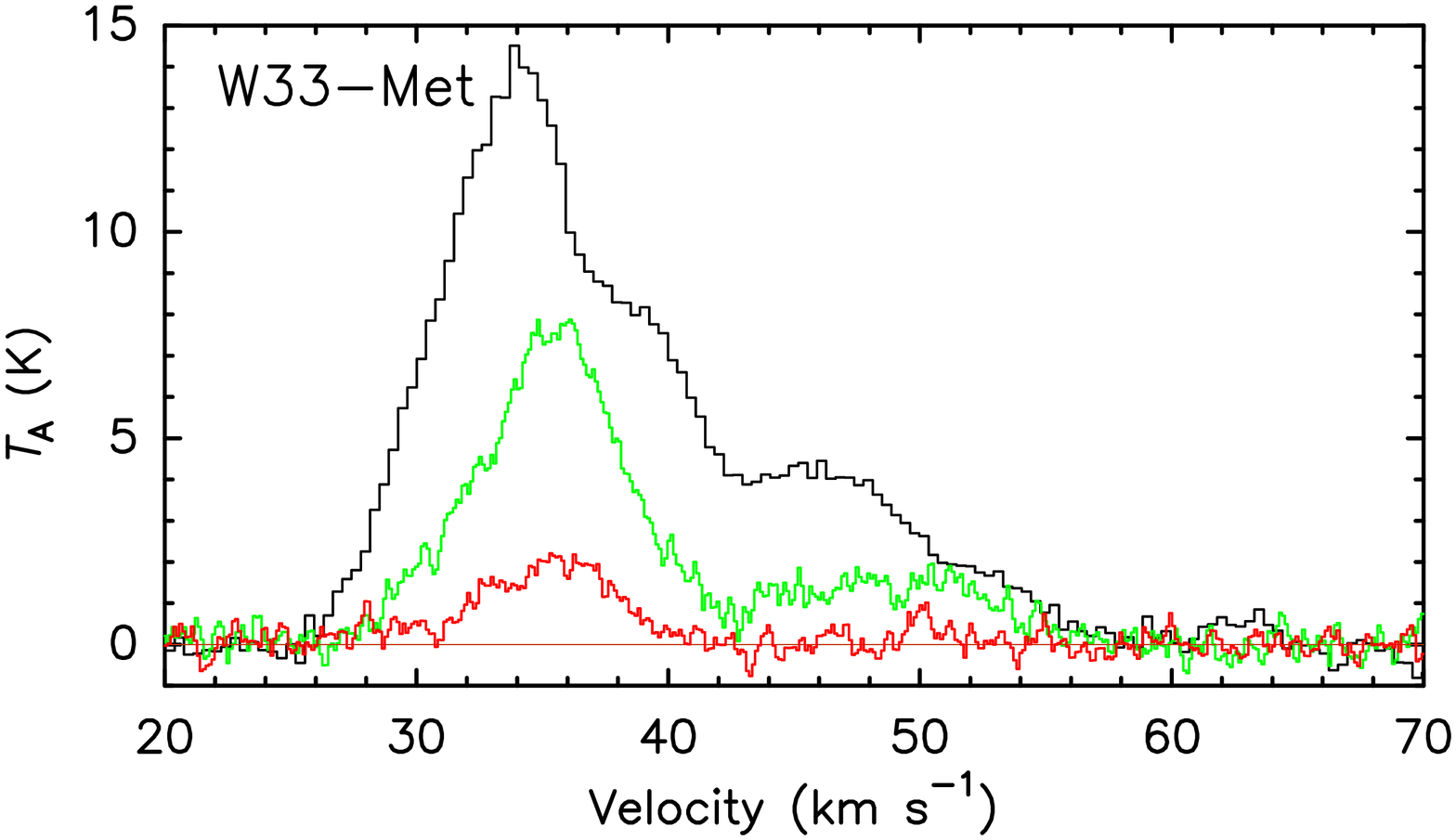}
\end{minipage}
\begin{minipage}[c]{0.5\textwidth}
  \centering
  \includegraphics[width=30mm,height=65mm,angle=-90.0]{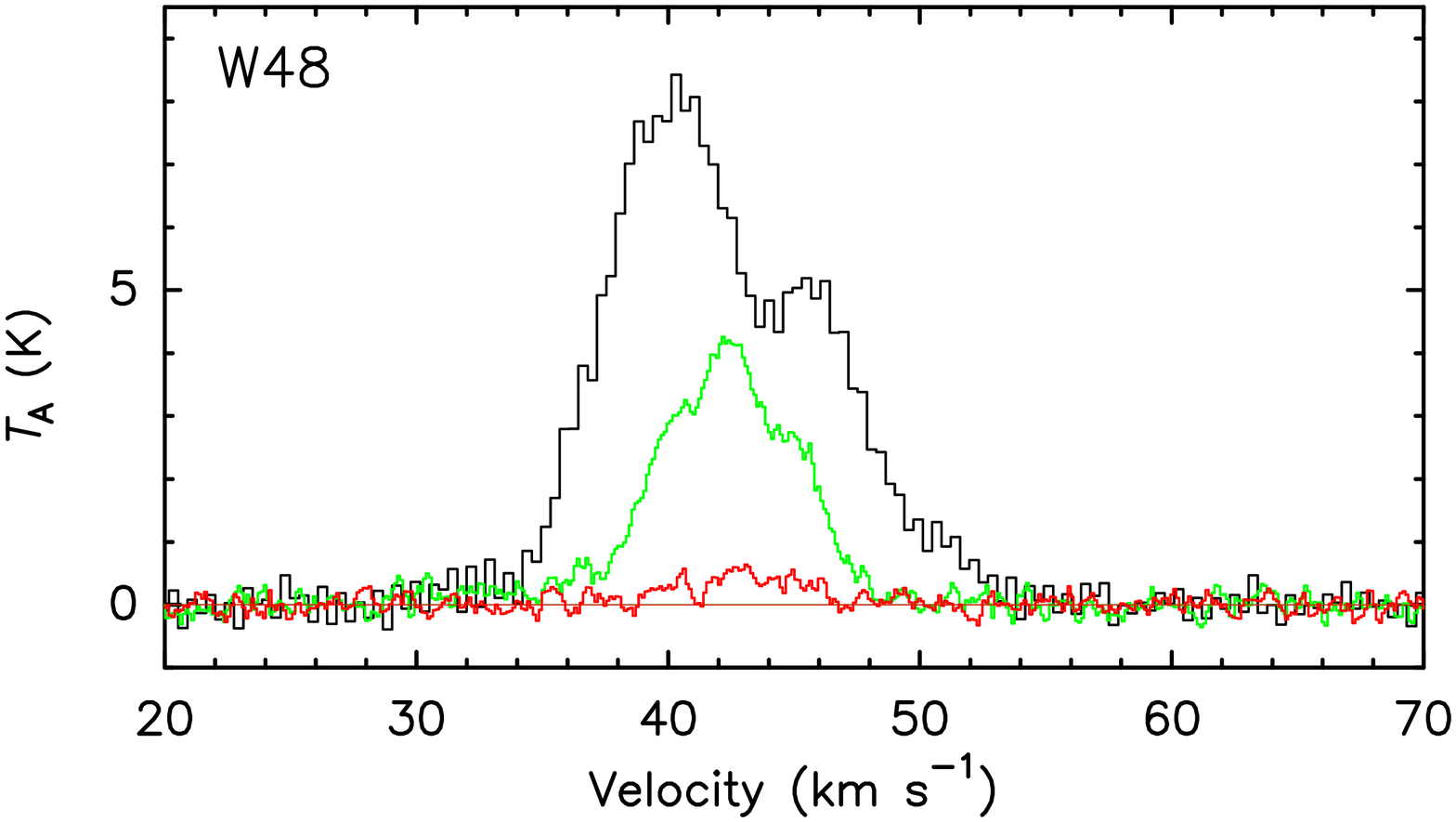}
\end{minipage}
\begin{minipage}[c]{0.5\textwidth}
  \centering
  \includegraphics[width=30mm,height=65mm,angle=-90.0]{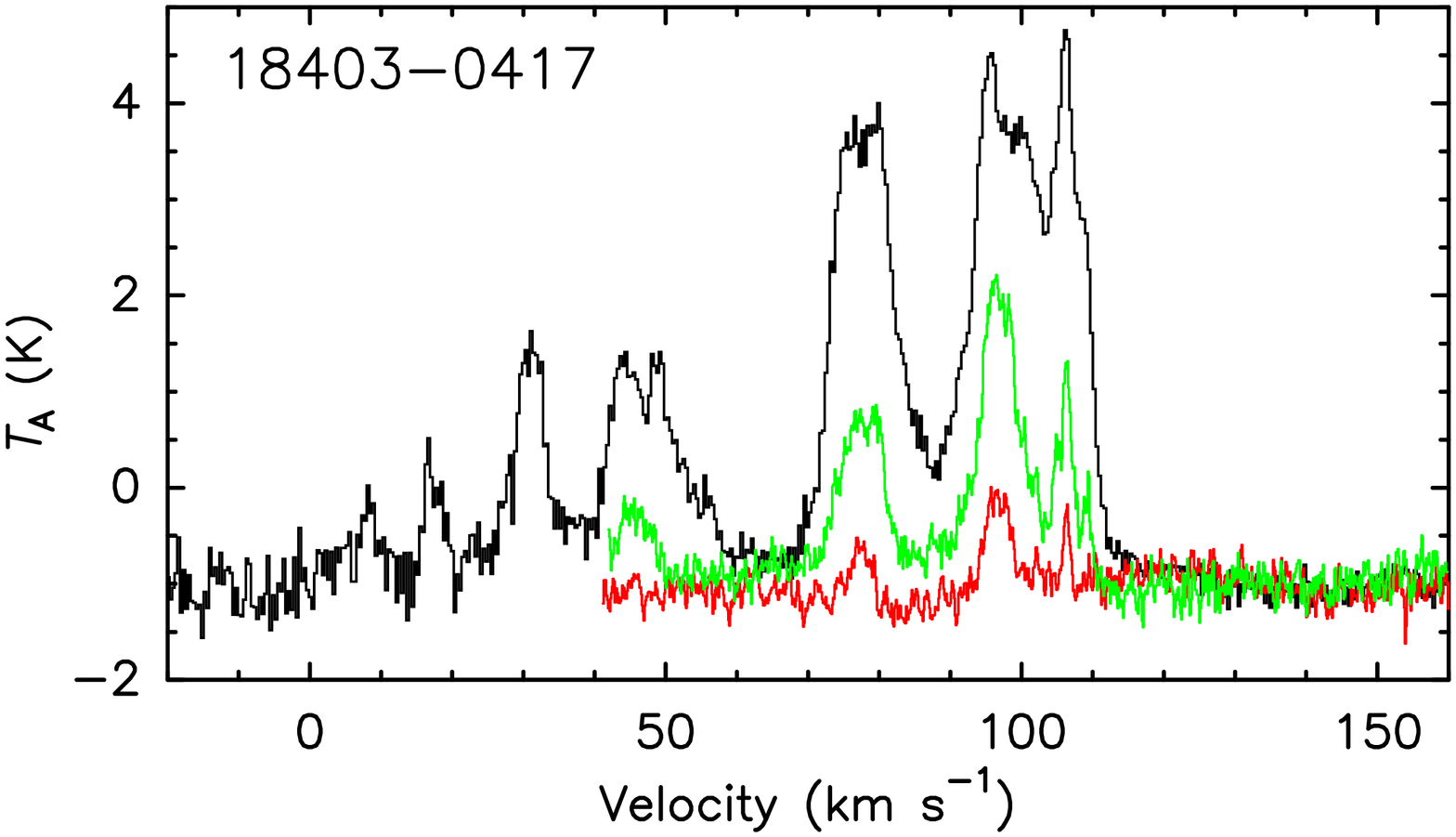}
\end{minipage}
\caption{Continued }
\end{figure}

\begin{figure}
  \centering
  \includegraphics[width=10cm]{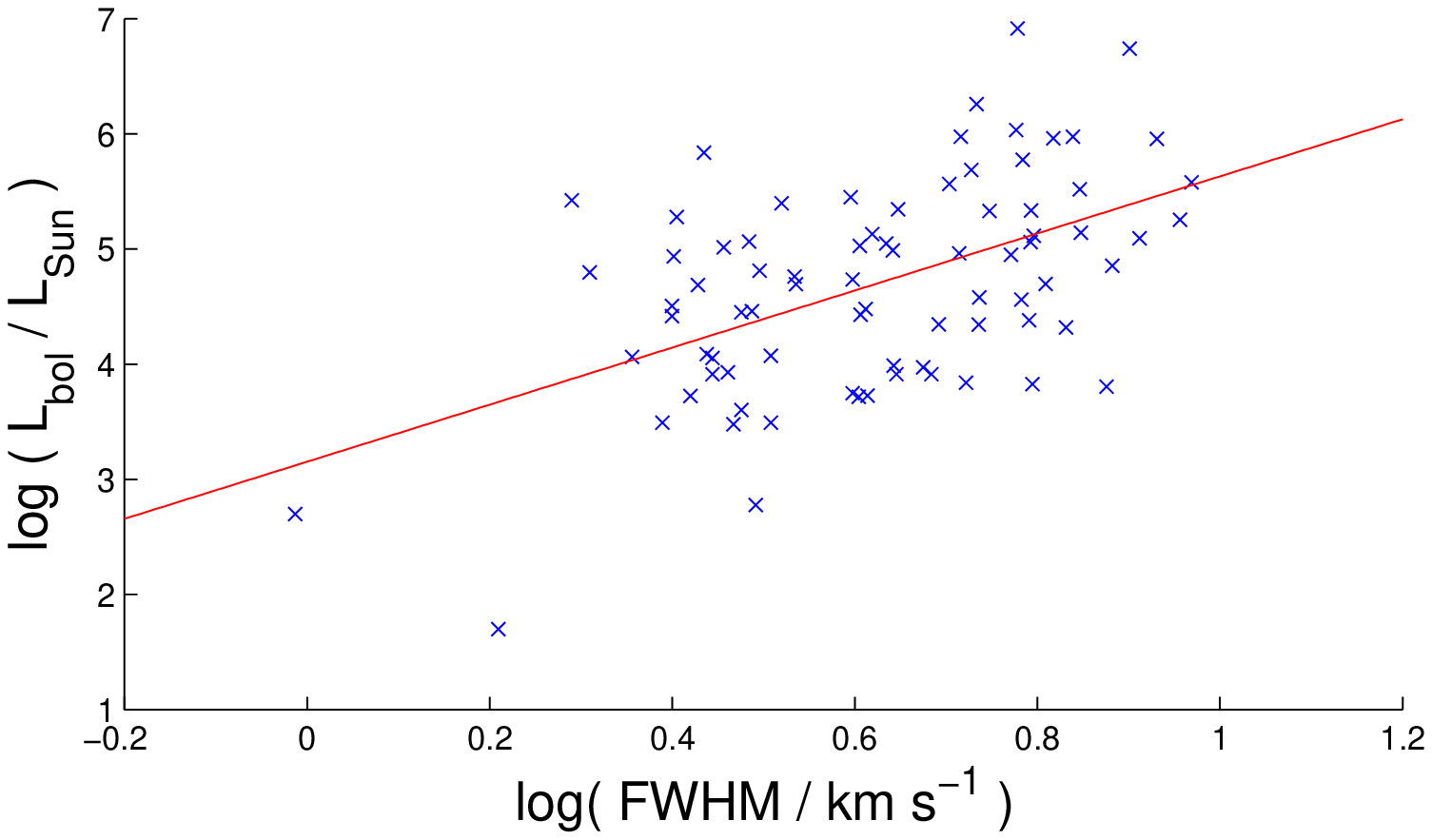}
\caption{The bolometric luminosity ($L_{bol}$) versus $^{13}$CO line
width. The linear fit of $L_{bol}$ and FWHM is:
$log(L_{bol}/L_{\odot})=2.5log(FWHM/km s^{-1})+3.2$; the correlation
coefficient r=0.52. }
\end{figure}

%
% two figures side-by-side with independent captions using package graphicx
%----------------------------------------------------- Figs 9 & 10:
%\vspace{-42.308mm}
\begin{figure}[h]
  \begin{minipage}[t]{0.495\linewidth}
  \centering
  \includegraphics[width=100mm,height=150mm,angle=-90.0]{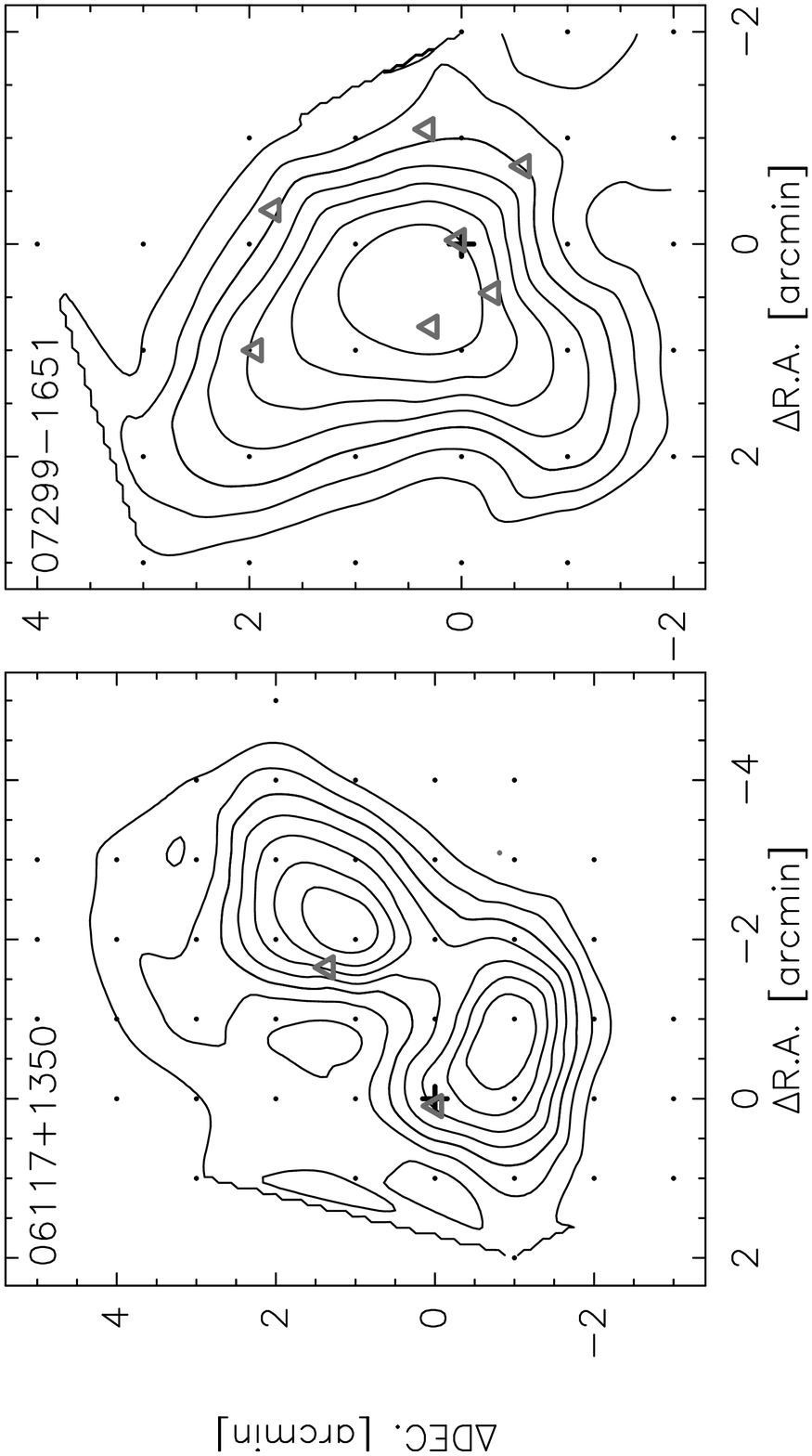}
  \vspace{-5mm}
  \end{minipage}%
  \label{Fig:fig23}
  \caption{{\small The contours of the integrated
intensities of $^{13}$CO lines. Left: IRAS 06117+1350, Right: IRAS
07299-1651. The positions of IRAS sources are marked with "+" and
the MSX sources are visualized with triangles.}}
\end{figure}
%% This demo can be adjusted to configure a figure with caption on left or right

\begin{table}[h]
  \caption[]{ Parameters of all the sources surveyed}
%%Please Capitalize the First Letter of Each Notional Word in table's caption
  \label{Tab:paraout}
  \begin{tiny}
  \begin{center}\begin{tabular}{ccccccccccc}
  \hline\noalign{\smallskip}
  Name &IRAS No. & $\alpha$(B1950)         & $\delta$(B1950)                            & l         & b          & log                          & log                         & Flux100  &  Type  &   Remark   \\
       &         & (h m s)                 & ($^{\circ}$ $^{\prime}$ $^{\prime\prime}$) & $^{\circ}$ & $^{\circ}$ & ($F_{25}~/~F_{12}$)  & ($F_{60}~/~F_{12}$) & (Jy)      &        &   Ref.    \\
  (1)  &(2)      & (3)                     & (4)                                        & (5)        & (6)        & (7)                          & (8)                         & (9)      &  (10) &   (11)  \\
\hline\noalign{\smallskip}
02455+6034                  &  02455+6034                 &  02 45 30.1                &  +60 34 34.5                              &  136.84       &   1.14     &   0.67                         &   1.51                      &    $3.395\times10^{2}$    &II   & Szymczak et al.~(\cite{szy00})   \\
05274+3345                & 05274+3345                  & 05 27 27.59             &  +33 45 37.32                                  &  174.20       & -0.08      &  1.00                         &    1.81                      &    $9.057\times10^{2}$       & II   & Szymczak et al.~(\cite{szy00})   \\
Orion's                      &  05327-0529                 &  05 32 44.8                &  -05 26 00.0                              &  209.01       & -19.41     &   1.03                         &   2.15                      &    $2.448\times10^{1}$    &I        & Haschick et al.~(\cite{has89})   \\
06056+2131                  &  06056+2131                 &  06 05 40.9                &  +21 31 32.4                              &  189.03       &   0.79     &   0.70                         &   1.55                      &    $2.563\times10^{3}$    &II       & Szymczak et al.~(\cite{szy00}) \\
G188.9+0.9                  &  06058+2138                 &  06 05 54.0                &  +21 39 09.4                              &  188.95       &   0.89     &   1.00                         &   1.83                      &    $1.666\times10^{3}$     &I       & Haschick et al.~(\cite{has89})   \\
06099+1800                  &  06099+1800                 &  06 09 57.90               &  +18 00 11.95                             &  192.60       & -0.05      &  0.54                          &   1.47                      &    $5.285\times10^{3}$      &II     & Szymczak et al.~(\cite{szy00})   \\
06117+1350                  &  06117+1350                 &  06 11 46.4                &  +13 50 32.7                              &  196.45       &  -1.68     &   0.90                         &   1.89                      &    $1.965\times10^{3}$    &II       & Szymczak et al.~(\cite{szy00})   \\
07299-1651                  &  07299-1651                 &  07 29 55.0                &  -16 51 47.2                              &  232.62       &   1.00     &   1.67                         &   2.37                      &    $1.269\times10^{3}$    &II       & Szymczak et al.~(\cite{szy00})   \\
G8.68-0.37                & 18032-2137                  & 18 03 21.69             &  -21 37 42.27                                  &  8.68         & -0.36      &  0.91                         &    2.00                       &   $5.221\times10^{3}$      &II      & Caswell et al.~(\cite{cas93})   \\
G10.47+0.03                 &  18056-1952                 &  18 05 40.54               &  -19 52 25.30                             &  10.47        &  0.03      &  1.13                          &   2.67                      &    $1.016\times10^{4}$     &II       & Caswell et al.~(\cite{cas93})   \\
G10.30-0.15                 &  18060-2005                 &  18 05 57.93               &  -20 06 25.58                             &  10.30        & -0.15      &  0.87                          &   1.63                      &    $1.201\times10^{4}$      &I     & Bachiller et al.~(\cite{bac90})   \\
18067-1927                & 18067-1927                  & 18 06 44.00             &  -19 27 02.90                                  &  10.96        & 0.02       &  0.87                            &  2.10                      &   $1.292\times10^{3}$        & II  & Szymczak et al.~(\cite{szy00})   \\
G10.6-0.4                   &  18075-1956                 &  18 07 30.6                &  -19 56 28.3                              &   10.62       &  -0.38     &   0.80                         &   2.61                    &      $2.137\times10^{4}$    &I        & Haschick et al.~(\cite{has89})   \\
G12.89+0.49               & 18089-1732                  & 18 08 56.49             &  -17 32 14.46                                  &  12.89        & 0.49       &  0.87                           &  2.31                      &    $3.148\times10^{3}$    &I           & Slysh et al.~(\cite{sly94})   \\
G12.03-0.04               & 18090-1832                  & 18 09 05.59             &  -18 32 44.17                                  &  12.02        & -0.03      &  0.71                         &    1.63                      &    $1.710\times10^{3}$    & II       & Caswell et al.~(\cite{cas93})   \\
18092-1842                & 18092-1842                  & 18 09 13.78             &  -18 42 19.77                                  &  11.90        & -0.14      &  0.10                            &  1.46                     &    $1.710\times10^{3}$    & II  & Szymczak et al.~(\cite{szy00})   \\
G11.90-0.14               & 18092-1842                  & 18 09 14.78             &  -18 42 20.84                                  &  11.90        & -0.14      &  0.10                         &    1.46                       &   $1.710\times10^{3}$    & II   & Caswell et al.~(\cite{cas93})   \\
18108-1759                & 18108-1759                  & 18 10 52.41             &  -17 59 35.92                                  &  12.71        & -0.13      &  0.38                         &    1.88                       &   $2.878\times10^{3}$    & II  & Szymczak et al.~(\cite{szy00})   \\
G12.68-0.18               & 18112-1801                  & 18 11 00.55             &  -18 02 47.51                                  &  12.68        & -0.19      &  0.64                         &    1.77                       &   $6.502\times10^{3}$    & II   & Caswell et al.~(\cite{cas93})   \\
G11.94-0.62                 &  18110-1854                 &  18 11 04.4                &  -18 54 19.8                              &   11.94       &  -0.62     &   1.21                         &   2.19                     &     $4.930\times10^{3}$    &I   & Bachiller et al.~(\cite{bac90})   \\
W33-Met                     &  18108-1759                 &  18 11 15.78               &  -17 56 52.62                             &  12.80        & -0.19      &  0.38                          &   1.88                     &     $2.878\times10^{3}$     &I   & Haschick et al.~(\cite{has90})   \\
W33A                        &  18117-1753                 &  18 11 44.56               &  -17 52 55.71                             &  12.91        & -0.26      &  1.11                          &   2.02                     &     $6.310\times10^{3}$     &I   & Haschick et al.~(\cite{has89})   \\
G12.91-0.26               & 18117-1753                  & 18 11 45.36             &  -17 53 09.77                                  &  12.91        & -0.27      &  1.11                         &    2.02                       &   $6.310\times10^{3}$    & II   & Caswell et al.~(\cite{cas93})   \\
18128-1640                & 18128-1640                  & 18 12 51.82             &  -16 40 00.54                                  &  14.10        & 0.09       &  0.38                        &     1.48                       &   $1.017\times10^{3}$    & II  & Szymczak et al.~(\cite{szy00})   \\
18144-1723                  &  18144-1723                 &  18 14 29.81               &  -17 23 22.70                             &  13.66        & -0.60      &  0.56                           &  1.61                     &     $1.137\times10^{3}$     & II & Szymczak et al.~(\cite{szy00})   \\
G14.33-0.64                 &  18160-1647                 &  18 16 00.9                &  -16 49 06.3                              &   14.33       &  -0.65     &   0.94                          &  2.24                     &     $2.819\times10^{3}$    &I       & Slysh et al.~(\cite{sly94})   \\
M17(3)                      &  18174-1612                 &  18 17 31.03               &  -16 12 49.82                             &  15.04        & -0.67      &  0.89                          &   1.71                     &     $6.589\times10^{4}$    &I   & Bachiller et al.~(\cite{bac90})   \\
G15.03-0.68                 &  18174-1612                 &  18 17 31.72               &  -16 13 06.87                             &  15.03        & -0.68      &  0.89                           &  1.71                     &     $6.589\times10^{4}$    &II    & Caswell et al.~(\cite{cas93})   \\
18181-1534                  &  18181-1534                 &  18 18 06.73               &  -15 34 36.38                             &  15.66        & -0.50      &  -0.26                          &  1.34                     &     $6.839\times10^{2}$      &II & Szymczak et al.~(\cite{szy00})   \\
G16.59-0.06               & 18182-1433                  & 18 18 20.41             &  -14 33 18.33                                  &  16.59        & -0.06      &  1.15                        &     2.23                      &    $1.071\times10^{3}$    & I      & Slysh et al.~(\cite{sly94})   \\
18220-1241                & 18220-1241                  & 18 22 02.53             &  -12 41 00.36                                  &  18.66        & 0.03       &  1.06                        &     2.23                      &    $1.095\times10^{3}$    & II  & Szymczak et al.~(\cite{szy00})  \\
G20.24+0.07                 &  18249-1116                 &  18 24 57.15               &  -11 16 48.95                             &  20.24        & 0.07       &  0.24                         &    1.53                     &     $3.696\times10^{2}$      & II & Caswell et al.~(\cite{cas93})   \\
18249-1116                & 18249-1116                  & 18 24 57.16             &  -11 16 41.95                                  &  20.24        & 0.07       &  0.24                        &     1.53                      &    $3.696\times10^{2}$     & II & Szymczak et al.~(\cite{szy00})   \\
18278-1009                & 18278-1009                  & 18 27 49.62             &  -10 09 19.38                                  &  21.56        & -0.03      &  0.51                       &      1.14                      &    $2.284\times10^{2}$      & II& Szymczak et al.~(\cite{szy00})   \\
G22.34-0.16               & 18297-0931                  & 18 29 47.29             &  -09 31 25.86                                  &  22.34        & -0.16      &  -0.26                       &     0.48                      &    $2.823\times10^{2}$     & II & Szymczak et al.~(\cite{szy02})   \\
G22.43-0.17               & 18297-0931                  & 18 29 59.07             &  -09 27 31.70                                  &  22.42        & -0.17      &  -0.26                       &     0.48                      &    $2.823\times10^{2}$     & II  & Caswell et al.~(\cite{cas93})   \\
18316-0602                  &  18316-0602                 &  18 31 39.0                &  -06 02 07.8                              &   25.65       &   1.05     &   0.78                        &    1.62                     &     $2.136\times10^{3}$    & II     & Slysh et al.~(\cite{sly99})   \\
18317-0859                & 18317-0859                  & 18 31 44.83             &  -08 59 41.32                                  &  23.04        & -0.34      &  0.18                         &    1.77                      &    $1.940\times10^{2}$      & II& Szymczak et al.~(\cite{szy00})   \\
G23.01-0.41               & 18318-0901                  & 18 31 56.26             &  -09 03 15.14                                  &  23.01        & -0.41      &  0.70                         &    2.50                      &    $4.393\times10^{3}$     & II  & Caswell et al.~(\cite{cas93})   \\
G23.01-0.41               & 18318-0901                  & 18 31 56.76             &  -09 03 18.18                                  &  23.01        & -0.42      &  0.70                         &    2.50                      &    $4.393\times10^{3}$     & I     & Slysh et al.~(\cite{sly94})   \\
18322-0721                & 18322-0721                  & 18 32 13.56             &  -07 21 42.32                                  &  24.54        & 0.31       &  -0.13                        &    1.25                     &     $2.935\times10^{2}$       & I& Szymczak et al.~(\cite{szy00})   \\
18326-0751                & 18326-0751                  & 18 32 38.38             &  -07 51 13.13                                  &  24.15        & -0.01      &  0.79                         &    2.09                      &    $5.740\times10^{2}$       & I& Szymczak et al.~(\cite{szy00})   \\
18335-0713                & 18335-0713                  & 18 33 30.13             &  -07 13 09.83                                  &  24.81        & 0.10       &  0.96                            & 1.98                      &    $6.936\times10^{3}$    & II  & Szymczak et al.~(\cite{szy00})   \\
MMI-117                     &  18335-0714                 &  18 33 30.36               &  -07 14 41.97                             &  24.79        &  0.08      &  0.96                          &   1.98                     &     $6.936\times10^{3}$     &I    & Val'tts et al.~(\cite{val07})  \\
18353-0628                  &  18353-0628                 &  18 35 23.8                &  -06 28 07.0                              &   25.70       &   0.03     &   0.81                          &  1.59                     &     $2.435\times10^{3}$   &II    & Szymczak et al.~(\cite{szy00})   \\
MMI-119                     &  18379-0546                 &  18 37 58.3                &  -05 46 28.0                              &   26.61       &  -0.22     &   0.91                          &  1.86                     &     $5.343\times10^{2}$   &I      & Val'tts et al.~(\cite{val07})  \\
MMI-121                     &  18391-0504                 &  18 39 11.55               &  -05 04 24.47                             &  27.37        & -0.16      &  -0.15                          &  0.59                     &     $1.633\times10^{3}$     &I    & Val'tts et al.~(\cite{val07})       \\
G28.85+0.50                 &  18396-0322                 &  18 39 35.13               &  -03 27 23.91                             &  28.85        & 0.50       &  0.37                          &   1.13                     &     $4.073\times10^{2}$ &II  & Blaszkiewicz \& Kus~(\cite{bla04})        \\

\noalign{\smallskip}\hline
  \end{tabular}\end{center}\end{tiny}
\end{table}

\setcounter{table}{0}

\begin{table}[h]
  \caption[]{ Continued.}
%%Please Capitalize the First Letter of Each Notional Word in table's caption
  \label{Tab:paraout}
  \begin{tiny}
  \begin{center}\begin{tabular}{ccccccccccc}
  \hline\noalign{\smallskip}
  Name&IRAS No. & $\alpha$(B1950)         & $\delta$(B1950)                            & l         & b          &  log                          & log                         & Flux100  &  Type  &   Remark   \\
      &         & (h m s)                 & ($^{\circ}$ $^{\prime}$ $^{\prime\prime}$) & $^{\circ}$& $^{\circ}$ &  ($F_{25}~/~F_{12}$)  & ($F_{60}~/~F_{12}$) & (Jy)      &        &   Ref.    \\
  (1) &(2)      & (3)                     & (4)                                        & (5)       & (6)        &  (7)                          & (8)                         & (9)      &  (10) &   (11)  \\
\hline\noalign{\smallskip}
G28.15+0.00               & 18402-0418                  & 18 40 02.45             &  -04 18 20.91                                &  28.15     & 0.00     &    0.33                      &      2.80           & $5.877\times10^{2}$                 & II  & Szymczak et al.~(\cite{szy02})        \\
G28.53+0.12               & 18403-0354                  & 18 40 19.50             &  -03 55 00.11                                &  28.53     & 0.12     &    -0.39                     &      1.04           & $2.437\times10^{2}$                 & II  & Szymczak et al.~(\cite{szy02})        \\
18403-0417                & 18403-0417                  & 18 40 19.58             &  -04 17 01.13                                &  28.20     &-0.05     &    0.91                      &      1.86           & $3.937\times10^{3}$                 & II  & Szymczak et al.~(\cite{szy00})        \\
18416-0420                  &  18416-0420                 &  18 41 39.81               &  -04 20 59.88                           &  28.29     &-0.38      &   0.95                       &     1.57          &  $4.358\times10^{3}$     & II              & Szymczak et al.~(\cite{szy00})   \\
G28.83-0.25              & 18421-0349                  & 18 42 11.63             &  -03 49 01.14                                &   28.83     &-0.25     &    1.04                      &      2.11           & $1.877\times10^{3}$                  & II  & Caswell et al.~(\cite{cas93})        \\
G29.86-0.05                 &  18434-0242                 &  18 43 24.69               &  -02 48 40.33                           &  29.86     &-0.05        & 0.89                       &     1.54          &  $1.167\times10^{4}$    & II               & Szymczak et al.~(\cite{szy02})        \\
G29.95-0.02                 &  18434-0242                 &  18 43 27.0                &  -02 42 45.5                            &   29.95    & -0.02       &  0.89                      &     1.54          &  $1.167\times10^{4}$    & II                & Caswell et al.~(\cite{cas93})        \\
G30.69-0.06               & 18449-0207                  & 18 44 58.94             &  -02 04 27.03                                &  30.70     &-0.06        & 0.64                      &      2.26          &  $4.025\times10^{3}$                     & I  & Slysh et al.~(\cite{sly94})        \\
MMI-125                  & 18449-0207                  & 18 44 58.96             &  -02 04 26.97                                &   30.70     &-0.06        & 0.64                      &      2.26           & $4.025\times10^{3}$                   & I  & Val'tts et al.~(\cite{val07})       \\
MMI-127                     &  18449-0115                 &  18 44 59.2                &  -01 16 10.6                            &   31.41    &  0.31       &  1.11                      &     2.43          &  $2.815\times10^{3}$    & I                 & Val'tts et al.~(\cite{val07})       \\
G30.8-0.1                   &  18449-0158                 &  18 45 11.06               &  -01 57 56.89                           &  30.82     &-0.06        & 0.80                       &     1.81          &  $1.908\times10^{4}$     & I               & Haschick et al.~(\cite{has89})        \\
G31.28+0.06               & 18456-0129                  & 18 45 36.80             &  -01 29 51.71                                &  31.28     &0.06         & 1.26                      &      2.34           & $3.693\times10^{3}$                  & II  & Caswell et al.~(\cite{cas93})        \\
18456-0129                & 18456-0129                  & 18 45 39.70             &   -01 29 48.92                               &  31.29     &0.05         & 1.26                      &      2.34           & $3.693\times10^{3}$                 & II  & Szymczak et al.~(\cite{szy00})        \\
G32.74-0.07?                &  18487-0015                 &  18 48 47.51               &  -00 15 48.25                           &  32.74     &-0.08        & 0.36                       &     1.89          &  $1.071\times10^{3}$     & II               & Caswell et al.~(\cite{cas93})        \\
18497+0022                &  18497+0022                 &  18 49 46.43              &   +00 22 05.59                             &  33.41     &-0.002       & 0.37                       &     1.61          &  $2.064\times10^{3}$     &II               & Szymczak et al.~(\cite{szy00})       \\
G33.74-0.15               & 18509+0035                  & 18 50 53.88             &  +00 35 14.80                                &  33.74     &-0.15        & 0.19                      &      2.68           & $1.757\times10^{2}$                 & II  & Szymczak et al.~(\cite{szy02})        \\
G33.64-0.21               & 18509+0027                  & 18 50 55.54             &  +00 28 11.68                                &  33.64     &-0.21        & -0.09                         &  0.95           & $5.536\times10^{2}$             & II  & Blaszkiewicz \& Kus.~(\cite{bla04})        \\
G35.05-0.52                 &  18546+0139                 &  18 54 37.21               &  +01 35 01.02                           &  35.05     &-0.52        & -0.26                          & 0.19          &  $1.487\times10^{3}$     & I              & Bachiller et al.~(\cite{bac90})        \\
G35.79-0.17               & 18547+0223                  & 18 54 45.12             &  +02 23 41.49                                &  35.79     &-0.17        & 0.11                         &   1.67           & $2.739\times10^{2}$                 & II  & Szymczak et al.~(\cite{szy02})       \\
G35.19-0.74                 &  18556+0136                 &  18 55 40.43               &  +01 36 33.55                           &  35.20     &-0.74        & 1.71                          &  2.66          &  $1.124\times10^{3}$     & II               & Caswell et al.~(\cite{cas93})        \\
G37.60+0.42                 &  18559+0416                 &  18 55 59.7                &  +04 16 26.3                            &   37.60    &  0.42       &  0.18                         &  1.36          &  $2.451\times10^{2}$    & II               & Szymczak et al.~(\cite{szy02})        \\
G37.02-0.03               & 18566+0330                  & 18 56 30.24             &  +03 33 30.11                                &  37.02     &-0.03        & -0.24                         &  0.24           & $1.553\times10^{2}$                 & II  & Szymczak et al.~(\cite{szy02})        \\
18577+0358                & 18577+0358                  & 18 57 45.20             &  +03 58 19.84                                &  37.53     &-0.11        & 0.86                         &   1.88           & $1.867\times10^{3}$             & II  & Blaszkiewicz \& Kus.~(\cite{bla04})        \\
G39.10+0.48               & 18585+0538                  & 18 58 32.19             &  +05 38 09.59                                &  39.10     &0.48         & -0.03                         &  1.57           & $1.734\times10^{2}$                 & II  & Szymczak et al.~(\cite{szy02})       \\
G35.20-1.73                 &  18592+0108                 &  18 59 12.72               &  +01 09 14.56                           &  35.20     &-1.73       &  0.95                          &  1.95          &  $1.396\times10^{4}$    & II                & Caswell et al.~(\cite{cas93})     \\
W48                         &  18592+0108                 &  18 59 13.82               &  +01 09 13.49                           &  35.20     &-1.74       &  0.95                          &  1.95          &  $1.396\times10^{4}$    &I                 & Haschick et al.~(\cite{has89})        \\
18592+0108                  &  18592+0108                 &  18 59 14.6                &  +01 08 46.4                            &   35.20    & -1.75      &   0.95                         &  1.95          &  $1.396\times10^{4}$    &II                & Szymczak et al.~(\cite{szy00})      \\
19031+0621                & 19031+0621                  & 19 03 09.87             &  +06 21 31.12                                &  40.27     &-0.20       &  1.26                         &   2.48           & $3.658\times10^{2}$                  &II  & Szymczak et al.~(\cite{szy00})        \\
19078+0901                  &  19078+0901                 &  19 07 51.8                &  +09 01 10.5                            &   43.17    &  0.00      &   1.01                         &  1.86          &  $3.615\times10^{4}$   &II                 & Szymczak et al.~(\cite{szy00})        \\
MMI-134                  & 19092+0841                  & 19 09 14.96             &  +08 41 34.73                                &   43.04     &-0.45       &  0.51                         &   1.82           & $5.105\times10^{2}$                   & I  & Val'tts et al.~(\cite{val07})       \\
19095+0930                & 19095+0930                  & 19 09 30.29             &  +09 30 41.71                                &  43.79     &-0.13       &  1.39                         &   2.51           & $2.744\times10^{3}$                 & II  & Szymczak et al.~(\cite{szy00})        \\
G43.8-0.1                 & 19099-0934                  & 19 09 31.26             &  -09 30 51.10                                &  26.83     &-8.88       &  0.14                            &  0.26         & $2.480\times10^{0}$                   &I  & Haschick et al.~(\cite{has89})        \\
MMI-138                     &  19120+1103                 &  19 12 04.6                &  +11 04 22.5                            &   45.47    &  0.05      &   0.91                         &  1.83           & $7.890\times10^{3}$    &I                  & Val'tts et al.~(\cite{val07})       \\
19186+1440                  &  19186+1440                 &  19 18 39.8                &  +14 40 58.0                            &   49.41    &  0.33      &   0.53                         &  1.76           & $1.710\times10^{2}$    &II                & Blaszkiewicz \& Kus.~(\cite{bla04})      \\
19211+1432                  &  19211+1432                 &  19 21 10.45               &  +14 32 18.69                           &  49.57     &-0.27       &  0.41                          &  1.42           & $6.230\times10^{1}$       &II             & Szymczak et al.~(\cite{szy00})        \\
MMI-143                  & 19213+1424                  & 19 21 19.66             &  +14 25 14.97                                 &  49.49     &-0.36       &  1.01                         &   -1.95         &  $2.676\times10^{4}$                   & I  & Val'tts et al.~(\cite{val07})      \\
G49.49-0.37                 & 19213+1424                    & 19 21 22.31             &  +14 25 07.88                            &  49.49     &-0.37       &  1.01                           & -1.95          & $2.676\times10^{4}$                  & II  & Caswell et al.~(\cite{cas93})        \\
19216+1429                  & 19216+1429                    & 19 21 37.3              &  +14 29 51.9                             &   49.59    & -0.39      &   0.96                          & 1.72           & $1.586\times10^{3}$                 & II  & Szymczak et al.~(\cite{szy00})        \\
19303+1651                  & 19303+1651                    & 19 30 20.3              &  +16 51 04.5                             &   52.66    & -1.09      &   0.64                          & 1.77           & $1.238\times10^{2}$             & II  & Blaszkiewicz \& Kus.~(\cite{bla04})        \\
19388+2357                  & 19388+2357                    & 19 38 52.7              &  +23 57 35.7                             &   59.83    &  0.67      &   0.93                          & 2.30           & $4.334\times10^{2}$                    & II  & Slysh et al.~(\cite{sly99})        \\
MMI-145                     & 19410+2336                    & 19 41 04.3              &  +23 36 42.0                             &   59.78    &  0.06      &   0.88                          & 1.83           & $1.631\times10^{3}$                   & I  & Val'tts et al.~(\cite{val07})       \\
20081+3122                  & 20081+3122                    & 20 08 09.9              &  +31 22 38.6                             &   69.54    & -0.98      &  1.72                         &   3.11           & $3.119\times10^{3}$                 & II  & Szymczak et al.~(\cite{szy00})        \\
ON1                         & 20081+3122                    & 20 08 09.92             &  +31 22 41.59                            &  69.54     &-0.98       &  1.72                           & 3.11           & $3.119\times10^{3}$                  & I  & Haschick et al.~(\cite{has90})        \\
20290+4052                  & 20290+4052                    & 20 29 03.1              &  +40 52 15.0                             &   79.74    &  0.99      &   0.98                          & 2.05           & $1.367\times10^{2}$             & II  & Blaszkiewicz \& Kus.~(\cite{bla04})        \\
20350+4126                  & 20350+4126                    & 20 35 04.9              &  +41 26 02.3                             &   80.87    &  0.42      &   1.07                          & 1.95           & $3.272\times10^{3}$                 & II  & Slysh et al.~(\cite{sly99})  \\
21074+4949                & 21074+4949                  & 21 07 28.47             &  +49 49 46.09                                &  90.92     &1.51        &  1.02                          &  2.25          &  $7.148\times10^{2}$                 & II  & Szymczak et al.~(\cite{szy00})        \\
GL2789                     & 21381+5000                    & 21 38 11.3              &  +50 00 45.3                             &   94.60    & -1.80      &   0.28                           &0.51           &  $4.546\times10^{2}$                    & II  & Slysh et al.~(\cite{sly99})        \\
R146                        & 21426+6556                    & 21 42 40.1              &  +65 52 56.9                             &  105.47    &  9.83      &  -0.17                           &0.74           & $2.119\times10^{1}$                 & I  & Kalenskii et al.~(\cite{kal94})       \\
S156A                      & 23030+5958                    & 23 03 05.03             &  +59 58 09.38                            &  110.11    & 0.04       &  0.81                            &1.59           &  $1.833\times10^{3}$                & II  & Bachiller et al.~(\cite{bac90})       \\
NGC7538S                    & 23116+6111                    & 23 11 36.11             &  +61 10 33.04                            &  111.53    & 0.76       &  0.87                            &1.46           & $1.414\times10^{4}$                  & I  & Haschick et al.~(\cite{has89})        \\
\noalign{\smallskip}\hline
  \end{tabular}\end{center}
  \end{tiny}
\end{table}

\begin{table}[h]
  \caption[]{ Sources with resolved components}
%%Please Capitalize the First Letter of Each Notional Word in table's caption
  \label{Tab:paraout}
  \begin{tiny}
  \begin{center}\begin{tabular}{cccccccccccc}
  \hline\noalign{\smallskip}
                              & $T_{A}^{\phantom{A}*}$      & $V_{LSR}$    & FWHM         &R                            &  D         &$L_{bol}$                      &  $T_{ex}$           & $\tau_{13}$                   & $N({}^{13}CO)$      &$N(H_{2})$                   & Pro\\
  Name    & (K)               & ($km s^{-1}$)& ($km s^{-1}$)&(kpc)  &(kpc)                 &$(L_{\odot})$         &  (K)       &                               & $(10^{16}cm^{-2})$  &$(10^{22}cm^{-2})              $&       \\
  (1)     & (2)               & (3)          &(4)           &(5)   &(6)                   &(7)                          & (8)        & (9)                           & (10)                &(11)                           &(12)\\
\hline\noalign{\smallskip}
02455+6034                    & 1.3(9)              & -42.51(0)        & 2.78(1)      & 12.19 & 4.52                 & $1.1\times10^{4}$    & 16.04                         & 0.22     &  0.9               &  0.8   & d  \\
                              & 1.2(8)              & -37.40(0)        & 2.78(3)      & 11.62 & 3.86                 & $8.2\times10^{3}$    & 13.22                         & 0.27     &  0.8               &  0.7   & d  \\
OrionS                        & 6.5(1)              &   7.80(4)        & 4.73(7)      & 9.39  & 1.06                 & $9.4\times10^{3}$    & 47.92                         & 0.26     &  14.2             &   1.2& e  \\
06056+2131                    & 5.8(0)              &   2.66(9)        & 2.99(5)      & 9.38  & 0.89                 & $4.0\times10^{3}$    & 19.99                         & 0.81     &  5.3             &    4.7  &    \\
G188.9+0.9                    & 4.4(4)              &   3.06(8)        & 3.22(6)      & 9.52  & 1.03                 & $3.1\times10^{3}$    & 18.78                         & 0.61     &  3.8             &    3.4  & a  \\
06117+1350                    & 2.1(5)              &  17.80(1)        & 3.42(8)      & 12.25 & 3.86                 & $5.8\times10^{4}$    & 20.52                         & 0.22     &  1.7             &    1.6  & f  \\
07299-1651                    & 3.4(7)              &  16.50(3)        & 2.89(3)      & 9.57  & 1.62                 & $8.5\times10^{3}$    & 15.04                         & 0.64     &  2.4             &    2.1  & a \\
G10.47+0.03                   & 2.8(1)              &  67.65(2)        & 9.31(2)      & 3.08  & 5.69                 & $3.8\times10^{5}$    & 13.18                         & 0.61     &  5.8             &    5.2  & g  \\
G10.6-0.4                     & 6.5(3)              &  -2.65(0)        & 6.00(7)      & 9.26  & 17.48                & $8.2\times10^{6}$    & 20.99                         & 0.89     &  12.7             &   11.4 & g  \\
                              & 1.4(3)              &  29.08(3)        & 5.05(9)      & 4.92  & 3.69                 & $3.7\times10^{5}$    & 7.77                          & 0.69     &  8.5             &    7.6  & d  \\
G11.94-0.62                   & 2.5(7)              &  35.78(9)        & 2.86(1)      & 4.72  & 3.94                 & $1.0\times10^{5}$    & 8.71                          & 1.39     &  2.0             &    1.8  & c,g  \\
                              & 2.6(6)              &  39.48(0)        & 3.05(2)      & 4.5   & 4.17                 & $1.2\times10^{5}$    & 7.51                          & 4.45     &  5.4             &    4.9  & g  \\
18144-1723                    & 0.5(0)              &  16.73(0)        & 6.23(0)      & 6.45  & 2.13                 & $6.7\times10^{3}$    & 9.92                          & 0.13     &  0.5               &  0.5   & g  \\
                              & 0.6(3)              &  41.65(7)        & 6.18(8)      & 4.68  & 4.04                 & $2.4\times10^{4}$    & 6.88                          & 0.32     &  0.7               &  0.6   &    \\
                              & 1.4(6)              &  48.15(8)        & 2.99(0)      & 4.36  & 4.39                 & $2.8\times10^{4}$    & 7.15                          & 0.89     &  1.0               &  0.9    &    \\
G14.33-0.64                   & 4.6(3)              &  21.52(3)        & 5.44(6)      & 6.1   & 2.51                 & $2.2\times10^{4}$    & 12.54                         & 1.61     &  8.3              &   7.4   & g  \\
G15.03-0.68                   & 8.1(9)              &  20.00(1)        & 5.34(6)      & 6.31  & 2.29                 & $4.9\times10^{5}$    & 32.03                         & 0.61     &  17.0             &   15.2 & e,g\\
18181-1534                    & 1.1(7)              &  -5.44(3)        & 3.31(2)      & 9.54  & 17.44                & $2.5\times10^{5}$    & 7.8                           & 0.52     &  0.7                & 0.7   & g  \\
                              & 2.6(2)              &  16.90(6)        & 2.45(5)      & 6.65  & 1.94                 & $3.1\times10^{3}$    & 10.13                         & 0.94     &  1.5             &    1.4  & e  \\
18316-0602                    & 3.4(7)              &  42.10(5)        & 4.09(5)      & 5.9   & 3.06                 & $3.0\times10^{4}$    & 11.16                         & 1.21     &  3.8             &    3.4  & e  \\
                              & 1.1(0)              &  47.59(4)        & 6.06(9)      & 5.66  & 3.36                 & $3.6\times10^{4}$    & 8.01                          & 0.46     &  1.2             &    1.1  & g  \\
18353-0628                    & 1.3(3)              &  54.12(6)        & 5.45(8)      & 5.41  & 3.7                  & $3.8\times10^{4}$    & 8.46                          & 0.52     &  1.4            &     1.2  & g  \\
                              & 1.3(4)              &  97.32(5)        & 5.18(4)      & 4.15  & 5.75                 & $9.2\times10^{4}$    & 9.31                          & 0.44     &  1.3            &     1.1  & f  \\
                              & 0.6(4)              & 105.88(4)        & 4.03(7)      & 3.97  & 6.19                 & $1.1\times10^{5}$    & 6.09                          & 0.43     &  0.5                & 0.4   & g  \\
MMI-119                       & 0.4(5)              &  99.18(8)        & 4.04(0)      & 4.18  & 5.87                 & $2.7\times10^{4}$    & 6.35                          & 0.26     &  0.3               &  0.3   & g   \\
                              & 1.0(2)              & 108.28(2)        & 2.51(4)      & 3.99  & 6.41                 & $3.2\times10^{4}$    & 6.18                          & 0.77     &  0.6               &  0.5   & g   \\
G29.95-0.02                   & 4.6(0)              &  98.68(2)        & 8.53(9)      & 4.43  & 6.1                  & $9.0\times10^{5}$    & 17.75                         & 0.71     &  10.6             &   9.5  & d  \\
MMI-127                       & 2.4(9)              &  96.94(9)        & 6.25(6)      & 4.56  & 6.17                 & $1.3\times10^{5}$    & 8.53                          & 1.39     &  4.3              &   3.8  & e  \\
G32.74-0.07                   & 0.8(7)              &  10.29(3)        & 0.97(5)      & 7.92  & 0.7                  & $0.5\times10^{3}$    & 6.15                          & 0.62     &  0.2              &   0.2   &    \\
                              & 1.2(9)              &  36.96(6)        & 5.27(2)      & 6.51  & 2.54                 & $6.9\times10^{3}$    & 7.59                          & 0.64     &  1.4             &    1.2  & g  \\
                              & 0.7(0)              &  78.24(1)        & 2.51(4)      & 5.1   & 4.94                 & $2.6\times10^{4}$    & 5.56                          & 0.61     &  0.4             &    0.3   & g  \\
G37.60+0.42                   & 1.5(0)              &  89.37(2)        & 2.27(0)      & 5.07  &                      &           & 6.55                           & 1.23    &  0.9            &     0.8   & c  \\
18592+0108                    & 2.3(4)              &  39.35(4)        & 2.54(0)      & 6.51  & 2.66                 & $1.9\times10^{5}$    & 13.53                         & 0.46     &  1.2              &   1.1  &    \\
                              & 3.8(6)              &  42.89(3)        & 4.44(0)      & 6.37  & 2.87                 & $2.2\times10^{5}$    & 13.53                         & 0.92     &  4.4              &   3.9  &    \\
19078+0901                    & 3.0(1)              &  12.04(5)        & 6.78(8)      & 7.97  & 0.76                 & $2.0\times10^{4}$    & 21.05                         & 0.32     &  5.1              &   4.6  & e  \\
                              & 2.3(1)              &   2.78(8)        & 7.96(5)      & 8.47  & 12.35                & $5.5\times10^{6}$    & 16.96                         & 0.32     &  4.0              &   3.6  & g  \\
MMI-138                       & 3.1(0)              &  59.12(4)        & 7.02(8)      & 6.19  & 4.71                 & $3.3\times10^{5}$    & 13.13                         & 0.7      &  5.0              &   4.5  & f  \\
19186+1440                    & 1.2(7)              & -21.10(7)        & 2.68(5)      & 9.93  & 13.08                & $4.9\times10^{4}$    & 8.89                          & 0.45     &  0.6               &  0.6   & g  \\
19216+1429                    & 0.8(5)              &  52.39(3)        & 4.38(8)      & 6.5   & 4.88                 & $9.8\times10^{4}$    & 7.79                          & 0.35     &  0.7             &    0.6 & g  \\
                              & 0.4(8)              &  61.16(1)        & 8.16(4)      & 6.24  &                      &           & 6.32                            & 0.28   &  0.7               &  0.6   &    \\
19303+1651                    & 1.2(8)              &  59.73(6)        & 3.96(1)      & 6.36  &                      &           & 8.76                             & 0.46  &  14.3            &    12.7 & b,i\\
19388+2357                    & 3.0(3)              &  34.50(6)        & 2.74(5)      & 7.26  &                      &           & 8.23                            & 3.34   &  4.3             &    3.8 &   \\
MMI-145                       & 4.9(2)              &  22.67(2)        & 3.22(4)      & 7.68  & 2.05                 & $1.2\times10^{4}$    & 18.37                         & 0.74     &  4.4              &   4.0  & a  \\
20081+3122                    & 4.2(8)              &  11.09(7)        & 4.02(3)      & 8.17  & 1.13                 & $5.2\times10^{3}$    & 11.77                         & 1.64     &  5.6              &   5.0  &   \\
20290+4052                    & 2.0(1)              &  -1.62(4)        & 2.93(8)      & 8.7   & 3.91                 & $3.0\times10^{3}$    & 9.12                          & 0.99     &  1.6             &    1.4  & g  \\
20350+4126                    & 3.5(0)              &  -2.87(0)        & 3.13(0)      & 8.75  & 3.83                 & $6.5\times10^{4}$    & 13.69                         & 0.96     &  3.3              &   3.0  & c  \\
GL2789                       & 2.8(7)              & -43.86(8)        & 2.52(5)      & 10.89 & 6.16                 &  $8.6\times10^{4}$    & 15                            & 0.6      &  1.9              &   1.7  & c  \\
R146                          & 3.0(0)              &  -9.73(9)        & 1.62(9)      & 9.07  &                      &           & 15.38                          & 0.61    &  1.3              &   1.2  &   \\
S156A                        & 3.0(6)              & -51.51(2)        & 4.16(0)      & 11.63 & 5.54                 &  $1.3\times10^{5}$    & 27.38                         & 0.26     &  4.3              &   3.8  & f  \\

\noalign{\smallskip}\hline
  \end{tabular}\end{center}\end{tiny}
  \tablecomments{0.86\textwidth}{a: {\sl wings}, b: {\sl red wing}, c: {\sl blue wing}, d: {\sl red
shoulder}; e: {\sl red asymmetry}, f: {\sl blue asymmetry}, g: {\sl
flat top}, h: {\sl two or three components}.}
\end{table}

\setcounter{table}{1}

\begin{table}[h]
  \caption[]{ Continued.}
%%Please Capitalize the First Letter of Each Notional Word in table's caption
  \label{Tab:paraout}
  \begin{tiny}
  \begin{center}\begin{tabular}{cccccccccccc}
  \hline\noalign{\smallskip}
                              & $T_{A}^{\phantom{A}*}$      & $V_{LSR}$    & FWHM         &R                            &  D         &$L_{bol}$                      &  $T_{ex}$           & $\tau_{13}$                   & $N({}^{13}CO)$      &$N(H_{2})$                   & Pro\\
  Name    & (K)               & ($km s^{-1}$)& ($km s^{-1}$)&(kpc)  &(kpc)                 &$(L_{\odot})$         &  (K)       &                               & $(10^{16}cm^{-2})$  &$(10^{22}cm^{-2})              $&       \\
  (1)     & (2)               & (3)          &(4)           &(5)   &(6)                   &(7)                          & (8)        & (9)                           & (10)                &(11)                           &(12)\\
\hline\noalign{\smallskip}
05274+3345                    &  5.1(9)                 & -3.66(5)     &  2.63(6)     &  10.41 & 1.92                & $5.3\times10^{3}$    & 16.42                         & 1      &  4.0              &  3.6 &   \\
06099+1800                    &  6.9(1)                 &  7.15(2)     &  3.07(6)     &  10.21 & 1.74                & $2.9\times10^{4}$    & 29.03                         & 0.56   &  7.5              &  6.7 &   \\
G8.68-0.37                    &  1.4(2)                 & 18.53(9)     &  2.04(6)     &  5.44  & 3.12                & $6.3\times10^{4}$    & 9.79                          & 0.43   &  0.5              &  0.5  &    \\
                              &  3.8(3)                 & 36.65(8)     &  7.04(4)     &  3.98  & 4.64                & $1.4\times10^{5}$    & 13.81                         & 0.87   &  6.7              &  6.1 & e  \\
G10.30-0.15                   &  3.7(1)                 & 12.86(8)     &  5.90(6)     &  6.43  & 2.12                & $8.9\times10^{4}$    & 14.94                         & 0.71   &  5.4              &  4.8 & b  \\
G12.03-0.04                   &  0.9(3)                 & 98.05(1)     &  3.43(3)     &  2.62  & 6.38                & $5.0\times10^{4}$    & 6.77                          & 0.53   &  0.6              &  0.6   & g   \\
                              &  1.7(3)                 &110.61(5)     &  3.96(9)     &  2.41  & 6.69                & $5.4\times10^{4}$    & 9.48                          & 0.58   &  1.4              &  1.2 & g  \\
W33-Met                       &  6.9(7)                 & 35.43(3)     &  6.44(0)     &  4.89  & 3.78                & $5.0\times10^{4}$    & 19.27                         & 1.2    &  15.7             &  14.0& g  \\
                              &  1.1(2)                 & 48.56(1)     &  7.62(6)     &  4.2   & 4.53                & $7.2\times10^{4}$    & 8.61                          & 0.41   &  1.5              &  1.4 &    \\
M17(3)                        &  9.1(3)                 & 19.98(6)     &  5.41(0)     &  4.12  & 4.42                & $1.8\times10^{6}$    & 32.58                         & 0.69   &  20.1             &  18.0& e  \\
G20.24+0.07                   &  1.6(2)                 & 71.06(4)     &  4.39(8)     &  4.35  & 4.77                & $9.7\times10^{3}$    & 6.65                          & 1.35   &  2.0              &  1.8 & a  \\
MMI-117                       &  3.4(9)                 &109.22(3)     &  6.20(4)     &  3.83  & 6.32                & $1.2\times10^{5}$    & 11.64                         & 1.1    &  5.7              &  5.0 & c  \\
G28.85+0.50                   &  1.4(0)                 & 85.13(2)     &  4.42(6)     &  4.68  & 5.2                 & $8.2\times10^{3}$    & 6.9                           & 0.92   &  1.4              &  1.3 & e  \\
18416-0420                    &  2.1(2)                 & 47.61(3)     &  4.31(8)     &  5.83  & 3.26                & $1.1\times10^{5}$    & 8.28                          & 1.11   &  2.2              &  2.0 & c  \\
                              &  1.5(7)                 & 84.78(5)     &  3.94(3)     &  4.65  & 5.17                & $2.8\times10^{5}$    & 6.3                           & 1.59   &  1.9              &  1.7 & a  \\
G29.86-0.05                   &  4.9(7)                 & 99.89(3)     &  5.98(3)     &  2.24  & 6.67                & $1.1\times10^{6}$    & 18.77                         & 0.72   &  8.3              &  7.5 & c  \\
G30.8-0.1                     &  3.6(8)                 & 95.93(4)     &  9.04(0)     &  4.55  & 5.99                & $1.8\times10^{5}$    & 14.82                         & 0.71   &  8.2              &  7.3 & g,a\\
                              &  1.3(9)                 &115.63(0)     &  1.95(7)     &  4.14  &                     &           & 8.46                        & 0.55     &  0.5              &  0.5  & c  \\
G35.19-0.74                   &  4.5(9)                 & 33.88(3)     &  4.92(2)     &  6.75  & 2.31                & $2.2\times10^{4}$    & 13.24                         & 1.34   &  6.8              &  6.1 & g\\
G39.10+0.48                   &  1.2(4)                 & 22.96(1)     &  3.10(0)     &  7.36  & 1.55                & $0.6\times10^{3}$    & 6.65                          & 0.84   &  0.9              &  0.8  & e,g  \\
G35.20-1.73                   &  4.2(6)                 & 41.99(4)     &  5.59(8)     &  6.41  & 2.82                & $2.1\times10^{5}$    & 13.83                         & 1.04   &  6.5              &  5.8 &  f  \\
W48                           &  3.8(6)                 & 42.37(7)     &  6.21(5)     &  6.39  & 2.84                & $2.2\times10^{5}$    & 13.34                         & 0.95   &  6.2              &  5.6  & d,f  \\
19211+1432                    &  2.0(4)                 & 52.32(3)     &  7.51(4)     &  6.5   & 4.85                & $6.4\times10^{3}$    & 11.96                         & 0.47   &  3.1              &  2.8 & c,e\\
                              &  0.5(5)                 & 63.10(1)     &  4.83(4)     &  6.19  &                     &           & 7.01                          & 0.26   &  0.5              &  0.4  &   \\
G49.49-0.37                   &  5.2(5)                 & 50.31(9)     &  6.08(8)     &  6.56  & 4.37                & $5.9\times10^{5}$    & 15.62                         & 1.14   &  9.7              &  8.6 & g  \\
                              &  7.6(0)                 & 60.61(9)     &  6.90(8)     &  6.26  &                     &           & 28.42                          & 0.66  &  19.0             &  17.0& g  \\
                              &  2.5(3)                 & 68.13(2)     &  5.20(8)     &  6.05  &                     &           & 13.14                          & 0.53  &  2.8              &  2.5 & g  \\
ON1                           &  4.3(3)                 & 11.13(4)     &  4.11(5)     &  8.17  & 1.14                & $5.3\times10^{3}$    & 11.9                          & 1.63   &  5.8              &  5.2 & e  \\
NGC7538S                      &  5.0(0)                 &-56.14(8)     &  6.57(9)     &  12.05 & 5.98                & $9.1\times10^{5}$    & 27.4                          & 0.48   &  12.2             &  11.0& f  \\
                              &  1.4(7)                 &-48.95(9)     &  2.72(0)     &  11.47 & 5.19                & $6.9\times10^{5}$    & 15.19                         & 0.26   &  0.9              &  0.8  & g  \\
\noalign{\smallskip}\hline
  \end{tabular}\end{center}\end{tiny}
\end{table}

%________________________________________ Table 1: paraout
\begin{table}[h]
  \caption[]{ Physical parameters of the $^{13}CO$ Molecular Cores}
%%Please Capitalize the First Letter of Each Notional Word in table's caption
  \label{Tab:paraout}
  \begin{tiny}
  \begin{center}\begin{tabular}{ccccccccccc}
  \hline\noalign{\smallskip}
  \ Name & $\alpha$(B1950)         & $\delta$(B1950)                                                     & $V_{LSR}$                           & FWHM                                                &D                                               &$T_{ex}$      & R        & N($H_{2}$)               & $M_{core}   $       & $M_{vir}$       \\
  \            & (h m s)                 & ($^{\circ}$ $^{\prime}$ $^{\prime\prime})$                          & $(km s^{-1})$                       & $(km s^{-1})$                                       &(kpc)                                           &(K)           & (pc)     & $(10^{22}cm^{-2})$       & $(M_{\odot})$       & $(M_{\odot})$           \\
 \hline\noalign{\smallskip}
 06117+1350-SE             & 06 11 46.4              & +13 50 32.7                                          &  16.48(2)                            &    3.34(9)                                            &  3.86                                          &  18.47       & 1.45   & 2.1                 & $4.0\times10^{3}$                            & $3.4\times10^{3}$                       \\
 06117+1350-NW             & 06 11 46.4              & +13 50 32.7                                          &  16.68(3)                            &    2.95(7)                                            &  3.86                                          &  17.29       & 1.42   & 2.0                 & $3.6\times10^{3}$                            & $2.6\times10^{3}$                       \\
 07299-1651                & 07 29 55.0              & -16 51 47.2                                          &  16.50(3)                            &    2.89(3)                                            &  1.62                                          &  15.04       & 0.79   & 2.2                 & $1.2\times10^{3}$                            & $1.4\times10^{3}$                       \\
  \noalign{\smallskip}\hline
  \end{tabular}\end{center}\end{tiny}
\end{table}

\label{lastpage}


\begin{thebibliography}{99}
%% you can type \apj for ApJ, \aap for A&A, \apss for Ap&SS, etc. Please consult
%% the macro raa.cls. You can also find them in aasguide.tex (AASTeX for ApJ, AJ, PASP)
%% Please follow the formats of RAA's references list as demonstrated below:



\bibitem[1990]{bac90} Bachiller R., Menten, K. M., G\'{o}mez-Gonz\'{a}lez, J., Barcia A.,
1990, A\&A, 240, 116

\bibitem[2004]{bla04} Blaszkiewicz, L., \& Kus, A. J., 2004, A\&A, 413, 233

%\bibitem[1996]{bro96} Bronfman L., Nyman L.-oA., May J., 1996, \aaps, 115, 81

\bibitem[1986]{cas86} Casoli, F., Dupraz, C., Gerin, M., Combes, F., Boulanger, F., 1986, A\&A,169,281

\bibitem[1993]{cas93} Caswell, J. L., Gardner, F. F., Norris, R. P., Wellington, K. J., McCutcheon, W. H., Peng, R. S., 1993, MNRAS, 260, 425

\bibitem[1991]{gar91} Garden, R. P., Hayashi, M., Gatley, I., Hasegawa, T., Kaifu, N., 1991,
\apj, 374, 540

\bibitem[2000]{gui00} Guilloteau, S. \& Lucas, R., 2000, in Imaging at Radio through
Submillimeter Wavelengths, eds. J. G. Mangum., \& S. Radford, ASP
Conf. Ser., 217, 299

\bibitem[1989]{has89} Haschick, A. D., \& Baan, W. A., 1989, \apj, 339, 949

\bibitem[1990]{has90} Haschick, A. D., Menten, K. M., \& Baan, W. A., 1990,
\apj, 354, 556

\bibitem[1994]{kal94} Kalenskii, S. V., Liljestr\"{o}m, T., Val'tts, I. E., Vasil'kov, V. I.,
Slysh, V. I., Urpo, S., 1994, A\&AS, 103, 129

\bibitem[2003]{min03} Minier, V., Ellingsen, S, P., Norris, R. P., Booth, R. S., 2003, \aap, 403,
1095

\bibitem[1983]{mye83} Myers, P., Linke, R. A., Benson, P. J., 1983, \apj, 264, 517

\bibitem[1992]{plu92} Plume, R., Jaffe, D. T., Evans, N. J. II, 1992, \apjs ,78,505

\bibitem[2009]{pur09} Purcell, C. R., Longmore, S. N., Burton, M. G., Walsh, A. J., Minier, V.,
 Cunningham, M. R. \& Balasubramanyam, R., 2009, MNRAS, 394, 323

\bibitem[1987]{shu87} Shu, F. H., Adams, F. C., \& Lizano, S., 1987, AR\aap, 25, 23

\bibitem[1994]{sly94} Slysh, V. I., Kalenskii, S. V., Val'tts, I. E., Otrupcek, R., 1994,
MNRAS, 268, 464

\bibitem[1999]{sly99} Slysh, V. I., Val'tts, I. E., Kalenskii, S. V., Voronkov, M. A.,
Palagi, F., Tofani, G., Catarzi, M., 1999, A\&AS, 134, 115

\bibitem[2005]{sob05} Sobolev, A. M., et al, 2006, arxiv:astro-ph/0601260

\bibitem[2000]{szy00} Szymczak, M., Hrynek, G., Kus A. J., 2000,
A\&AS, 143, 269

\bibitem[2002]{szy02} Szymczak, M., Kus, A. J., Hrynek, G., Kepa, A., Pazderski, E., 2002,
A\&A, 392, 277


\bibitem[2000]{ung00} Ungerechts, H., Umbanhowar, P., Thaddeus, P., 2000, \apj, 537, 221

\bibitem[2007]{val07} Val'tts, I. E., \& Condon, G. M., 2007, \textit{Astronomy reports}, 51,
519

\bibitem[2009]{Wan09} Wang K., Wu Y., Ran L., Yu W., Miller M., 2009, \aap, 507,369

%\bibitem[1998]{wal98} Walsh, A. J., Burton, M. G., Hyland, A. R., Robinson, G., 1998,
MNRAS, 301, 640

\bibitem[1989]{woo89} Wood, D. O. S., \& Churchwell, E., 1989, \apj, 340, 265

\bibitem[2003]{wu03} Wu, Y., Wu, J., Wang, J., 2003, Chinese Physics Letters, 20,
1409

\bibitem[2004]{wu04} Wu, Y., Wei, Y, Zhao, M., Shi, Y., Yu, W., Qin, S., Huang, M., 2004, \aap, 426,
503

\bibitem[2006]{wu06} Wu, Y., Zhang, Q., Yu, W., Miller, M., Mao, R., Sun, K., Wang, Y., 2006,
\aap, 450, 607


\end{thebibliography}
\end{document}